\documentclass[letterpaper,11pt,fleqn]{article}
\pdfoutput=1
\usepackage{jheppub}

\usepackage{graphicx}
\usepackage[figuresright]{rotating}
\usepackage{bm,amsmath,amssymb}
\usepackage[mathscr]{eucal}
\usepackage{color}
\usepackage{multirow}
\usepackage{subcaption}

\long\def\comment#1{ }
\newcommand{\eqn}[1]{Eq.~\eqref{#1}}
\newcommand{\beq}{\begin{equation}}
\newcommand{\eeq}{\end{equation}}
\newcommand{\nn}{\nonumber\\}

\newcommand{\dif}{{\rm{d }}}
\newcommand{\del}{\partial}
\newcommand{\ktmin}{k_{\perp,\textrm{min}}}
\newcommand{\ombr}{\omega_{\textrm{br}}}
\newcommand{\as}{\alpha_s}
\newcommand{\abar}{\bar{\alpha}_s}
\newcommand{\amed}{{\alpha}_{s,\text{med}}}

\def\om{\omega}
\def\qhat{\hat{q}}
\def\th{\theta}

\newcommand{\rme}{{\rm e}}

\newcommand{\rmI}{{\rm I}}

\usepackage{color}
\definecolor{darkgreen}{rgb}{0,0.5,0}
\definecolor{darkblue}{rgb}{0,0,0.7}
\definecolor{darkred}{rgb}{0.5,0,0.0}
\definecolor{darkorange}{rgb}{0.8,0.4,0.0}

\makeatletter
\g@addto@macro\bfseries{\boldmath}
\makeatother

\title{Nuclear modification factors for jet fragmentation}

\author[a]{P.~Caucal,}
\author[a]{E.~Iancu,}
\author[b]{A.H.~Mueller}
\author[a]{and G.~Soyez}
\affiliation[a]{Institut de Physique Th\'{e}orique, Universit\'{e} Paris-Saclay, CNRS, CEA, F-91191, Gif-sur-Yvette, France}
\affiliation[b]{Department of Physics, Columbia University, New York, NY 10027, USA}

\emailAdd{paul.caucal@ipht.fr}
\emailAdd{amh@phys.columbia.edu}
\emailAdd{edmond.iancu@ipht.fr}
\emailAdd{gregory.soyez@ipht.fr}

\abstract{Using a recently-developed perturbative-QCD approach for jet
  evolution in a dense quark-gluon plasma, we study the nuclear
  modification factor for the jet fragmentation function.
  The qualitative behaviour that we find is in agreement with the
  respective experimental observations in Pb+Pb collisions at the LHC:
  a pronounced nuclear enhancement at both ends of the spectrum.
  Our Monte Carlo simulations are supplemented with analytic
  estimates which clarify the physical interpretation of the results.
  The main source of theoretical uncertainty is the sensitivity of our
  calculations to a low-momentum cutoff which mimics confinement. To
  reduce this sensitivity, we propose a new observable, which
  describes the jet fragmentation into subjets and is
  infrared-and-collinear safe by construction. We present Monte Carlo
  predictions for the associated nuclear modification factor together
  with their physical interpretation.  }

\begin{document}
\maketitle

\section{Introduction}

One important source of information about the dense partonic matter
--- the {\it quark-gluon plasma} --- created in the intermediate
stages of ultrarelativistic heavy ion collisions at RHIC and the LHC
comes from studies of jets propagating through this dense medium and
of the associated modifications of the jet structure and
properties. Generically known as ``jet quenching'', these
modifications cover a large variety of phenomena and observables, from
more inclusive ones, like the energy loss by the jet (measured e.g.\
by the nuclear modification factor $R_{AA}$), to more detailed ones
which probe the pattern of the in-medium jet fragmentation
(e.g. jet-substructure observables and the fragmentation function) or
the medium response to the jet (which influences the jet shapes).

On the theory side, various approaches and physical scenarios have been proposed. 
They generally adopt a perturbative QCD (pQCD) picture for
the high-virtuality part of the parton showers, but differ in their
treatment of the interactions between the jet and the medium, and of
the medium itself.
Even the approaches assuming a
weak QCD coupling throughout most stage do still involve some non-perturbative aspects, 
like the geometry of the medium and of the interaction region, or the transition from partonic 
to hadronic degrees of freedom at very low virtualities. Besides, there are several pQCD-based
approaches, which differ in their assumptions about the dominant medium effects and the
best-suited approximation schemes. Notable differences concern the
description of the medium-induced
radiation --- triggered by the collisions between the partons in the jet and those in the medium --- 
and its interplay with the vacuum-like parton branchings triggered by the  virtualities.

It is therefore crucial to identify observables which probe different
aspects of the in-medium dynamics and can thus be used to test the
physical ingredients and assumptions underlying the various
theoretical scenarios. In this paper, we focus on one such observable,
the nuclear modification of the jet fragmentation function, for which
there are interesting data at the LHC \cite{Aaboud:2018hpb}, but few
dedicated conceptual studies (see
however~\cite{Spousta:2015fca,Casalderrey-Solana:2016jvj,Tachibana:2017syd,KunnawalkamElayavalli:2017hxo,Chen:2017zte,Casalderrey-Solana:2018wrw,Casalderrey-Solana:2019ubu}).
The theoretical framework that we
use to address this (and related) observable(s) is the pQCD approach recently developed
in Refs.~\cite{Caucal:2018dla,Caucal:2019uvr}, in which vacuum-like emissions (VLEs) and
medium-induced emissions (MIEs) are factorised from each other via controlled approximations
at weak coupling. This simple description is manifestly probabilistic, hence
allowing for an efficient Monte-Carlo implementation.
In~\cite{Caucal:2019uvr}, we already successfully applied it to two observables measured at the LHC: 
the jet $R_{AA}$ (the nuclear modification factor for inclusive jet production) 
and the $z_g$-distribution (reflecting the jet substructure in terms of relatively hard splittings). 
 
At a first sight, the fragmentation function looks like an ideal
observable to study the jet structure in terms of parton showers and
its modifications by the interactions with the medium.
Indeed, the experimental results~\cite{Aaboud:2018hpb} in PbPb
collisions at the LHC show an interesting pattern with a strong
nuclear enhancement of the jet fragmentation into hadrons visible at
both ends of the spectrum, that is, at both small $x\ll 1$ and largish
$x\gtrsim 0.5$
(with $x\simeq p_T/p_{T,\textrm{jet}}$ the longitudinal momentum
fraction of a hadron inside the jet.)
One should however be cautious as the jet fragmentation function is
not a well-defined (``infrared and collinear safe'') quantity in pQCD.
This means that  its theoretical
predictions are strongly sensitive to non-perturbative (confinement)
physics like the modelling of the hadronisation mechanism.

Another potential drawback of the fragmentation function, already
recognised in the
literature~\cite{Spousta:2015fca,Casalderrey-Solana:2018wrw}, is that the
nuclear enhancement seen in the LHC data at $x\gtrsim 0.5$ is not
necessarily an evidence for new physics in the jet fragmentation at
large $x$, but merely a consequence of the overall energy loss by the
jet together with the bias introduced by the initial spectrum for jet
production via hard (nucleon-nucleon) scatterings. In that sense, the
physics of the in-medium jet fragmentation at large $x$ is strongly
correlated with that of the jet $R_{AA}$ --- a correlation that we
confirm in this paper.

The small-$x$ part of the in-medium fragmentation
  function is further affected by the fact that, in practice, one cannot 
distinguish the soft hadrons produced by the fragmentation of the jet itself from those from the
medium which are dragged by the wake of the jet and are co-moving with it.
This effect, know as the ``medium response'' should be included in any
realistic theoretical comparisons with the data at small $x$
(see
e.g.~\cite{Casalderrey-Solana:2016jvj,Tachibana:2017syd,KunnawalkamElayavalli:2017hxo,Chen:2017zte}).
This is however not the case of our current framework in which the
medium is simply described as a "brick" with a uniform
value for the jet quenching parameter $\hat q$, the rate for
transverse momentum broadening via elastic collisions.
The absence of hadronisation in our framework further limits our
accuracy in the small-$x$ region, even though this can to some extend
be probed by varying the transverse momentum cut-off of our partonic
cascade.
In view of these limitations, our current study should be viewed as
merely exploratory and we shall not perform a direct comparison
between our results and the data~\cite{Aaboud:2018hpb} for the nuclear
modification of the jet fragmentation.


Despite these simplifications, one should still hope
  that our framework captures (most of) the qualitative features of
  the nuclear effects on the jet fragmentation and, in particular,
  those that are mainly driven by the medium effects included in our
  parton showers.
 The results for the nuclear modifications of the fragmentation
function that we obtain in this paper are indeed encouraging.
They show that despite the large uncertainties associated with the
poorly-controlled soft-physics effects, one can still use this
observable for physical considerations and provide a physical
interpretation of some of their dominant qualitative features.

First of all, we find that our Monte Carlo results for the nuclear effects on the
jet fragmentation function show the same qualitative behaviour as the respective 
LHC data~\cite{Aaboud:2018hpb}.
Furthermore, the relative simplicity of our approach allows us to
present semi-analytic calculations, based on piecewise approximations,
which clarify the physical interpretation of the Monte Carlo results.
We are thus able to identify the various physical mechanisms
contributing to a given nuclear effect --- say, the enhancement in the
nuclear fragmentation function at small $x$ --- and quantify their
relative importance.

Our physical picture at weak coupling includes three main
medium-induced phenomena, all originating from multiple elastic
collisions off the medium constituents: transverse momentum
broadening, medium-induced radiation, and colour decoherence.
These phenomena lead to a variety of physical effects.
For instance, the energy lost by a jet is associated with
soft gluons which, after being produced via medium-induced multiple
branchings, are deviated at angles larger than the jet radius by
elastic collisions.  Vice-versa, the relatively hard medium-induced
emissions propagate at small angles, inside the jet, and hence
contribute to the final jet multiplicity, both directly and
indirectly via their subsequent radiations.
The analytic calculations in this paper, supported by numerical tests,
show that these phenomena are differently probed by the jet
fragmentation at small and large $x$.

The interplay between the various phenomena is often subtle. For example, one may
think that the nuclear enhancement observed in the jet fragmentation function at small $x$
is due to the copious production of soft gluons via medium-induced emissions.
This is however not right since the soft gluons produced
(via MIEs) inside the medium are efficiently deflected outside the jet
by elastic collisions and hence cannot contribute to the jet
multiplicity.
In reality, the nuclear excess in the jet
multiplicity at small $x$ is a combined effect of two phenomena: the colour decoherence,
which opens the angular phase-space for radiation outside the medium, and the
presence of additional sources for this radiation, as represented by relatively hard, 
intra-jet, MIEs.

We similarly discuss nuclear effects on the jet fragmentation at large $x\gtrsim 0.5$.
This refers to jets which suffer relatively little evolution, so the leading parton is 
unambiguously identified in the final state. As recognised
in the literature \cite{Spousta:2015fca}, these are typically quark-initiated jets, which are less suppressed
by the dense medium than the gluon-initiated jets. This argument takes into account the total
energy loss by a jet together with the bias introduced by its production spectrum, 
but it ignores possible nuclear modifications in the fragmentation mechanism
itself. To clarify this point, we perform analytic studies of the in-medium jet
fragmentation near $x=1$. We identify several medium effects which compete
with each other. Notably, the two MIE effects already mentioned --- energy loss
at large angles via soft emissions and  energy redistribution inside the jet via 
semi-hard MIEs --- act in opposite directions and almost compensate each other,
except possibly at $x>0.9$.
We thus conclude that the strong nuclear enhancement seen
in the LHC data for the fragmentation function at large $x>0.5$ is not teaching us much about the
jet fragmentation, but only about the jet global energy loss  and its interplay with the bias
 introduced by the steeply-falling initial spectrum.


Although our qualitative description of the LHC data for the jet fragmentation function in
Pb+Pb collisions looks satisfactory, it would be still interesting to allow
for more precise, quantitative, comparisons between theory and data. Besides the
current intrinsic limitations of our approach --- which could, at
least in principle, be improved in the future ---, 
such comparisons are hindered by the infrared sensitivity of the
fragmentation function. Motivated by that, we propose a new,
infrared-and-collinear-safe, observable which is directly probing the
jet fragmentation.
Instead of counting the hadrons inside the jet (in bins of $x$), this new observable counts
the {\it primary subjets} --- i.e. the subjets generated by partons directly
emitted by the leading parton --- which are {\it hard enough}, in the sense of having a
sufficiently large transverse momentum w.r.t. their emitter. This observable lies
on the same footing as other, perhaps more familiar, observables associated with the
jet substructure, such as the $z_g$-distribution. We present our Monte Carlo predictions
for this new observables together with their physical interpretation.
The associated nuclear effects are rather pronounced and our respective predictions are
under control both qualitatively and quantitatively.

The paper is organised as follows: in section~\ref{sec:general} we
provide a brief reminder of our physical picture, introduced in
Refs.~\cite{Caucal:2018dla,Caucal:2019uvr}. Section~\ref{sec:MC} gives
our Monte Carlo results for the fragmentation function and discusses
the physical mechanisms at play. We give additional details and
perform semi-analytic calculations in section~\ref{sec:x=1} for the
fragmentation function at large $x$ and in section~\ref{sec:smallx}
for small $x$.
Section~\ref{sec:frag-subjets} introduces and discusses our new
observable based on subjets and section~\ref{sec:conclusions} concludes.

\section{General picture and its Monte Carlo implementation}\label{sec:general}

We first provide a brief reminder of the physical picture, and the
corresponding implementation as a Monte-Carlo parton shower, as introduced
in Refs.~\cite{Caucal:2018dla,Caucal:2019uvr}, that we need to discuss
our new results on nuclear effects for the fragmentation function.

In essence, our picture includes two types of radiation: standard
{\em vacuum-like emissions} (VLEs) triggered by the parton virtuality,
as well as {\em medium-induced emissions} (MIEs) triggered by
collisions between the high-energy partons and the quark-gluon plasma.
Our description is correct to double-logarithmic accuracy within perturbative QCD,
including running-coupling and hard-collinear (DGLAP-like) branchings
for the VLEs. We make the assumption of a fixed (non-expanding) medium
of length $L$. MIEs are treated as multiple BDMPS-Z-like branchings, with
a jet-quenching parameter $\hat{q}$ 
that is fixed in time.

In the double-logarithmic approximation, we have shown \cite{Caucal:2018dla,Caucal:2019uvr}
 that the partonic cascade can be factorised in three steps:
\begin{enumerate}
\item a pure vacuum-like cascade with emission {\em inside the
    medium}: these corresponds to emissions of angle $\theta$ and
  energy $\omega$ satisfying $\omega^3\theta^4>2\hat{q}$ and $\theta>\theta_c \equiv
  \frac{2}{\sqrt{{\hat q}L^3}}$; these emissions have a formation time
  $t_f=2/\omega\theta^2$ much smaller than the medium size $L$; 
\item each parton resulting from the above pure-VLE cascade travels
  through the medium over a distance of order $L$ and can thus source
  MIEs;
\item the resulting partons (VLEs from the first step and MIEs from
  the second step) are the source to another cascade of VLEs outside
  the medium, i.e.\ in the region $\omega\theta^2<2/L$.
  For each of these cascades, the first emission can occur at any
  angle (i.e.\ is not constrained by angular ordering), a consequence
  of the colour decoherence following the interactions with the medium
  \cite{MehtarTani:2010ma,MehtarTani:2011tz,CasalderreySolana:2011rz,Mehtar-Tani:2014yea}.
  \end{enumerate}

\begin{figure}[t] 
\centering
\hspace*{0.5cm}%
{\includegraphics[width=0.45\textwidth]{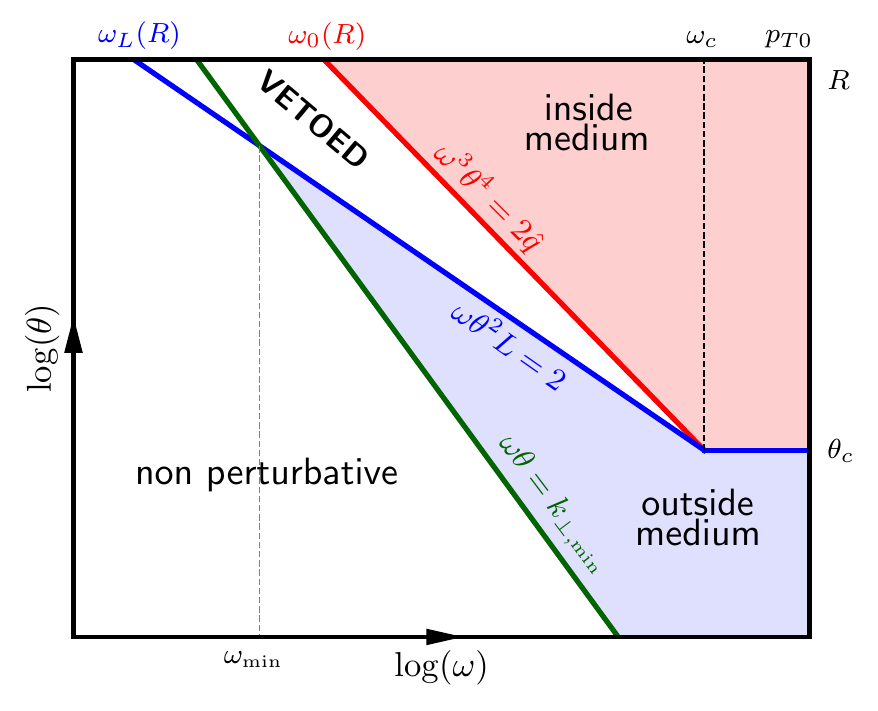}}\hfill%
\includegraphics[width=0.45\textwidth]{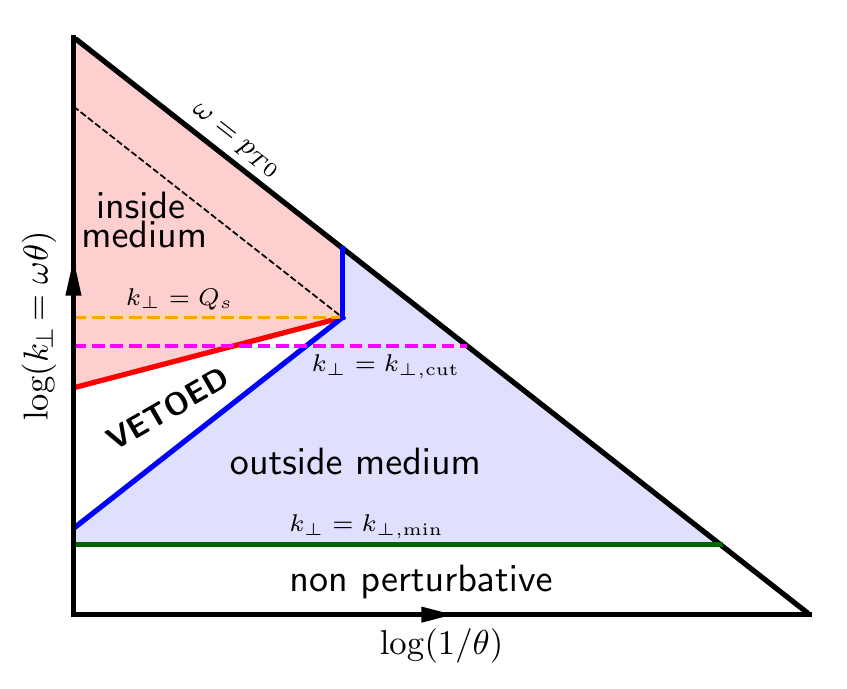}\hspace*{0.5cm}
\caption{\small The phase-space for vacuum-like gluon emissions by a jet propagating
through a dense QCD medium, in logarithmic units. In the left plot, the variables are the gluon energy $\omega$ and its emission angle $\theta$. In the right plot, we rather use the relative transverse momentum $k_\perp\simeq \omega\theta$ and the inverse of the angle $1/\theta$.}
 \label{Fig:LundPS}
\end{figure}

Our vacuum-like cascade is described as an angular-ordered shower,
starting from a maximal angle $\theta_\text{max}$ and keeping only
emissions with a relative transverse momentum w.r.t. their emitter
($k_\perp=\omega \theta$ for an emission of energy $\omega$ at an
angle $\theta$) above a cut-off $k_{\perp\text{min}}$.
For the third step of the factorised cascade, the first emission can
again happen up to angles $\theta_\text{max}$. 

To the accuracy of interest, the only medium effects on  the
VLEs occurring inside the medium can be formulated as kinematic boundaries
on the ($\om,\,\th$) phase-space.
This gives a vetoed region for VLEs which is represented pictorially
in Fig.~\ref{Fig:LundPS}.
Emissions with $\omega\theta^2>2/L$ and $\theta<\theta_c$ are formally
produced inside the medium but lose energy coherently with their
emitter \cite{MehtarTani:2010ma,MehtarTani:2011tz,CasalderreySolana:2011rz}.
 They can therefore be treated as if they happen outside the medium.

Medium-induced emissions can occur anywhere
inside the medium. They are generated with the following emission
rate \cite{Baier:2000sb,Jeon:2003gi,Blaizot:2013hx,Blaizot:2013vha}:
\begin{equation}\label{eq:BDPMS-rate}
  \frac{{\rm d}^2\Gamma_\text{med}}{{\rm d}z\,{\rm d}t}
  = \frac{\amed P(z)}{\sqrt{2}\pi}\frac{1}{t_\text{med}(x,z)},
\end{equation}
with $P(z)$ the splitting function and $t_\text{med}$ the formation time for
a MIE off a parent parton with energy $xE$. 
Both depend on the partonic channel under consideration. For, say, a $g\to gg$ channel one has
\begin{equation}\label{eq:tmed}
  t_\text{med}(x,z)
  =\,\sqrt\frac{2z(1-z)xE}{[1-z(1-z)]\hat{q}}\,
  \approx\, \sqrt{\frac{2zxE}{\hat{q}}},
\end{equation}
where the approximate equality holds for $z\ll 1$.
This spectrum is valid for soft emissions, $\omega<\omega_c$,
where $\omega=zxE$ is the energy of the emitted gluon and 
$\omega_c\equiv \hat{q}L^2/2$ is the most energetic such an emission,
corresponding to a formation time $t_\text{med}=L$.
Integrating~(\ref{eq:BDPMS-rate}) over a time of order $L$ we get the
BDMPS-Z spectrum for soft emissions 
\cite{Baier:1996kr,Baier:1996sk,Zakharov:1996fv,Zakharov:1997uu,Baier:1998kq}
\begin{equation}
\label{BDMPS}
\omega \frac{{\rm d}\mathcal{P}_\text{med}}{{\rm d}\omega} = \frac{\amed N_c}{\pi}
\sqrt{\frac{2\omega_c}{\omega}}\,\Theta(\omega_c-\omega)
\end{equation}
In our Monte Carlo simulations, the QCD coupling $\amed$ in
Eqs.~\eqref{eq:BDPMS-rate} and~\eqref{BDMPS} is kept
fixed.\footnote{On physical grounds, one expects that the right
  momentum scale for the running should be the transverse momentum
  $k_f^2=\sqrt{\hat q \om}$ acquired during formation. This energy
  dependence would complicate the MC implementation.}

After being produced at time $t$, MIEs propagate through the medium over
a distance $L-t$ and thus acquire a transverse momentum broadening via random
collisions. This is treated as a Gaussian distribution in $k_\perp$,
of width $\Delta k_\perp^2=\hat{q}(L-t)$.
A similar broadening applies to the VLEs, for which one can safely take $t\approx 0$ (since
$t\sim t_f\ll L$).

Physically, one can identify two main regimes in the cascade of MIEs:
\texttt{(i)} for $\omega_c\gg \omega \gg \omega_\text{br}\equiv (\amed
N_c/\pi)^2\omega_c$, the
probability for multiple emissions is small. This corresponds to
relatively rare semi-hard emissions at small angles (in particular at
angles which can remain inside a jet).
\texttt{(ii)} for $\omega\lesssim \omega_\text{br}$ multiple
branchings are important. This corresponds to a turbulent flow of soft
emissions at large angles (larger than the jet radius), which are the
main cause for energy loss by the jet \cite{Blaizot:2013hx,Blaizot:2013vha,Fister:2014zxa}.

In this picture, the energy lost by a jet is driven by two mechanisms:
first, the in-medium vacuum-like cascade creates a sequence of 
emissions within the jet, then, {\em each} of these emissions is the source of (soft)
MIEs with $\omega\lesssim\omega_\text{br}$ which propagate outside the jet.
The increase of the number of sources with the jet transverse momentum 
$p_{T,\textrm{jet}}$ is crucial for explaining
the almost-flat jet nuclear suppression factor $R_{AA}$ observed at 
high $p_{T,\textrm{jet}}$ at the LHC~\cite{Aaboud:2018twu}.

In fine, our Monte-Carlo for parton cascades in the medium contains
two ``non-physical'' parameters: $\theta_\text{max}$ which can be
viewed as an uncertainty on our collinear resummation, and
$k_{\perp\text{min}}$ which corresponds to a scale of order
$\Lambda_\text{QCD}$ (or $\sim 1$~GeV) at which hadronisation should
become important. It also has 3 ``physical'' parameters describing the
interaction with the medium: $\hat{q}$, $L$ and $\amed$. From these 3
parameters one can obtain the constants $\theta_c$ and $\omega_c$
(which, in particular, control the size of the veto region in
Fig.~\ref{Fig:LundPS}), and $\omega_{\text{br}}$ which control the
energy lost by a parton at large angles (and hence the jet energy
loss).

\begin{table}
  \centering
  \begin{tabular}{|l|ccc|ccc|}
    \hline
    & \multicolumn{3}{|c|}{parameters} 
    & \multicolumn{3}{|c|}{physics constants} \\
    \cline{2-7}
    Description
    & $\hat{q}$ [GeV$^2$/fm] & $L$ [fm] & $\amed$
    & $\theta_c$ & $\omega_c$ [GeV] & $\omega_\text{br}$ [GeV] \\
    \hline
    default
    & 1.5   & 4     & 0.24  & 0.0408 &  60 & 3.456 \\
    \hline
    & 1.5   & 3     & 0.35  & 0.0629 & 33.75 & 4.134 \\
    similar $R_{AA}$
    & 2     & 3     & 0.29  & 0.0544 & 45    & 3.784 \\
    & 2     & 4     & 0.2   & 0.0354 & 80    & 3.200 \\
       \hline
  \end{tabular}
  \caption{\small Table of medium parameters used in this paper. The default
    set of parameters is given in the first line. The next 3 lines
    are parameters which give a similar prediction for
    $R_\text{AA}$. The physics scales are defined as $\theta_c= {2}/{\sqrt{\hat q L^3}}$,
    $\om_c=\hat q L^2/2$, and $\ombr=\abar^2\om_c$, with $\abar=\alpha_{s,\textrm{med}}
    N_c/\pi$ and $N_c=3$.
    }\label{tab:parameters}
\end{table}

In Ref.~\cite{Caucal:2019uvr}, we found a series of parameters led to a
good description of the LHC data for the jet $R_{AA}$, as measured by
ATLAS~\cite{Aaboud:2018twu}. These parameters are listed in
Table~\ref{tab:parameters}.
It was also shown in~\cite{Caucal:2019uvr} that the above picture provides
a qualitatively-correct description of the $z_g$ distribution.

Our goal in this paper is to extend our study to the jet fragmentation
function.
The first set of parameters from Table~\ref{tab:parameters} will be our default choice throughout this
paper and the other three will be used to probe the sensitivity of the
fragmentation function to the medium parameters beyond what is
provided by the measurement of $R_{AA}$.

\section{Monte Carlo results for the in-medium fragmentation function}
\label{sec:MC}

In this section, we present our Monte Carlo results for the in-medium modification of the jet fragmentation function 
together with a discussion of their physical interpretation. This interpretation
is supported by the analytic calculations described in the next sections.

\subsection{Definitions and general set-up}
\label{sec:defin}

In order to describe $pp$ and PbPb collisions at the LHC, we consider
jets with an initial spectrum given by a $pp$ collision\footnote{For simplicity,
we have used the same hard-scattering spectrum for both
  the $pp$ baseline and the PbPb sample. This means that we neglect
  the effects of nuclear PDF, which can sometimes be as large as
  15-20 \% and can be added in a more phenomenologically-oriented
  study.} with centre-of-mass energy $\sqrt{s}=5.02$ TeV computed at
leading-order, i.e.\ with Born-level $2\to 2$ partonic hard
scatterings.
A key property of this initial parton (or dijet) spectrum is that it
is steeply falling with the partons'
transverse momentum $p_{T0}$:  $\dif N^{\textrm{hard}}/{\dif p_{T0}}\propto 1/p_{T0}^n$
with $n\gtrsim 5$. For each event, both final partons are showered using our Monte Carlo. 
Jets are reconstructed using the anti-$k_\perp$ algorithm~\cite{Cacciari:2008gp} as implemented in FastJet
v3.3.2~\cite{Cacciari:2011ma}.  The final jets are characterised by their transverse momentum 
$p_{T,\textrm{jet}}$, which is generally different from the initial momentum $p_{T0}$,
in particular for jets in PbPb collisions which suffer energy loss.
The $pp$ baseline is obtained by using the vacuum limit of our Monte Carlo.

We denote the final jet spectrum by
$\dif N_{\textrm{jets}}/{\dif p_{T,\textrm{jet}}}$ and use the upper
scripts ``med'' and ``vac'' to distinguish between jets in the medium
(PbPb collisions) and jets in the vacuum ($pp$ collisions),
respectively.  The jets can be initiated by either a quark or a
gluon. In practice, one often considers the jet yield integrated over
an interval in $p_{T,\textrm{jet}}$, that is,
\beq\label{pTrange}
N_{\textrm{jets}} (p_{T,{\rm min}}, p_{T,{\rm max}})\,=
\int_{p_{T,{\rm min}}}^{p_{T,{\rm max}}}\dif p_{T,\textrm{jet}}\,
\frac{\dif N_{\textrm{jets}}}{\dif p_{T,\textrm{jet}}}\,.
\eeq
For a given jet with transverse momentum $p_{T,\textrm{jet}}$, 
we characterise its fragmentation in terms of the longitudinal
momentum fraction
\begin{equation}\label{eq:def-x}
  x\equiv \frac{p_T\cos(\Delta R)}{p_{T,\textrm{jet}}},
\end{equation}
where $p_T$ is the transverse momentum of a constituent of the jet and
$\Delta R =\sqrt{(\Delta y)^2+(\Delta\phi)^2}$, with $\Delta y$ and
$\Delta\phi$ the differences between the jet axis and the particle
direction in rapidity and azimuth. Note that since our Monte Carlo
does not include hadronisation, the jet constituents are partons.

The jet fragmentation function $\mathcal{D}(x)$ 
and its nuclear modification factor $\mathcal{R}(x)$ are defined as
\begin{equation}
\label{frag-def}
 \mathcal{D}(x)=\frac{1}{N_{\textrm{jets}}}\frac{\dif N}{\dif x}\textrm{ , }\qquad 
 \mathcal{R}(x)=\frac{\mathcal{D}^{\textrm{med}}(x)}{\mathcal{D}^{\textrm{vac}}(x)}\,,
\end{equation}
with $N_{\textrm{jets}}$ the number of jets (in the considered
$p_{T,\textrm{jet}}$ range) and ${\dif N}/{\dif x}$ the number of
jet constituents with a given momentum fraction $x$. 

For later conceptual studies, we shall also consider ``monochromatic
jets'' produced by a well identified parton, quark or gluon, with a fixed initial transverse
momentum $p_{T0}$. In such a case, we denote the fragmentation function by
$D_i(x|p_{T0})$, where $i\in\{q,g\}$ refers to the flavour of the leading parton.
The corresponding medium/vacuum ratio is defined as 
${\cal {R}}_i(x|p_{T0})\equiv D^{\textrm{med}}_i(x|p_{T0})/D^{\textrm{vac}}_i(x|p_{T0})$. 

\subsection{Monte Carlo results and physical interpretation}
\label{sec:MCmain}

We now present our Monte Carlo results for the fragmentation function
and the associated nuclear modification factor.
We want to pay a special attention to their dependence on the two
``unphysical'' parameters of the Monte Carlo, $\theta_{\rm max}$ and
$k_{\perp,\text{min}}$, and to the 3 ``physical'' parameters,
$\hat q$, $L$ and $\amed$.
The dependence on the former can be viewed as an uncertainty in our
underlying parton-level theoretical description and a large
uncertainty would signal a strong dependence of the
observable on  non-perturbative effects such as hadronisation.
Conversely, the dependence on the ``physical'' medium parameters sheds
light on the role and importance of the medium effects at play.

\subsubsection{Variability with respect to the unphysical cutoffs}

\begin{figure}[t] 
  \centering
  \begin{subfigure}[t]{0.48\textwidth}
    \includegraphics[page=1,width=\textwidth]{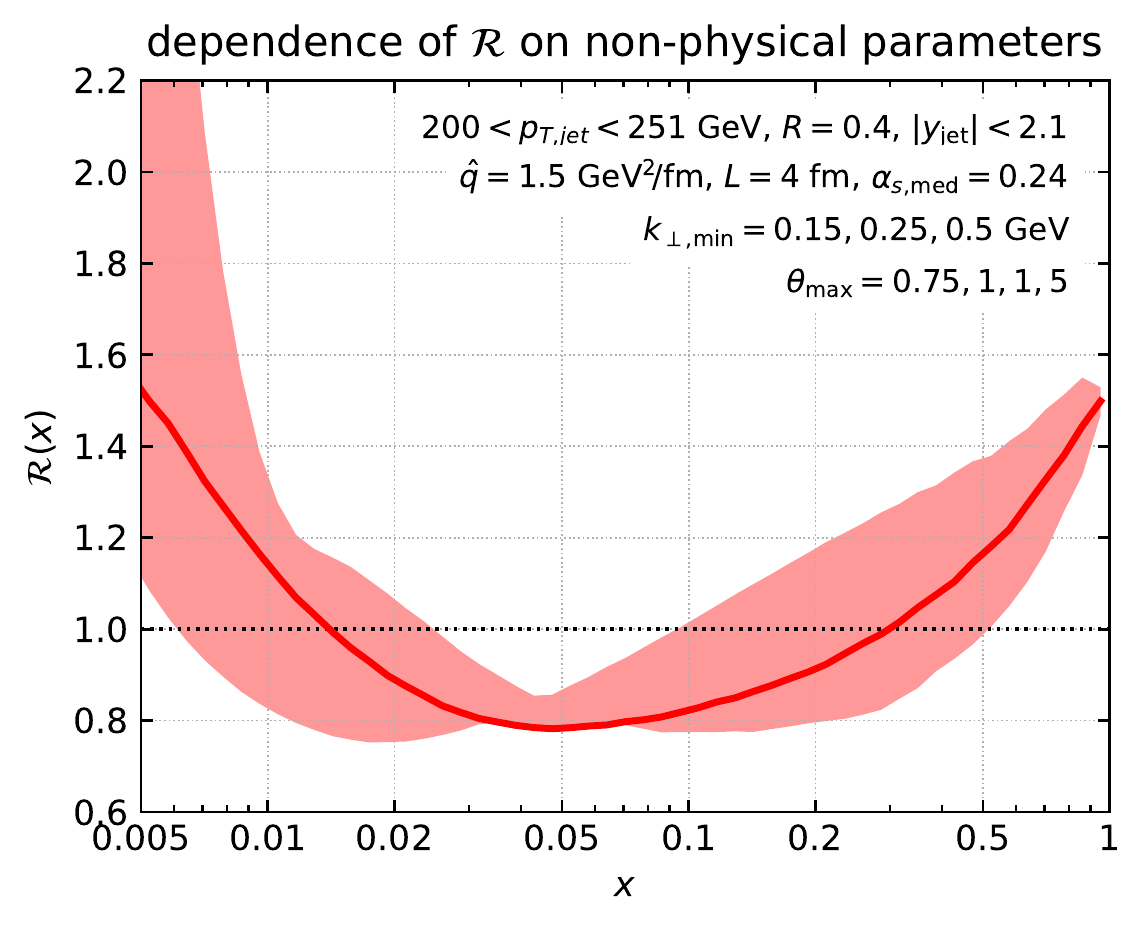}
    \caption{\small Variations in  $\theta_{\rm max}$ and $k_{\perp,\text{min}}$.}\label{Fig:MCunphys} 
  \end{subfigure}
  \hfill
  \begin{subfigure}[t]{0.48\textwidth}
    \includegraphics[page=1,width=\textwidth]{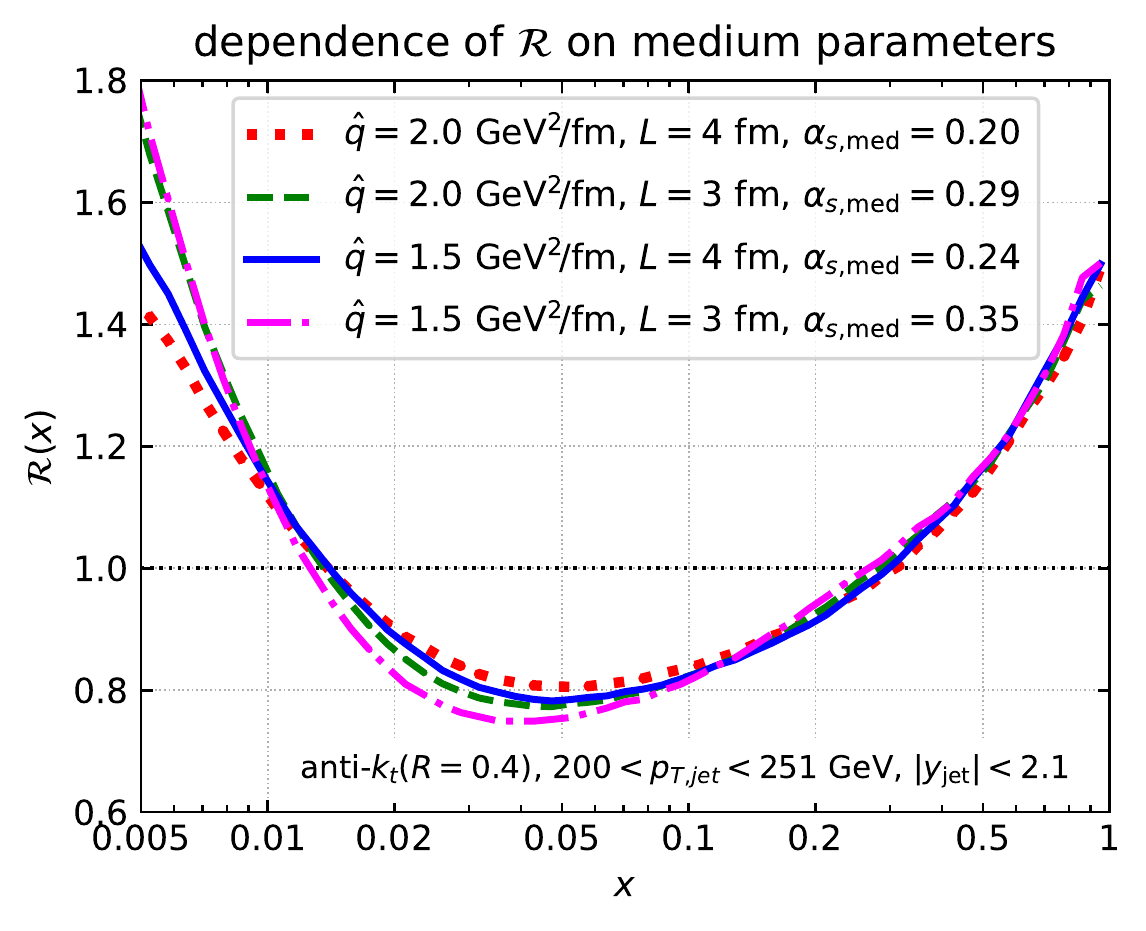}
    \caption{\small Variations in  $\hat q$, $L$ and $\amed$.}
    \label{Fig:MC-pheno}
  \end{subfigure}
  \caption{\small The variability of our MC results for the ratio $\mathcal{R}(x)$ w.r.t. changes
  in the ``unphysical'' (left) and ``physical'' (right) parameters. The 4 sets of values for
  the ``physical'' parameters are correlated in that they provide similarly good descriptions
  of the LHC data~\cite{Aaboud:2018twu} for the ``standard'' nuclear  modification factor for jets $R_{AA}$ (see the discussion in \cite{Caucal:2019uvr}).    }
  \label{fig-R-param-dependence}
\end{figure}

Fig.~\ref{Fig:MCunphys} displays the sensitivity of our MC  results for  $\mathcal{R}(x)$  
to variations of the ``unphysical'' parameters 
around their central values $\theta_{\rm max}=1$ and $k_{\perp,\text{min}}=0.25$~GeV,
for fixed values of $\hat q$, $L$ and $\amed$.

The first observation from Fig.~\ref{Fig:MCunphys} is reassuring: the
distribution shows a strong enhancement both at small $x$ and at large
$x$, with a nuclear suppression at intermediate values of $x$. This is
in qualitative agreement with experimental measurements (see e.g.~\cite{Aaboud:2018hpb}).

However, the variations w.r.t. the unphysical parameters appear to be
very large. We have checked that they were strongly dominated by
variations in $k_{\perp,\text{min}}$.
This should not come as a surprise since the fragmentation function,
measured directly on individual constituents, is not an
infrared-and-collinear (IRC) safe observable.
The sizeable variations in the small-$x$ region directly come from the
variations of the available phase-space for radiating soft gluons when
varying $k_{\perp,\text{min}}$.
The large variations in the radiation of soft particles directly
affect the spectrum of hard particles in the jet, hence the large
uncertainty in the large-$x$ region.
Only a proper description of hadronisation (including varying
hadronisation parameters) would (hopefully) reduce this uncertainty.
This should be kept in mind when studying the dependence of our
results on the medium parameters and when comparing our MC results in
this work with actual experimental data.

\subsubsection{Variability with respect to the (physical) medium parameters}

We now fix the unphysical parameters to their central value and study
how $\mathcal{R}(x)$ depends on the medium parameters $\hat{q}$, $L$, and
$\alpha_{s,\textrm{med}}$.
We first consider 4 different sets of values, given in
Table~\ref{tab:parameters} together with the angular and energy scales $\theta_c$,
$\om_c$ and $\ombr$ characterising the medium-induced radiation, as
discussed in Sect.~\ref{sec:general}.

The plot in Fig.~\ref{Fig:MC-pheno} shows our new results
for $\mathcal{R}(x)$ for the 4 sets of values for the
physical parameters. 
For large values of $x$, $x\gtrsim 0.1$, the small variations in
$\ombr$ (see Table~\ref{tab:parameters}) are compensated by relatively
large variations of $\om_c$ and $\theta_c$.
This is similar to what happens for $R_{AA}$, as discussed at length
in Ref.~\cite{Caucal:2019uvr}.
This suggests that for largish $x\gtrsim 0.1$, the nuclear effects on
jet fragmentation and on the inclusive jet production are strongly
correlated and in particular that they are both controlled by the jet
energy loss.  Such a correlation has been already pointed out in the
literature \cite{Spousta:2015fca,Casalderrey-Solana:2018wrw} and used to
provide a simple and largely model-independent argument for explaining
the enhancement  in the ratio $\mathcal{R}(x)$ at $x\gtrsim 0.5$,
as observed both in the LHC data~\cite{Aaboud:2018hpb} and in our MC results in
Fig.~\ref{Fig:MC-pheno}.  This argument will be revisited and completed
in the next subsection and also in Sect.~\ref{sec:x=1}.

Turning to smaller $x$ values, $x\le 0.01$, the situation becomes
different. There is a clear lift of degeneracy between the 4 sets of
values, with two of them --- corresponding to the smallest medium size
$L=3$~fm, but larger values for $\amed$ --- yielding results that are
significantly larger than those predicted by the two other sets (with
$L=4$).  In what follows, we provide physical explanations for these
trends.

\subsection{Behaviour at large $x$}
\label{sec:phys}

The behaviour at large $x$ is largely controlled by the physics of energy loss and its interplay
with the initial production spectrum, as we now explain.

A jet which, after crossing the medium, is measured with a
transverse momentum $p_{T,\textrm{jet}}$ has originally been produced
from a hard quark or gluon emerging from a hard process with a larger momentum $p_{T0}=
p_{T,\textrm{jet}}+\mathcal{E}(p_{T0})$, where $\mathcal{E}(p_{T0})$ is the energy
lost by the jet via MIEs at large angles  
(see Ref.~\cite{Caucal:2019uvr} for an extensive discussion of this quantity). 
While the energy lost by a {\it parton} with momentum
$p_T\gg \ombr$ saturates at a value $\epsilon\sim\ombr$, which is
independent of $p_T$ \cite{Blaizot:2013hx}, 
the average energy lost by a jet keeps increasing with $p_{T0}$, 
because of the rise in the phase space for VLEs and hence in the number of partonic sources for medium-induced radiation\footnote{Within our pQCD picture, 
this increase in the number of sources for medium-induced emissions explains the fact
that $R_{AA}$ increases only slowly with $p_{T,\textrm{jet}}$, including at large
$p_{T,\textrm{jet}}\gtrsim 500$~GeV \cite{Caucal:2019uvr}.}.

Due to the steeply-falling underlying $p_{T0}$ spectrum, cutting on the jet $p_T$ tends to select 
jets which lose less energy than average. In particular, this bias favours the 
``hard-fragmenting'' jets which contain a  parton with large $x$ (say, $x>0.5$). Such jets
correspond to rare configurations,  in which the radiation from leading parton is strongly limited 
in order to have a final $x$ fraction close to one.  
Since they contain only few partons, the hard-fragmenting jets  suffer very little energy loss,
of the order of the partonic energy loss $\epsilon\sim\ombr$. They are therefore less
suppressed than the average jets by the steeply-falling initial spectrum.   In other terms,
the medium acts as a filter which enhances the proportion of hard-fragmenting  jets
compared to the vacuum.

This bias has already consequences for the {\it inclusive} jet production, 
as measured by $R_{AA}$: 
the fraction of hard-fragmenting jets among the total number of jets 
(say, in a given bin in $p_{T,\textrm{jet}}$)
is larger in $AA$ collisions than in $pp$ collisions. The effects of this bias are
however expected to become even stronger for the jet distribution ${\dif N}/{\dif x}$ at large $x$, 
which by definition selects {\it only} hard-fragmenting jets. 
This stronger bias towards hard-fragmenting jets has been proposed as an explanation for the nuclear
enhancement in the fragmentation function 
observed in the LHC data~\cite{Aaboud:2018hpb} at large $x\gtrsim 0.5$.
This argument is very general: it applies to
a large variety of microscopic pictures for the jet-medium interactions, assuming either
weak coupling \cite{Milhano:2015mng,KunnawalkamElayavalli:2017hxo}, 
or strong coupling \cite{Chesler:2015nqz,Rajagopal:2016uip}, or a hybrid scenario 
\cite{Casalderrey-Solana:2016jvj,Casalderrey-Solana:2018wrw,Casalderrey-Solana:2019ubu}.
All these scenarios naturally predict that hard-fragmenting jets lose less energy
towards the medium than average jets, for the physical reason that we already mentioned:
hard-fragmenting jets contain less partonic sources for in-medium energy loss. 
This physical argument is manifest in both the pQCD~\cite{Milhano:2015mng,KunnawalkamElayavalli:2017hxo}
and the hybrid approaches~\cite{Casalderrey-Solana:2016jvj,Casalderrey-Solana:2018wrw,Casalderrey-Solana:2019ubu}, 
which explicitly include
a vacuum-like parton shower. It is also implicit in the strong coupling scenario  in~\cite{Chesler:2015nqz,Rajagopal:2016uip} which
is tuned such as to reproduce the angular distribution of jets in p+p collisions at the LHC
(itself well described by PYTHIA).



In this section, we argue that this is also the main explanation for the rise seen in our
results in Fig.~\ref{Fig:MC-pheno} at $x\gtrsim 0.5$. 
Within our pQCD approach
this is not entirely obvious since our scenario also
allows for nuclear modifications of the fragmentation process itself, via medium-induced emissions and energy loss effects. Similar ingredients
are {\it a priori} present in other scenarios, like JEWEL, but their relative importance
has not been explicitly studied to our knowledge. In Sect.~\ref{sec:x=1}, we shall
perform an extensive study of these effects, via both analytical and
numerical (MC) methods. Our conclusions are briefly anticipated towards the end of this section.


%
%

\begin{figure}
  \centering
  \begin{subfigure}[t]{0.48\textwidth}
    \includegraphics[width=\textwidth]{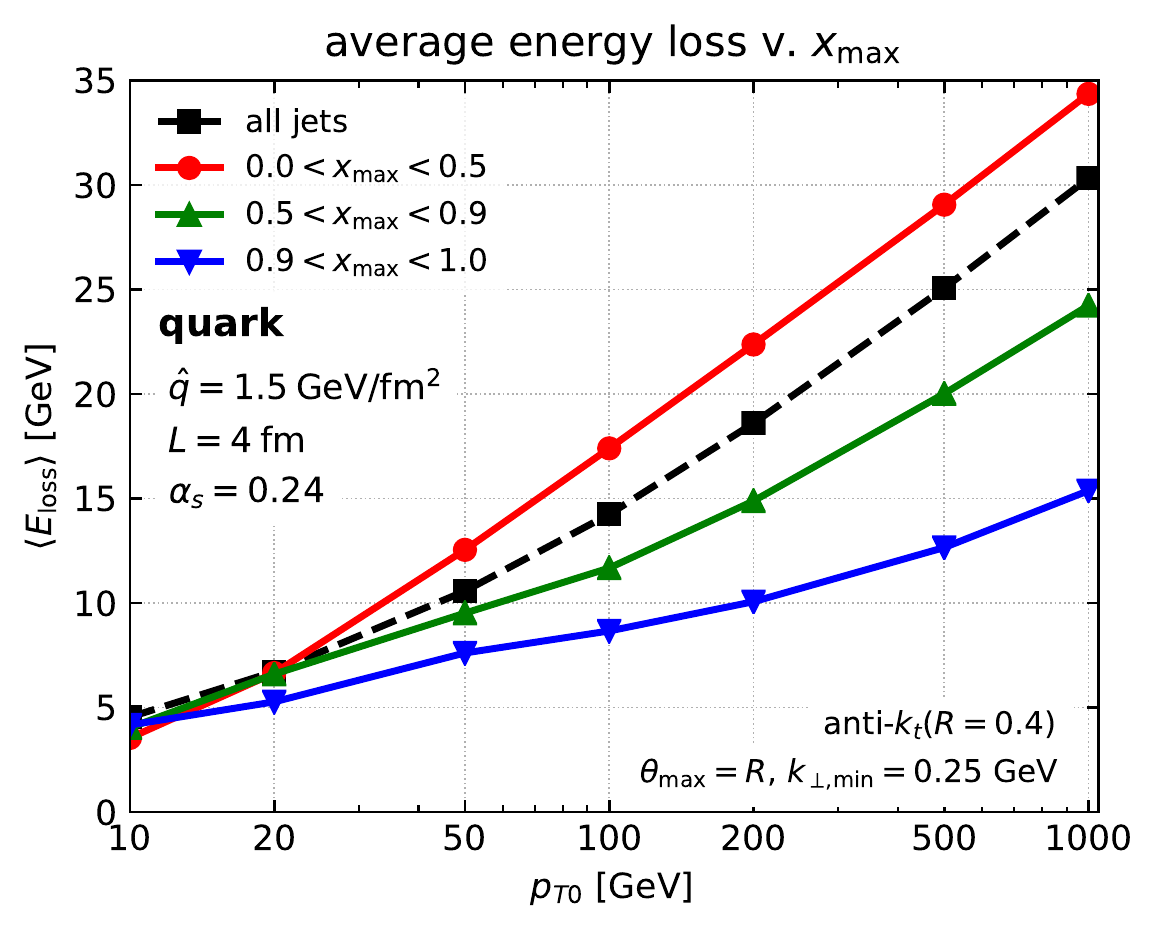}
    \caption{Energy loss as a function of the initial parton
      transverse momentum $p_{T0}$.}\label{fig:eloss-v-xmax}
  \end{subfigure}
  \hfill
  \begin{subfigure}[t]{0.48\textwidth}
    \includegraphics[width=\textwidth]{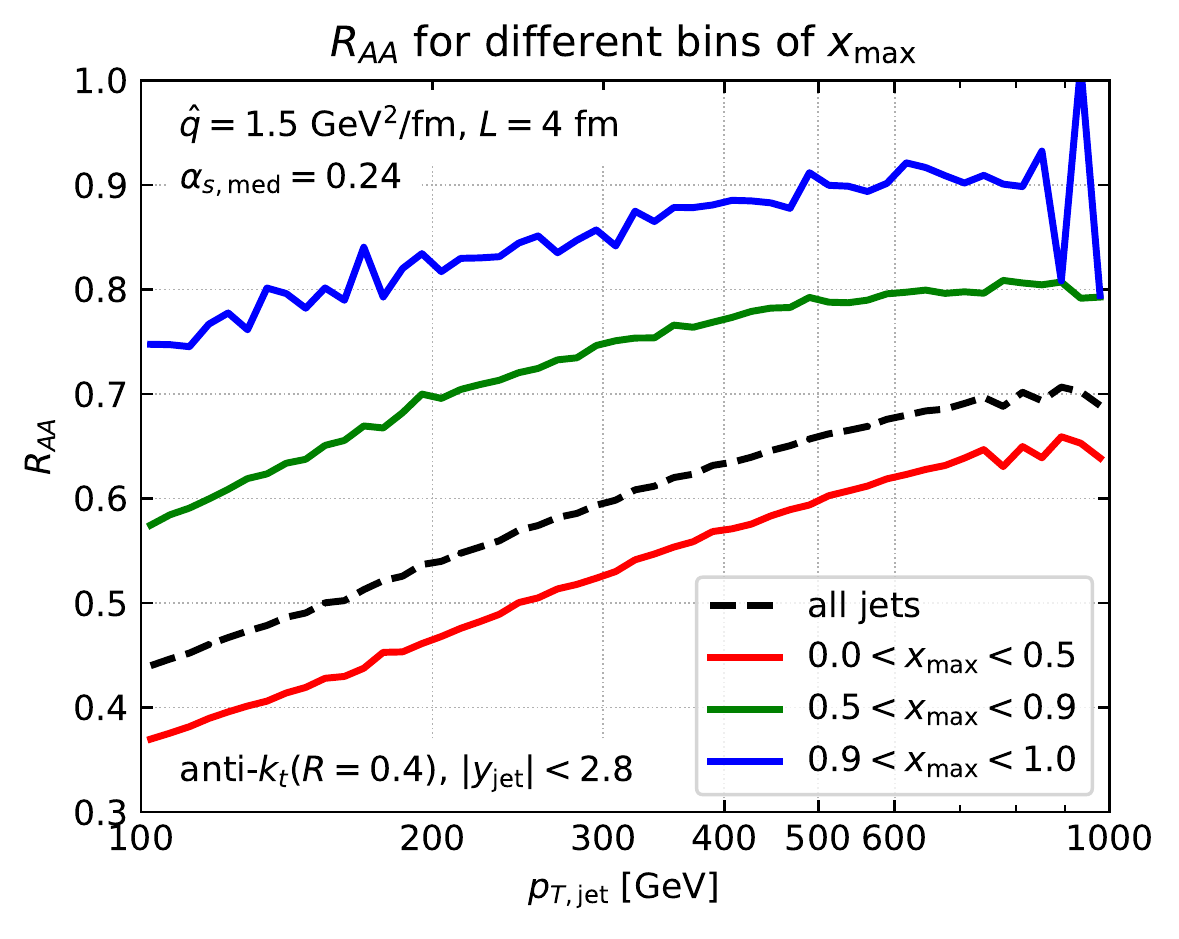}
    \caption{Jet nuclear modification factor $R_{AA}$ as a function of
      the jet $p_T$.}\label{fig:RAA-v-xmax}
  \end{subfigure}
  \caption{Energy loss and $R_{AA}$ for different bins of
    $x_\text{max}$, the momentum fraction of the jet harder
    constituent.}\label{fig:quenching-v-xmax}
\end{figure}
%

Before we discuss the fragmentation function {\it per se}, let us first
demonstrate that,  in our picture too,  a hard-branching jet loses less energy than
the average one. We have numerically verified this, by selecting (in our MC events)
jets for which the harder parton carries a momentum fraction
$x_\text{max}$ in a restricted window.
These results are presented in Fig.~\ref{fig:eloss-v-xmax} for the
energy loss of monochromatic jets and in Fig.~\ref{fig:RAA-v-xmax}
for the jet $R_{AA}$,  for the 3 bins in $x_\text{max}$ and (for comparison)
also for the inclusive jets.
Focusing first on the left figure, we find indeed that the energy lost by
jets with $x_\text{max}>0.9$, i.e.\ hard-fragmenting jets, is both
considerably smaller and also less rapidly growing with $p_{T0}$ then
for the average jets.\footnote{The MC results for  $x_\text{max}>0.9$ 
are only slightly larger
  than the energy loss expected on the basis
  of \eqn{eloss-flow} for a jet made of two partons. This will play
  an important role when discussing the large-$x$ behaviour in
  Sect.~\ref{sec:x=1}.}
As $x_\text{max}$ decreases, both the energy loss and its $p_{T0}$
growth increase.
This tendency is confirmed by the study of $R_{AA}$,
Fig.~\ref{fig:RAA-v-xmax}, where jets with a large $x_\text{max}$
show a smaller-than-average nuclear suppression.
 It would be interesting to experimentally measure the correlation between the jet  $R_{AA}$
and the momentum fraction $x_\text{max}$ and compare to our above predictions 
(see also \cite{Casalderrey-Solana:2018wrw} for a related observable, which compares
the nuclear suppression for high-$p_T$ hadrons and inclusive jets).
%

To have a more quantitative argument, let us focus on a
single bin in $p_T\equiv p_{T,\textrm{jet}}$ with a (vacuum)
Born-level $p_T$ spectrum.
The vacuum fragmentation function can then be easily estimated as
\beq\label{Dvac}
{\mathcal{D}^{\textrm{vac}}(x|p_T)}\simeq \frac{N_q(p_T)D^{\textrm{vac}}_q(x|p_{T})
+N_g(p_T)D^{\textrm{vac}}_g(x|p_{T})}{N_q(p_T)+N_g(p_T)}\,,
\eeq
where $N_i(p_T)\equiv \dif N^{\textrm{hard}}_i/{\dif p_{T}}\propto 1/p_{T}^n$
are the initial spectra for quarks ($i=q$) and gluons ($i=g$) and the fragmentation
functions for monochromatic jets have been introduced at the end of
Sect.~\ref{sec:defin}.
To write down the corresponding formula for jets in the medium, let us assume
that the only medium effect on the jet production is the energy
loss. One can thus write
\beq\label{Dmed}
{\mathcal{D}^{\textrm{med}}(x|p_T)}\simeq 
\frac{\sum\limits_{i\in\{q,g\}} N_i (p_{T}+\varepsilon_i (x))
D^{\textrm{med}}_i(x|p_{T}+\varepsilon_i(x))}
{\sum\limits_{i\in\{q,g\}} N_i (p_{T}+\mathcal{E}_i(p_T))}\qquad\mbox{for $x\simeq 1$}.
\eeq
The  quantity $\varepsilon_i(x)$  in the numerator is the energy loss
of a hard-fragmenting jet.
It depends on $x$ because the focus on large values 
$x>0.5$ selects special configurations in which jets are made with only few partons.
Its precise $x$--dependence is not important for what follows. Rather, it suffices
to know that $\varepsilon_i(x)$  is a partonic energy loss, of order $\ombr$, 
and to a good approximation is independent of the jet $p_T$.
The corresponding quantity in the denominator, $\mathcal{E}_i(p_T)$,
is the average energy loss by a jet with transverse momentum $p_T$.
It is much larger than $\varepsilon_i(x)$ and increases with $p_T$.
This difference between the {\it partonic} energy loss  $\varepsilon_i(x)$ in the numerator of 
 \eqn{Dmed} and the {\it average} energy loss $\mathcal{E}_i(p_T)$ in its denominator,
 together with the rapid decrease of $N_i (p_{T})$ when increasing $p_T$, are the origin
 of the nuclear bias towards hard-fragmenting jets at large $x$, discussed
 at the beginning of this section.\footnote{Strictly speaking, the ``average'' energy loss 
 $\mathcal{E}_i(p_T)$ in the denominator is influenced too by this bias, since it should be
 computed as an average over an inclusive sample of jets produced in $AA$ collisions.
 However, this bias is less important for the inclusive sample
 than for the large-$x$ distribution in the numerator of \eqn{Dmed}.}

On top of their bias towards less energy loss, hard-fragmenting jets also favour
quark-initiated jets. There are two reasons for
this~\cite{Spousta:2015fca,Caucal:2019uvr}:
\texttt{(i)} a quark radiates less than a gluon due to its reduced
colour charge ($C_F < C_A$), resulting in a larger probability to
contribute at large $x$, and
\texttt{(ii)} quark-initiated jets typically contain less partons than
gluon-initiated jets and hence lose less energy
($\varepsilon_q < \varepsilon_g$); this feature together with
 the steeply-falling $p_T$ spectrum favours their production
 in $AA$ collisions.
We can therefore only keep the quark contribution to the numerators of
Eqs.~\eqref{Dvac} and \eqref{Dmed} and write
\beq\label{RQ}
\mathcal{R}(x|p_T)\,\simeq\, \frac{f^\textrm{med}_q(x|p_T)}{f^\textrm{vac}_q(p_T)}
\,
\mathcal{R}_q(x|p_T)\,,
\eeq
with the following definitions:
\beq\label{fq}
 f^\textrm{vac}_q(p_T)\equiv \frac{N_q(p_T)}
{N_q(p_T)+N_g(p_T)}\,,\qquad
f^\textrm{med}_q(x|p_T)\equiv \frac{N_q (p_{T}+\varepsilon_q(x))}
{{\sum\limits_{i\in\{q,g\}} N_i (p_{T}+\mathcal{E}_i(p_T))}}\,.
\eeq
For jets in the vacuum, $ f^\textrm{vac}_q(p_T)$ is simply the 
fraction of quark-initiated jets. However, the corresponding quantity for jets in the medium 
is generally {\it not} a fraction, because of the different energy losses appearing in the
numerator and in the denominator of $f^\textrm{med}_q(x|p_T)$.

The condition of hard fragmentation ($x\sim 1$) only plays a role in the case of the
medium, where it distinguishes between the ``partonic'' energy loss  $\varepsilon_q(x)$ 
in the numerator and the jet energy loss $\mathcal{E}_q(p_T)$ in the
denominator.
As already discussed, the physical observation that $\varepsilon_q(x)\ll
\mathcal{E}_i(p_T)$ implies that the  fraction of hard-fragmenting jets in
the medium is larger than that in the vacuum, i.e., 
${f^\textrm{med}_q(x|p_T)}/{f^\textrm{vac}_q(p_T)}>1$, which in turn
causes $\mathcal{R}(x|p_T)$ to go above one
for $x\lesssim 1$.
As $x$ decreases, the energy loss of jets contributing at this value
of $x$ increases, becoming closer to $\mathcal{E}(p_T)$ and the
nuclear enhancement is less pronounced.

\eqn{RQ} also involves the medium/vacuum ratio $R_q(x|p_T)=
{D^{\textrm{med}}_q(x|p_{T})}/{D^{\textrm{vac}}_q(x|p_{T})}$ of the fragmentation functions
for quark-initiated, monochromatic, jets. This ratio encodes the nuclear modifications
of the fragmentation process itself and is perhaps the most
interesting quantity one would like to extract from observables
like $\mathcal{R}(x)$ as it encodes internal properties of the jet
rather than its global energy loss.
One of the goals of this paper is therefore to identify medium effects
on the nuclear modification factor $\mathcal{R}(x)$ beyond global jet
energy-loss effects.

Specifically, in Sect.~\ref{sec:x=1} we shall discuss three types of nuclear effects
on the fragmentation function $D^{\textrm{med}}_q(x|p_{T0})$, which act in
opposite directions and almost compensate each other.
First, the presence of a vetoed
region in the phase-space for in-medium VLEs reduces the probability for the leading
parton to radiate a (vacuum-like) soft gluon and thus increases the
probability to find that parton at large $x$.
Then, the energy lost by a two-parton system (after a
vacuum-like emission) also goes in this direction.\footnote{A similar effect
was discussed in Ref.~\cite{Caucal:2019uvr} in relation with the $z_g$
distribution.}
Finally, the MIEs which are hard enough to remain
inside the jet (i.e.\ with energies $\om > \ombr$) redistribute the
energy within the jet and thus decreases the probability to find the
leading parton with a fraction $x$ close to one.
Our numerical studies show that these effects are individually not so
small (at least for $x$ large enough, such that
$1-x \lesssim \om_c/p_{T0}$), but their net effect on $\mathcal{R}$ is
much smaller than the strong enhancement due to the
factor $f^\textrm{med}_q(x|p_T)/f^\textrm{vac}_q(p_T)$.

In summary, for relatively large $x$, the observable $\mathcal{R}(x)$ is not sensitive
to the details of the in-medium fragmentation function, but merely to the bias in the
distribution of hard-branching jets as introduced by the deeply
falling initial $p_T$  spectrum.

\subsection{Behaviour at small $x$}\label{sec:MCsmallx}

Let us now consider the situation at small $x\lesssim 0.01$, where our numerical results in
Fig.~\ref{Fig:MC-pheno} show a pronounced medium enhancement of the fragmentation
function, in qualitative agreement with the experimental
observations~\cite{Aaboud:2018hpb}.
These results also exhibit a (partial) lift of the degeneracy between the various sets of values 
for the medium parameters, suggesting a weaker correlation between
$\mathcal{R}(x)$ and  the jet nuclear modification factor $R_{AA}$.
This section provides explanations for these observations within our
framework.

We first note that,
for the considered range in $p_{T,\textrm{jet}}$, $x\lesssim 0.01$ corresponds to momenta
$p_T\lesssim 2$~GeV for the emitted partons, which are smaller than the characteristic medium
scale $\ombr$ for multiple branching.
In our framework, such soft emissions are dominated by 
VLEs outside the medium since MIEs with energies $\om\lesssim \ombr$
would fragment into very soft gluons propagating at angles larger than the jet
radius (i.e.\ outside the jet).
The medium enhancement of VLEs outside the medium has two main
origins: \texttt{(i)} the
violation of angular ordering by the first emission outside the
medium, which opens the angular phase-space beyond what is allowed in
the vacuum \cite{Mehtar-Tani:2014yea,Caucal:2018dla}, and \texttt{(ii)} the presence of MIEs with $\om> \ombr$
which remain inside the jet and can radiate VLEs outside the medium~\cite{Caucal:2019uvr}.
Our (analytic and numerical) studies in Sect.~\ref{sec:smallx} show that
both effects contribute to explaining the enhancement visible in the
MC results.

The above interpretation of the nuclear enhancement at small $x$ as
additional VLEs outside the medium does explain the differences
between the various choices of medium parameters seen in Fig.~\ref{Fig:MC-pheno}.
A smaller value for $L$ increases the energy phase-space for the
parton cascades developing outside the medium because the energy of the first emission
outside the medium, $\omega\sim 2/(L\theta^2)$, with an emission angle
$\theta\le R$, increases with $1/L$.
Furthermore, a larger value of $\amed$ enhances the rate for MIEs and
hence the number of sources for VLEs outside the medium.

Even though our MC results at small $x$ show the same qualitative trend as  
the relevant LHC data \cite{Aaboud:2018hpb}, one must remain cautious when
interpreting this agreement. Indeed, our current formalism lacks some
important physical ingredients, which are known to influence the soft region of
the fragmentation function: the hadronisation and the medium response
to the energy and momentum deposited by the jet. Whereas one may expect the 
effects of hadronisation to at least partially compensate when forming the medium-to-vacuum
ratio $\mathcal{R}(x)$, the medium-response effect --- i.e.\ the fact
that the experimentally reconstructed jets also include soft particles 
originating from the wake of moving plasma trailing behind the jet 
(and not only from the jet itself) --- is clearly missing in our approach and its inclusion
should further enhance the ratio $\mathcal{R}(x)$ at small $x$. Indeed, we know from other
approaches~\cite{Casalderrey-Solana:2016jvj,Tachibana:2017syd,KunnawalkamElayavalli:2017hxo,Chen:2017zte}, where the medium response is the only (or at least the main) mechanism  
for producing such an enhancement, that this effect by itself is comparable with the
enhancement seen in the data (see also \cite{Chien:2015hda} for a different
picture). 

Of course, it is of utmost importance to complete our formalism with a more
realistic description of the medium, including its feedback on the jet.
(We shall return to this point in the concluding section.)
Before such a more complete calculation is actually performed, it 
is difficult to anticipate what should be the combined effect of both mechanisms
on the behaviour of $\mathcal{R}(x)$ at small $x$.

%
%

\subsection{Dependence on the jet $p_T$}

\begin{figure}[t] 
  \centering
  \includegraphics[page=1,width=0.48\textwidth]{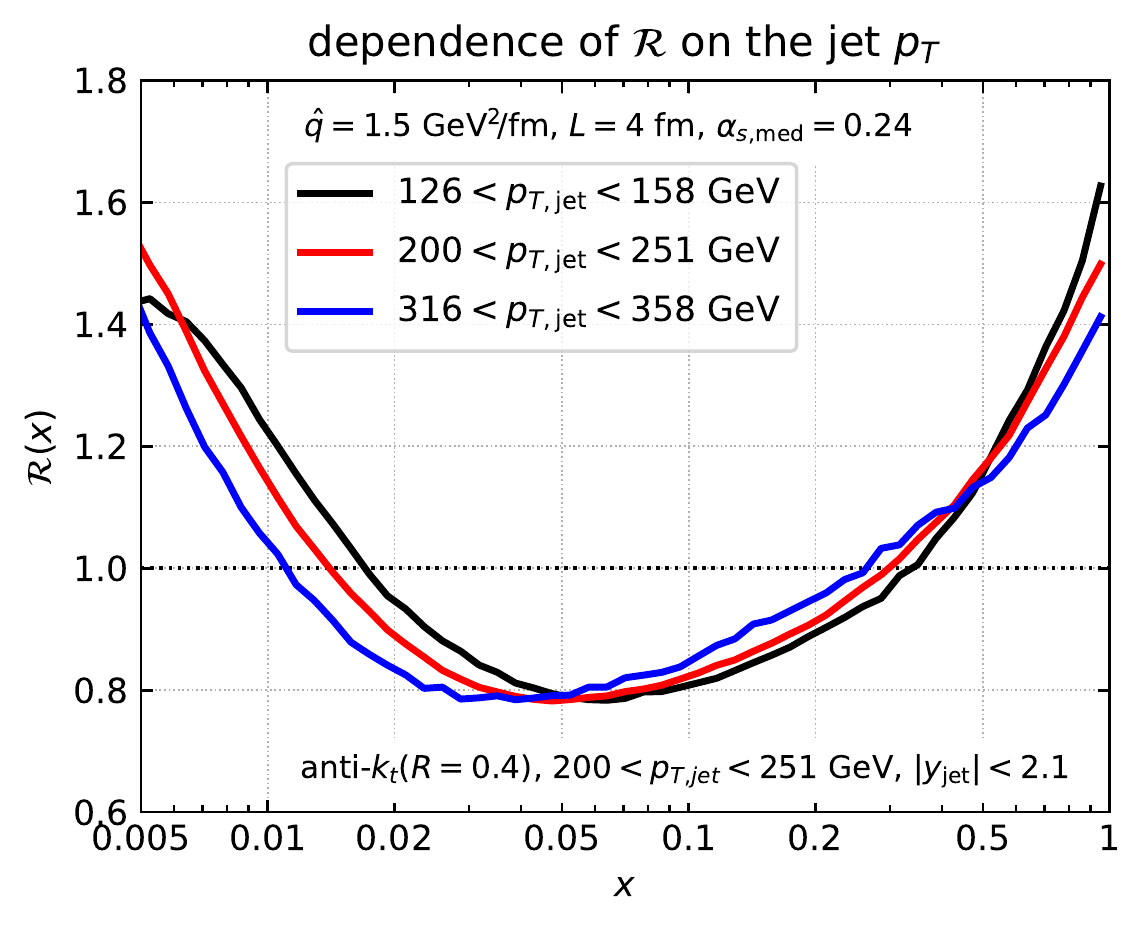}
  \hfill
  \includegraphics[page=2,width=0.48\textwidth]{ff-ptdep.pdf}
  \caption{\small Our MC results for the nuclear modification factor $\mathcal{R}(x)$ 
 shown as a function of the energy fraction $x$ of a jet constituent (left) and of its
  transverse momentum $p_T$  (right), for 3 bins of the jet $p_{T,\textrm{jet}}$.
  }
\label{Fig:MC-ptdep} 
\end{figure}

Our Monte Carlo predictions for the nuclear modification $\mathcal{R}(x)$
are shown in Fig.~\ref{Fig:MC-ptdep} for three bins of
$p_{T,\textrm{jet}}$ and for the default set of (medium and
unphysical) parameters, cf. the first line in Table~\ref{tab:parameters}.
Following the experimental analysis by ATLAS~\cite{Aaboud:2018hpb}, we
have separately plotted our results as a function of $x$ (left plot)
and of the parton $p_T$ (right plot).
The left-hand plot shows only a mild dependence of $\mathcal{R}(x)$ on
$p_{T,\textrm{jet}}$ for $x\gtrsim 0.1$ when increasing. In
view of Eq.~\eqref{RQ}, this suggests a weak
$p_{T,\textrm{jet}}$-dependence for the ratio
${f^\textrm{med}_q(x|p_{T,\textrm{jet}})}/{f^\textrm{vac}_q(p_{T,\textrm{jet}})}$,
which is likely correlated to the similarly weak dependence observed
for $R_{AA}$.
At small $x$, the $x$ scale below which the ratio is larger than
$1$ decreases with $p_{T,\textrm{jet}}$, but the corresponding $p_T$
scale increases with $p_{T,\textrm{jet}}$. These trends are in
qualitative agreement with the respective ATLAS
results~\cite{Aaboud:2018hpb}.

\section{Analytic insight for $x$ close to one}
\label{sec:x=1}

With this section, we start our analytic investigations of the nuclear effects on the jet
fragmentation function. 
Since our main goal is to discuss the effects beyond the jet-spectrum energy-loss
factor
${f^\textrm{med}_q(x|p_{T,\textrm{jet}})}/{f^\textrm{vac}_q(p_{T,\textrm{jet}})}$
in \eqref{RQ}, we mostly work with monochromatic jets with a given
initial transverse momentum $p_{T0}$.
We therefore focus on the jet fragmentation function $D_i(x|p_{T0})$
with $i\in\{q,g\}$, which can be conveniently computed as a derivative
of the {\it cumulative} fragmentation distribution
\begin{equation}
\label{def-sigma}
\Sigma_i(x|p_{T0})\equiv \int_x^1 \dif x' D_i(x'|p_{T0})\,.
\end{equation} 

We consider separately the two limiting cases where $x$ is either very
close to one ($1-x\ll 1$), discussed in this section, or very small
($x\ll 1$), discussed in the next section.
For $x\simeq 1$ the integral in the r.h.s. of \eqn{def-sigma} is the
probability to find the leading parton with an energy fraction
$x'\ge x$.

\subsection{Brief summary of the vacuum results}
\label{sec:vac}

Before addressing the nuclear effects, we briefly recall the main
results for jet fragmentation in the vacuum (see
e.g. \cite{Catani:1992ua}).
For simplicity, we identify the jet opening
angle $R$ with the maximal angle $\theta_\text{max}$ allowed for the
first emission.
Due to angular ordering, (most of) the emitted partons will
remain inside the jet, hence $p_{T,\textrm{jet}}=p_{T0}$ and $x=\omega/p_{T0}$, with
$\omega$ the energy\footnote{We often refer to the transverse momentum $p_T$ of a parton in the jet as its
``energy'' and use the notation $\om\equiv p_T$.} of a parton
inside the jet.

When $x\sim 1$, the perturbative expansion of the cumulative fragmentation distribution receives contributions 
enhanced by two types of logarithms: \texttt{(i)} the collinear logarithm $L_0\equiv\ln(p_{T0}R/\ktmin)$ generated
by integrating over emission angles in the range $\ktmin/p_{T0}
<\theta<R$, with $\ktmin$ the lower transverse-momentum cut-off of the
parton shower, and \texttt{(ii)} the soft logarithm $L\equiv\ln\frac{1}{1-x}$ generated by integrating over soft
gluon emissions with energy fractions $z$ in the range $1-x<z<1$.
The explicit logarithmic dependence on the shower cut-off $\ktmin$ is
a consequence of the fact that the jet fragmentation function is not IRC-safe.
One has $L_0\ge L$, since all emissions
must obey $z\theta p_{T0}>\ktmin$ for any $z\ge 1-x$ and any $\theta
\le R$.
The resummation of the contributions enhanced by factors $L$ or $L_0$
can be organised as the following perturbative series
\begin{equation}
\label{log-expansion}
 \ln(\Sigma_i(x|p_{T0}))=Lg_{1,i}(\alpha_sL,\alpha_sL_0)+g_{2,i}(\alpha_sL,\alpha_sL_0)+O(\alpha_s^{n+1}\ln^n)
\end{equation}
with $\alpha_s\equiv\as(p_{T0}R)\ll1$. $L g_{1,i}$ and $g_{2,i}$
resum respectively all the leading-log (LL) terms
$\alpha_s^n\ln^{n+1}$ and the next-to-leading-log (NLL) terms
$\alpha_s^n\ln^{n}$ with $n\ge1$, where $\ln$ means either $L$, or
$L_0$. We use this perturbative result at NLL accuracy to compute both the vacuum benchmark
$D^{\textrm{vac}}_i(x|p_{T0})$ and the contribution of the VLEs to the medium fragmentation function
$D^{\textrm{med}}_i(x|p_{T0})$.

The  LL piece is the standard double-logarithmic (DL) contribution in which successive emissions are 
strongly ordered both in energy fraction $z$ and in emission angle
$\theta$. It includes the effects of the running of the coupling,
$\alpha_s\to \alpha_s(k_\perp)$ with $k_\perp$ the
transverse momentum of each emission w.r.t. its emitter, and of the lower
momentum cutoff $k_\perp>\ktmin$.
For simplicity and easier physical interpretation of our results, we
quote in the main text expressions assuming a fixed coupling.
Results including running-coupling effects are presented in
Appendix.~\ref{sec:rc-effects}.
All the figures presented in the paper have been obtained using the
expressions which include running-coupling effects.

At LL accuracy, one can assume that a single emission, the one with
the larger momentum fraction $z$, dominates the jet fragmentation
function near $x=1$, with all other emissions having much smaller
values of $z$.\footnote{At LL, all softer emissions are unresolved by
  $D_i(x|p_{T0})$ and therefore cancel between real and virtual
  corrections.}
The probability \eqref{def-sigma} for the leading parton to carry a
momentum fraction $x'\ge x$ is the probability for having no emissions
with an energy fraction larger than $1-x$:
 \begin{equation}
\label{sigma-vac-LL}
  \Sigma^{\textrm{vac,LL}}_{i}(x|p_{T0})=\exp\left(-\frac{2 C_i}{\pi}\int_{1-x}^1 \frac{\dif z}{z}\int_{0}^{R}\frac{\dif \theta}{\theta}\alpha_s(k_\perp= z\theta p_{T0})\,\Theta(k_\perp-\ktmin) \right).
\end{equation}
Defining $u\equiv \alpha_{0}L$ and $v\equiv \alpha_{0}L_0$ ($v> u$)
one easily gets
\begin{equation}
\label{g1}
 L g_{1,i}^{\text{vac}} = \frac{\alpha_s C_i}{\pi}\left[(L_0-L)^2 - L_0^2\right].
\end{equation}
which is negative, as expected.
The (NLL) calculation of $g_{2,i}$ is more complicated. It is sensitive to multiple emissions
and to the non-singular pieces of quark/gluon splitting function. One finds 
\begin{equation}
\label{g2}
g_{2,i}^{\text{vac}} =
\gamma_E\frac{\partial L g_{1,i}}{\partial L}
-\ln\bigg[\Gamma\Big(1-\frac{\partial L g_{1,i}}{\partial L}\Big)\bigg]
-\frac{2 \alpha_s C_iB_i}{\pi} L_0\,,
\end{equation}
with $\Gamma$ the Euler function, $B_q=\tfrac{-3}{4}$ and
$B_g=-\tfrac{11CA-2 n_f}{12 C_A}$, with $n_f$ the number of active
quark flavours.
A brief derivation of this expression is given in Appendix~\ref{app:NLL}.

For gluon jet, we have also included the effect of flavour changes due
to $g\to q\bar q$ splittings through which the leading parton in a
gluon-initiated jet becomes a quark. Although this effect is formally
suppressed by powers of $1-x$ and therefore subleading, it has a
sizeable numerical impact.
This is because the large Sudakov suppression, Eq.~(\ref{g1}), comes
with a factor $C_i$. A $g\to q\bar q$ splittings therefore replaces a
suppression enhanced by a factor $C_A$ by one only proportional to
$C_F$, at the expense of a contribution proportional to $\alpha_s
(1-x)$ from the splitting itself.
This significantly improves our description of the large-$x$
fragmentation of gluon jets in the vacuum and additional details are
given in Appendix~\ref{app:NLL}.

\begin{figure}[t] 
  \centering
  \includegraphics[page=1,width=0.48\textwidth]{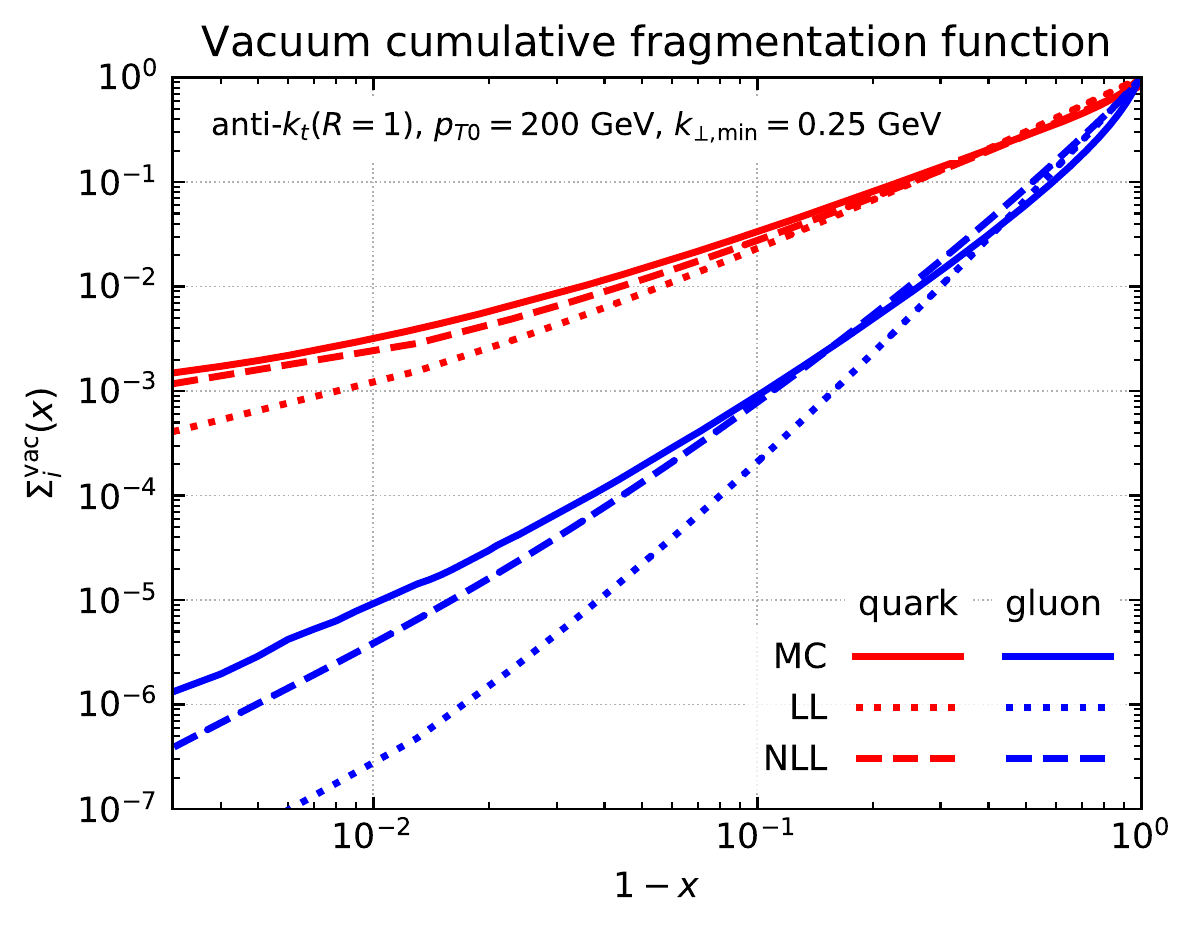}
  \caption{\small 
   The cumulative fragmentation function $\Sigma_i(x|p_{T0})$
   for quark ($i=q$) and gluon ($i=g$) initiated monochromatic jets in the vacuum.
   Our MC calculations are shown with solid lines, and the two
  analytic  approximations, LL and NLL, by dotted and dashed lines,
  respectively.}\label{Fig:sigma-vac} 
\end{figure}

In Fig.~\ref{Fig:sigma-vac}, we show the cumulative fragmentation
distribution in the vacuum for quark and gluon jets as given by our MC
compared to the analytic calculation from Eqs.~\eqref{log-expansion}, 
\eqref{g1} and \eqref{g2}.
While the LL description captures already the main trend of the
distribution, NLL corrections bring a sizeable quantitative improvement.
The main conclusion from this figure is that the
fragmentation function near $x=1$ is much larger for
quark-initiated jets than for gluon-initiated jets.

\subsection{Nuclear effects on the fragmentation function near $x=1$}

\label{sec:x=1nuc}

To discuss medium-induced effects, it is sufficient to work in the LL
approximation where jet fragmentation function near $x=1$ is dominated by a single, relatively soft,
gluon emitted by the leading parton.
From this two-parton system we then have to take three effects into account:
\texttt{(1)} emissions in the vetoed region of Fig.~\ref{Fig:LundPS} are forbidden,
\texttt{(2)} the leading parton and the emitted gluon can both lose energy via MIEs at large angles,
\texttt{(3)} the gluon emission can be a MIE remaining inside the jet.
We consider the effect of the vetoed region before the other two.

\subsubsection{Effect of the vetoed region}
\label{sec:vetoed}

The effect of the vetoed region in Fig.~\ref{Fig:LundPS} can be implemented
as a $\Theta$-function excluding this particular region from the phase-space for VLEs.
At LL accuracy, this amounts to having an extra factor
\begin{equation}\label{step-veto}
 \Theta_{\textrm{veto}}=1-\Theta(\sqrt{2\hat{q} zp_{T0}}-k_\perp^2)\Theta(k_\perp-2zp_{T0}L^{-1}),
\end{equation}
in the integrand of~\eqref{sigma-vac-LL}.
The first (second) $\Theta$-function in the r.h.s.\
of~\eqref{step-veto} corresponds to the upper (lower) boundary of the
vetoed region.
For a fixed-coupling approximation, we find assuming for simplicity $1-x\le2/(Lp_{T0}R^2)$ (see
Appendix~\ref{sec:rc-effects} for the result including running coupling)
\begin{equation}\label{eq:g1-veto}
  L g_{1,i}^{\textrm{veto}}(\alpha_s L, \alpha_sL_0)
  = L g_{1,i}^{\text{vac}}(\alpha_s L, \alpha_sL_0)
  + \frac{2\alpha_sC_i}{3\pi}\ln^2\frac{R}{\theta_c}\,.
\end{equation}
NLL corrections, $g_{2,i}^{\textrm{veto}}$, can be
obtained using~\eqref{g2}.
In particular, the hard-collinear term proportional to $B_i$ is not
modified by the veto region and therefore cancels in the medium/vacuum
ratio.

Our analytic estimate for the ratio $R_i(x|p_{T0})$ is shown in
Fig.~\ref{fig:large-x-analytic}~left in green
  for $p_{T0}=200$ GeV.
For comparison, we also show the corresponding MC result, which only
includes VLEs (the green curve in Fig.~\ref{fig:large-x-MC}). These
results agree well with each other and they both predict a nuclear
enhancement near $x=1$.
This enhancement can be easily understood on the basis of \eqref{eq:g1-veto},
which implies
\begin{equation}
\ln\frac{\Sigma^{\textrm{veto,LL}}_{i}(x)}{\Sigma^{\textrm{vac,LL}}_{i}(x)}
=\frac{2\alpha_{0}C_i}{3\pi}\ln^2
\frac{R}{\theta_c}\,> 0\,,
\end{equation}
meaning $\Sigma^{\textrm{med}}_{i}(x) \simeq \Sigma^{\textrm{veto}}_{i}(x) > \Sigma^{\textrm{vac}}_{i}(x)$
and hence $R_i(x)>1$ when $x\to 1$. Indeed, the presence of the vetoed region reduces the phase-space
allowed for the decay of the leading parton.

\subsubsection{Effect of medium-induced emissions} 
\label{sec:MIE}

The medium-induced emissions (MIEs), as triggered by the interactions with
the plasma constituents, affect differently the total jet momentum
$p_{T,\textrm{jet}}$ and the energy $\om_\textrm{LP}$ carried by its
leading parton. This implies a nuclear modification
$\mathcal{R}(x)$ at large $x\equiv
\om_\textrm{LP}/p_{T,\textrm{jet}}$.

For convenience, we focus on the case where $x$ is not
{\it too} close to one, such that $\ombr/p_{T0}\ll 1-x\ll 1$, with $\ombr\sim \alpha_s^2\hat q L^2$
the characteristic scale  for multiple branchings.
For jets with $p_T\ge 200$~GeV, a phenomenological region $0.80
\lesssim x \lesssim 0.95$ translates into
$(1-x)p_{T0}\gtrsim 10$~GeV
which is indeed larger than $\ombr\sim 4$~GeV (cf.\ Table~\ref{tab:parameters}).

Within this regime, the medium-induced emissions which control the energy
loss by the leading parton are relatively hard,  with energies $\om\gg \ombr$.
Thus, they remain inside the jet and can be accurately computed
in the single emission approximation. This situation 
is similar to the one discussed for
jets in the vacuum at double-logarithmic accuracy: 
the parton distribution near $x=1$ is controlled by a single intra-jet
emission, with an energy of the order of $(1-x)p_{T0}$. This emission
can be either vacuum-like, or medium-induced. This
``semi-hard'' emission is accompanied by an arbitrary number of soft
MIEs, with energies $\om\lesssim \ombr$, which propagate outside the
jet and take energy away from the jet constituents.
The in-medium fragmentation function near $x=1$ can therefore be
evaluated as:
\begin{align}  \label{Elossred}
 D^{\textrm{med}}_i(x|p_{T0})\,\simeq\,
  \int\dif\omega \,\Delta_i^{\text{VLE}}  (\omega) \,\Delta_i^{\text{MIE}}  (\omega)\,
  \left[\frac{\del\mathcal{P}_{i, \text{vac}}}{\del \omega} 
  +\frac{\del\mathcal{P}_{i, \text{med}}}{\del \omega}\right]
 \delta\left(x-\frac{p_{T0}-\omega-\varepsilon_i}
{p_{T0}-\mathcal{E}_i}\right).
\end{align}
In this expression,
${\del\mathcal{P}_{i, \text{vac}}}/{\del \omega}$ is the differential
probability for emitting a soft gluon with energy $\omega$ at any
emission angle $\theta$ (with $\ktmin/\om< \theta < R$) and
$\Delta_i^{\text{VLE}} (\omega)$ is the Sudakov factor forbidding VLEs
with energies larger than $\om$ (including the condition
\eqref{step-veto} for the vetoed region), i.e.\
\begin{equation}\label{Pvac}
 \Delta_i^{\text{VLE}}  (\omega) =
 \Sigma^{\text{veto}}\left(1-\frac{\omega}{p_{T0}}\right)
 \qquad \text{ and } \qquad
   \frac{\del\mathcal{P}_{i, \text{vac}}}{\del \omega} 
=\frac{\dif \ln\Delta_i^{\text{VLE}}}{\dif \omega} 
\simeq \frac{2 \alpha_sC_i}{\pi}\frac{1}{\om}\, \ln\left(\frac{\om R}{\ktmin}\right),
\end{equation}
where the second expression for ${\del\mathcal{P}_{i, \text{vac}}}/{\del\omega}$, shown only
for illustration, 
holds for the case of a fixed coupling $\alpha_s$ and ignores the constraints 
introduced by the vetoed region.

Furthermore, ${\del\mathcal{P}_{i, \text{med}}}/{\del \omega}$ and
$\Delta_i^{\text{MIE}} (\omega)$ are the corresponding quantities for
the semi-hard MIE inside the jet ($\theta_c <\theta < R$).
Its energy is restricted to $\bar\om< \om<\om_c$, where
$\om_c=\hat q L^2/2$ and $\bar\om$ is a cutoff of order $\ombr$,
separating between ``semi-hard'' and ``soft'' MIEs.\footnote{The
  precise value of this cutoff is not important: as we will show below
  the energy integration is controlled by the $\delta$-function, and
  since the energy losses are relatively small one roughly has
  $\omega \simeq (1-x)p_{T0}\gg \bar\om\sim \ombr$.}
In this regime, one can safely use the single emission approximation,
i.e. (compare to \eqn{BDMPS})
\begin{align}\label{Pmed}
 \frac{\del\mathcal{P}_{i, \text{med}}}{\del
 \omega_m} \simeq \frac{\amed C_i}{\pi}\sqrt{\frac{2\omega_c}{\omega^3_m}}\,,
 \qquad\quad \Delta_i^{ \text{MIE}} (\om_m) 
 =\exp\left(-\int_{\omega_m}^{\omega_c}\dif\omega\,  \frac{\del\mathcal{P}_{i, \text{med}}}{\del
 \omega} \right)\,.
\end{align}
Next, $\varepsilon_i$ and $\mathcal{E}_i$ refer to the energy loss via soft
MIEs outside the jet ($\theta > R$), for the leading parton and for
the jet as a whole, respectively.
Finally, the $\delta$-function in \eqn{Elossred} encodes the fact that, in our
present approximation, the energy of the leading parton is the
energy $p_{T0}$ of the parton initiating the jet minus the energy of
the semi-hard emission and the partonic energy loss $\varepsilon_i$,
while the energy of the jet is $p_{T,\textrm{jet}}= p_{T0}-\mathcal{E}_i$.

For more clarity, we study separately the two types of medium effects included in
\eqn{Elossred}, namely energy loss at large angles and energy redistribution via intra-jet MIEs.

\subsubsection{Energy loss at large angles}
\label{sec:eloss}

To study the  energy loss effects alone, we temporarily neglect the contribution of
the intra-jet MIEs to \eqn{Elossred}, which then  simplifies to (with
$\omega_s$ the energy of the soft VLE)
\begin{align}  \label{VL+Eloss}
 D^{\textrm{med}}_i(x|p_{T0})\Big |_{\text{e-loss}}= 
  \int\dif\omega_s \frac{\del\mathcal{P}_{i, \text{vac}}}{\del
 \omega_s}\,\Delta_i^{\text{VLE}}  (\omega_s) \,
 \delta\left(x-\frac{p_{T0}-\omega_s-\varepsilon_i}
{p_{T0}-\mathcal{E}_i}\right).
\end{align}

In the absence of VLEs, a single parton with initial energy $\omega_0$ 
loses energy by radiating MIEs at large angles ($\theta\gtrsim
\theta_c/\abar^2$).
This is associated with the ``turbulent'' component of the medium-induced cascades,
associated with very soft partons of
energies $\omega\lesssim \ombr$, which are deflected at large angles
via collisions with the plasma.
The average energy loss is estimated by~\cite{Blaizot:2013hx}
\begin{equation}\label{eloss-flow}
  \varepsilon_i(\omega_0) =\omega_0 \big[1-\rme^{-v_0\ombr/\omega_0}\big],
  \qquad\text{ with} \quad
  \ombr
  = \left(\frac{\amed}{\pi}\right)^2C_AC_i\,\frac{\hat q L^2}{2}\,.
\end{equation}
$v_0$ is a number which can be either obtained via analytic
approximations~\cite{Baier:2000sb,Blaizot:2013hx,Fister:2014zxa}
(e.g.\ one finds $v_0\simeq 4.96$ for $\om_0< \om_c$), or
extracted from MC calculations.
$\varepsilon_i$ depends on the flavour index $i$ and on the distance $L$
travelled by the parton through the medium.
For energetic partons with $\omega_0\gg \ombr$ --- the most relevant case
here ---, this energy loss saturates at a value $ \varepsilon_i=v_0\ombr$ independent 
of $\omega_0$.

For a full jet, the energy loss receives contributions of the form of
\eqn{eloss-flow} from both the leading parton (LP) and each of the
(vacuum-like or medium-induced) {\it intra-jet} emissions ($\theta<R$)
 which are radiated within the medium,
 i.e.\ in the ``inside'' region in Fig.~\ref{Fig:LundPS}.
For a hard-fragmenting jet made of only two partons (the LP and a relatively soft
VLE, as in \eqn{VL+Eloss}), we have to consider two options. 
If the VLE is emitted outside the medium, i.e.\
either with $\theta<\theta_c$ or with $t_{\rm f}=2/(\om \theta^2)>L$,
only the LP loses energy and we have
$\mathcal{E}_i= \varepsilon_i$.\footnote{For $\theta<\theta_c$, the two
  partons lose energy coherently, so one can see the energy loss as coming only from the 
  LP~\cite{MehtarTani:2010ma,MehtarTani:2011tz,CasalderreySolana:2011rz,Mehtar-Tani:2017ypq}.}
If the VLE occurs inside the medium, both partons lose energy and we
have $\mathcal{E}_i=\varepsilon_i+\varepsilon_g$, with $\epsilon_g$ the
energy lost by the VLE.\footnote{In this case, $t_{\rm f}\ll L$ so
  the VLE travels a length or order $L$ through the medium.}

\begin{figure}[t] 
  \centering
  \begin{subfigure}[t]{0.48\textwidth}
    \includegraphics[page=1,width=\textwidth]{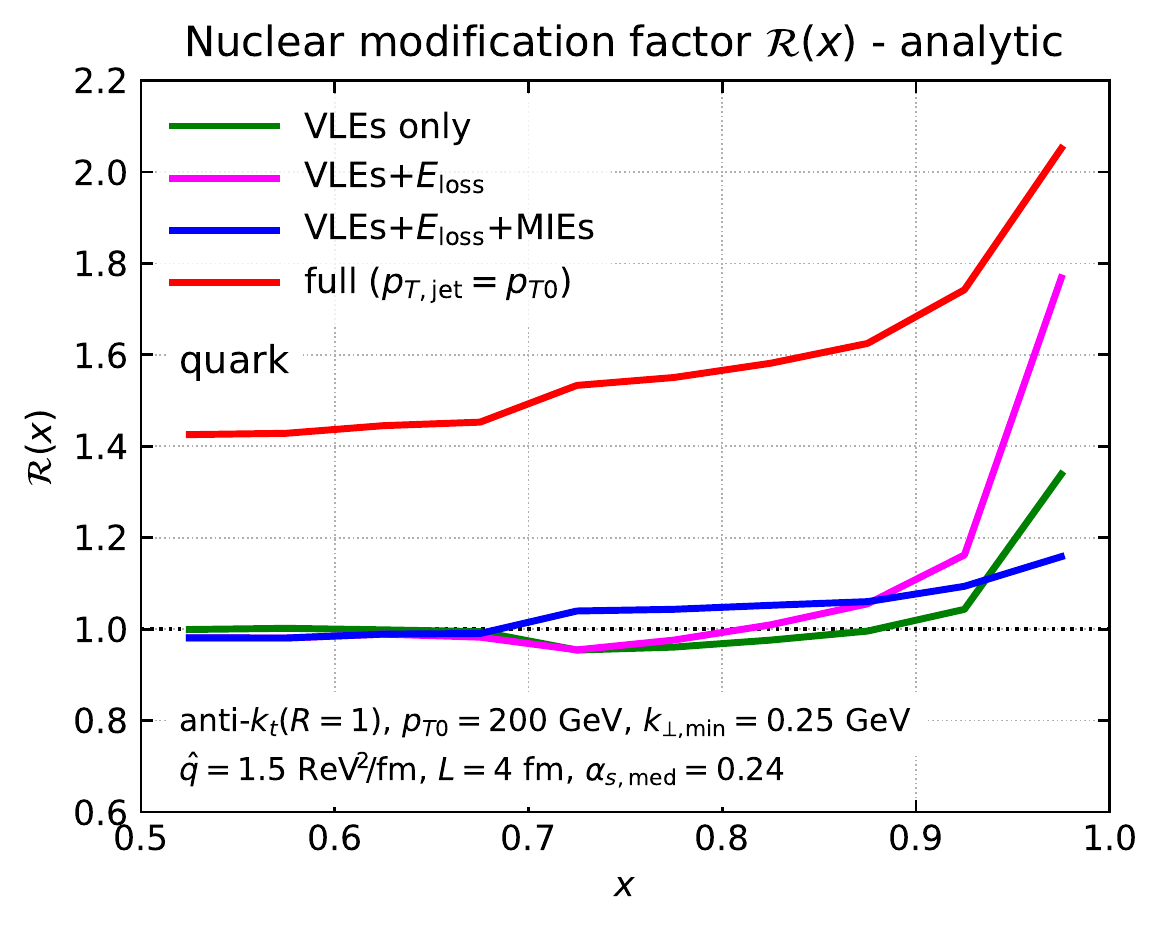}
    \caption{\small Semi-analytic estimates.}\label{fig:large-x-analytic}
  \end{subfigure}
  \hfill
  \begin{subfigure}[t]{0.48\textwidth}
    \includegraphics[page=2,width=\textwidth]{plot-frag-ratio-largex.pdf}
    \caption{\small Monte-Carlo simulations.}\label{fig:large-x-MC}
  \end{subfigure}
  \caption{\small Nuclear effects on the fragmentation function at
    large $x$ for monochromatic jets.
    Three increasingly more physical scenarios are considered: \texttt{(i)}
    VLEs only (only the nuclear effects from the vetoed region are
    included), \texttt{(ii)} adding energy loss via soft MIEs at large
    angles (not shown on the right plot), and \texttt{(iii)} further
    adding semi-hard MIEs inside the jet.
    Additionally, we show the ``full'' curve in red which includes the bias
    introduced by the initial hard spectrum and is manifestly the
    dominant effect.}
\label{Fig:analytic} 
\end{figure}

For a VLE inside the medium, the $\delta$-function in \eqn{VL+Eloss} can be equivalently rewritten as
\beq
\delta\left(1- x-\frac{\omega_s - (\mathcal{E}_i-\varepsilon_i)}{p_{T0}-\mathcal{E}_i}\right)
\simeq
\delta\left(1- x- z+ \frac{ (1-z)\mathcal{E}_i-\varepsilon_i}{p_{T0}}\right)\,,
\eeq
with $z\equiv\om_s/p_{T0}$ the splitting fraction of the VLE. We have used the fact that the energy loss is 
relatively small, $\mathcal{E}_i\ll p_{T0}$.
The effect of the in-medium energy loss is a small increase of the splitting fraction, from its initial value
in the vacuum, $z_{\textrm{vac}}=1-x$, to
\beq\label{z-Eloss}
z = 1-x+ \frac{ (1-z)\mathcal{E}_i-\varepsilon_i}{p_{T0}}
\simeq 1-x+\frac{x\varepsilon_g-(1-x)\varepsilon_i}{p_{T0}}
\simeq 1-x+\frac{\varepsilon_g}{p_{T0}} > z_\text{vac}.
\eeq
In the second equality we have used $\mathcal{E}_i=\varepsilon_i+\varepsilon_g
$ and $z\simeq 1-x$.
For the third equality we have used $x\simeq 1$
and $\varepsilon_g\ge \varepsilon_i$, making clear that the dominant effect is
the energy loss by the soft gluon.\footnote{Interestingly,
  for a VLE outside the medium (cf.\ Fig.~\ref{Fig:LundPS}), we can
  set $\varepsilon_g\to 0$ to get $z = 1-x-(1-x)\varepsilon_i/{p_{T0}}$
with $1-x\ll 1$. The energy loss effect is therefore much smaller than
for an in-medium VLE and with an opposite sign.}

The fact that $z>z_\text{vac}\equiv 1-x$ means
that the probability $P(z)\propto 1/z$ of its emission is smaller, so
there is an {\em enhancement} in the
probability for the leading parton to survive at large $x$. This effect is reinforced by the 
associated  Sudakov factor: when $\om_s=zp_{T0} > (1-x)p_{T0}$, there is a reduction in  
the phase-space for emissions by the leading parton and therefore
$\Delta_i^{\text{VLE}}  (\omega_s) >\Delta_i^{\text{VLE}}  ((1-x)p_{T0}) $.

The purple curve in Fig.~\ref{Fig:analytic}-left shows
a calculation of $\mathcal{R}_q(x|p_{T0})$
based on \eqn{VL+Eloss} together with  $\mathcal{E}_q=\varepsilon_q+\varepsilon_g$ 
and with \eqn{eloss-flow} for the partonic energy loss. 
Compared to the green curve in the same figure, which
includes solely the effect of the vetoed region, the purple curves indeed shows a larger
enhancement near $x=1$.

\subsubsection{Energy redistribution via a hard MIE}
\label{sec:redis}

A semi-hard MIE with energy $\omega \gg \ombr$ and which remains inside
the jet can modify the fragmentation function
$D^{\textrm{med}}_i(x|p_{T0})$
near $x=1$ in two ways. On one hand, it brings a positive contribution
via the term proportional to $\partial\mathcal{P}_{i,\text{med}}/\partial\omega$
in \eqn{Elossred}.
On the other hand, the additional Sudakov factor
$\Delta_i^{\text{MIE}}(\om)$ induces an extra suppression.
These two effects are competing with each other. It turns out that
the second effect is stronger, resulting in a {\it decrease} of
$D_i^\text{med}(x|p_{T0})$ near $x=1$ as compared to the 
vacuum, and hence a decrease of the medium/vacuum ratio
$\mathcal{R}_i(x|p_{T0})$.

We can actually estimate these two contributions to \eqn{Elossred}.
To that aim, we can neglect the effects of the energy loss
at large angles.\footnote{Indeed, in this case, the intra-jet MIE is the dominant medium
effect, whereas the energy loss at large angles is a subdominant effect since
  $\mathcal{E}_i\sim\varepsilon_i\sim\ombr$ are much smaller than $\om\simeq
  (1-x)p_{T0}$.}
Using the $\delta$-function to 
perform the integral over $\omega$ we find
\begin{align}  \label{hardMIE}
 D^{\textrm{med}}_i(x|p_{T0})\Big |_{\text{MIE}}& = p_{T0}
   \left[\frac{\del\mathcal{P}_{i, \text{vac}}}{\del \omega} 
  +\frac{\del\mathcal{P}_{i, \text{med}}}{\del \omega}\right]
\Delta_i^{\text{VLE}}  (\omega) \,\Delta_i^{\text{MIE}}  (\omega)\Big |_{\om=
(1-x)p_{T0}}.
\end{align}
We need to show that the ``medium'' Sudakov effect on the VLE (first term in the
square bracket) is larger in absolute value than the direct
contribution from MIEs (second term in the square
bracket):
\beq\label{ineq}
\frac{\del\mathcal{P}_{i, \text{vac}}}{\del \omega} \left[1-\Delta_i^{\text{MIE}} 
  (\omega)\right] > \frac{\del\mathcal{P}_{i, \text{med}}}{\del \omega}
\, \Delta_i^{\text{MIE}}  (\omega)\,.
\eeq
At leading-order accuracy for the MIE, one can set
$\Delta_i^{\text{MIE}} \simeq 1$ in the r.h.s.\ of the above inequality, whereas in
the l.h.s. one must also keep the linear term in its Taylor expansion:
\begin{align} 1-\Delta_i^{\text{MIE}} 
  (\omega)\,\simeq\,
\frac{2\alpha_s C_i}{\pi}\sqrt{\frac{2\omega_c}{\om}}\,.
\end{align}
Using a fixed-order approximation for the vacuum emission probability
(cf.\ \eqn{Pvac}), together with \eqn{Pmed} for the medium-induced,
one finds after simple algebra that \eqn{ineq} is equivalent to
\beq\label{log}
\frac{4 \alpha_sC_i}{\pi} \,\ln\left(\frac{(1-x)p_{T0}R}{\ktmin}\right) \,>\,1\,.
\eeq 
This is satisfied both parametrically and numerically under our
working assumptions that collinear logarithms are large.
For the parameters used in Fig.~\ref{Fig:analytic}, namely $p_{T0}=200$~GeV, $R=1$, and
$\ktmin=0.25$~GeV, and with $x=0.9$ and $\alpha_s=0.3$, one finds that
the l.h.s.\ of \eqn{log} is about $5.3$.

These considerations are confirmed by the explicit 
numerical integration of~\eqn{Elossred}.
The blue curve in Fig.~\ref{fig:large-x-analytic} includes all the medium effects
discussed in this section (the vetoed region, the energy loss at large
angles and the effects of semi-hard MIEs).
Comparing it to the purple curve which does not include the effects of
semi-hard MIEs, we see that the latter reduce the ratio
$\mathcal{R}_i(x|p_{T0})$ near $x=1$, as expected.
This plot also shows that the three medium effects appear to be of
similar magnitude and to almost compensate each other, leaving only
a modest enhancement at $x\gtrsim 0.9$.
This pattern is in very good agreement with what we see from our MC
simulations, Fig.~\ref{fig:large-x-MC}.
Whereas the details of this compensation depend on the specific
parameters used in our calculation, we have checked using our MC that
such a competition between comparable but opposite effects is
a relatively robust prediction from our pQCD scenario.

One can view this conclusion as a little bit deceptive since it shows
that the fragmentation function has a reduced sensitivity to nuclear
effects associated with the internal dynamics of the jets.

\subsection{Bias introduced by the steeply falling jet spectrum} 
\label{sec:bias}

In Section~\ref{sec:phys} we have argued (see
also~\cite{Spousta:2015fca,Casalderrey-Solana:2018wrw}) that the strong
enhancement of $\mathcal{R}(x)$ seen at large $x$ in the ATLAS Pb+Pb
data~\cite{Aaboud:2018hpb} is a consequence of the bias introduced by
the steeply-falling jet spectrum, which favours jets which lose only
little energy, notably hard-fragmenting quark-initiated jets.
In this section, we present a more detailed (numerical) argument, 
based on simple 2-parton jets, which supports
 \eqn{RQ} proposed in Section~\ref{sec:phys} to quantify this effect.

\eqn{RQ} relies on the ``fraction'' $f_i(x|p_T)$ of hard-fragmenting jets 
with one constituent having an energy of at least $xp_T$.
In practice, we define (cf.\ \eqn{fq})
\begin{equation}
\label{quenching-factor}
f^\textrm{vac}_q(p_T)=\frac{\frac{\dif\sigma_q}{\dif p_{T}}}
{\sum\limits_{i\in\{q,g\}}\frac{\dif\sigma_i}{\dif p_T}}\,,\qquad
f^\textrm{med}_q(x|p_T)=\frac{\frac{\dif\sigma_q}{\dif p_{T0}}\big |_{p_T+\mathcal{E}^{\textrm{n=2}}_q}}{\sum\limits_{i\in\{q,g\}}\frac{\dif\sigma_i}{\dif p_{T0}}\big|_{p_T+\mathcal{E}_i(p_{T0})}}\,,
\end{equation}
where  $\dif\sigma_i/\dif p_{T0}\propto p_{T0}^{-n_i}$ is the initial
jet spectrum.
$n_q=5$ and $n_g = 5.6$ give a decent description over the kinematic
range covered in this paper.
$\mathcal{E}_i(p_{T0})$ is the average energy loss by a jet with
initial transverse momentum $p_{T0}$ and is numerically extracted from
MC simulations~\cite{Caucal:2019uvr}. $\mathcal{E}^{\textrm{n=2}}_q$
is the energy lost by a simple two-parton jet (a leading quark of
energy fraction $x\sim 1$ and a relatively soft gluon of energy
fraction $1-x$). The dominant
contribution (cf.\ Sect.~\ref{sec:eloss}) comes from events where the
quark and gluon lose energy
independently of each other\footnote{Strictly speaking, the energy
  argument of $\varepsilon_g$ and $\varepsilon_q$ should be $zp_{T0}$
  and $(1-z)p_{T0}$, respectively, with $z$ the gluon splitting
  fraction, cf.  \eqn{z-Eloss}, but to the accuracy of interest one
  can replace $z\simeq 1-x$ and $p_{T0}\simeq p_T$.}:
$\mathcal{E}^{\textrm{n=2}}_q= \varepsilon_q(xp_{T0}) +
\varepsilon_g((1-x)p_{T0})$, with $\varepsilon_g$ and $\varepsilon_q$
given by \eqn{eloss-flow}.

By combining \eqn{quenching-factor} for the fractions of
hard-fragmenting jets with our previous calculations of the ratio
$\mathcal{R}_q(x|p_T)$ for monochromatic jets, we can provide a
semi-analytic estimate for the physical observable
$\mathcal{R}(x|p_T)$ using \eqn{RQ}.
This is shown by the red curve in Fig.~\ref{fig:large-x-analytic}, that
should be compared to the corresponding MC result in
Fig.~\ref{fig:large-x-MC}.
The two red curves are both in good agreement with each other and with
the general trend seen in the LHC data~\cite{Aaboud:2018hpb}.
For $x$ very close to 1 (mainly the last
bin in our plots), the pattern observed in our MC calculations is
a combination of the bias induced by the jet spectrum and of the
medium effects on the internal jet dynamics $\mathcal{R}_q(x|p_T)$,
with a strong domination of the former.
The current experimental uncertainties in this region of $x$ are too
large to draw a stronger conclusion, notably concerning  the relative importance of 
the nuclear effects associated with $\mathcal{R}_q(x|p_T)$, i.e.\ with the
medium modifications of jet fragmentation itself.

\section{Small-$x$ enhancement: colour decoherence and medium-induced radiation}\label{sec:smallx}

We argued in Sect.~\ref{sec:MCsmallx} that the nuclear enhancement of the
fragmentation function at small-$x$, $x\lesssim 0.02$, is driven by
two main phenomena: \texttt{(i)} colour decoherence, which enlarges the angular
phase-space for emissions outside the medium, and \texttt{(ii)}
medium-induced radiation producing additional partonic sources for
these outside-medium emissions.
This section provides analytic studies backing up this picture.
For simplicity we mostly treat VLEs at fixed coupling and in the
double-logarithmic approximation (DLA). We then present MC
calculations which hold beyond DLA.

\subsection{Analytic estimates}

Our aim is to compute the double-differential gluon distribution in a
jet of initial transverse momentum (or energy) $p_{T0}$, initial
flavour $i$ and radius $R$
 \begin{equation}
 T_i(\omega,\theta^2|p_{T0},R^2)=\omega\theta^2\frac{\dif^2 N_i}{\dif \omega\dif \theta^2}\,.
\end{equation}
The fragmentation function can be obtained from $T_i$ by integrating
over all the angles in the jet (with $\theta_{\textrm{min}}=
\ktmin/\omega$)
\begin{equation}\label{DfromT}
\om D_{i}(\omega,\theta^2|p_{T0},R^2)  =
 \int_{\theta^2_{\textrm{min}}}^{R^2}\frac{\dif \theta^2_1}{\theta^2_1}
\, T_{i}(\omega,\theta^2_1|p_{T0},R^2)
\end{equation}

\paragraph{Vacuum case.}  In pQCD, the leading contribution to the
multiplicity of soft gluons in a jet comes from double-logarithmic
emissions in a fixed-coupling approximation~\cite{Dokshitzer:1991wu},
i.e.\ via successive VLEs in our context~\cite{Caucal:2018dla}.
In this limit, successive gluon emissions are strongly ordered in both energy and emission
angle and one finds
\begin{equation}
\label{Tvac}
 T_{i}^{\textrm{vac}}(\omega,\theta^2|p_{T0},R^2)=\frac{\as C_i}{\pi}\,
 \rmI_0\left(2\sqrt{\abar 
\ln\frac{p_{T0}}{\omega}\,\ln\frac{R^2}{\theta^2}}\right)
+\om\theta^2\delta(p_{T0}-\om)\delta(R^2-\theta^2)
\end{equation}
where $\abar = \alpha_sC_A/\pi$ and $ \rmI_0(x)$ is the modified Bessel function of rank 0 which increases
exponentially for $x\gg 1$. The second term in the r.h.s.\ represents
the leading parton and the first term is associated with subsequent
gluon emissions. The vacuum fragmentation function is then found to be
\begin{equation}
\label{DvacDLA}
\om D_{i}^{\textrm{vac}}(\omega,\theta^2|p_{T0},R^2)
= \delta(p_{T0}-\om)+\,\frac{C_i}{C_A}\,\sqrt{\frac{2\abar \ln\frac{\om R}{\ktmin}}
{\ln\frac{p_{T0}}{\omega}}}
\ \rmI_1\left(2\sqrt{2\abar 
\ln\frac{p_{T0}}{\omega}\,\ln\frac{\om R}{\ktmin}}\right).
\end{equation}

\paragraph{VLEs in the medium.} In the presence of the medium, the DLA
calculation is modified by two effects~\cite{Caucal:2018dla}: the
presence of a vetoed phase-space for VLEs inside the medium (cf.\
Fig.~\ref{Fig:LundPS}), and the colour decoherence allowing for the
violation of angular ordering by the first emission outside the
medium.
At DL accuracy, MIEs can be formally neglected and their discussion is
postponed to later in this section.
It is helpful to split the medium fragmentation function
$T^{\textrm{med}}_{i}$ in two contributions
(see~\cite{Caucal:2018dla}):
\begin{equation}
\label{fullT}
 T^{\textrm{med}}_{i}(\omega,\theta^2|p_{T0},R^2)=\Theta_{\textrm{in}}(\omega,\theta^2)T_{i}^{\textrm{vac}}+\Theta_{\textrm{out}}(\omega,\theta^2)T_{i,\textrm{out}}
\end{equation}
where the step functions $\Theta_{\textrm{in/out}}$ enforces that an
emission $(\omega,\theta^2)$ belongs to the ``inside'' or ``outside''
region, in the sense of Fig.~\ref{Fig:LundPS}.
The first term,
$\Theta_{\textrm{in}}(\omega,\theta^2)T_{i}^{\textrm{vac}}$,
corresponding to the in-medium contribution, is unmodified compared to
the vacuum.
The outside-medium, $T_{i,\textrm{out}}$, contribution can be
expressed as the product of a vacuum-like cascade inside the medium,
up to an intermediate point $(\omega_1,\theta^2_1)$, followed by a
first emission outside the medium at $(\omega_2,\theta^2_2)$ (possibly
violating angular ordering), and by a standard vacuum cascade from
$(\omega_2,\theta^2_2)$ to the final point $(\omega,\theta^2)$:
\begin{align}
\label{master-eq}
 T_{i,\textrm{out}}(\om,\theta^2|p_{T0},R^2)=  \abar\int_\omega^{p_{T0}}\frac{\dif \omega_1}{\omega_1}
 \int_{\theta^2_c}^{R^2}\frac{\dif \theta^2_1}{\theta^2_1}\,\Theta_{\textrm{in}}(\omega_1,\theta^2_1) \int_\omega^{\omega_1}\frac{\dif \omega_2}{\omega_2}\int_{\theta^2}^{R^2}\frac{\dif \theta^2_2}{\theta^2_2}\,\Theta_{\textrm{out}}(\omega_2,\theta^2_2)&\nn
 T_{i}^{\textrm{vac}}(\om_1,\theta_1^2|p_{T0},R^2)T_{g}^{\textrm{vac}}(\om,\theta^2|\omega_2,\theta^2_2)
 &
\end{align}
The integral over $\theta_2^2$ is not constrained by the angle
$\theta_1^2$  of the previous emission due to absence of angular
ordering for the first emission outside the medium.

The two angular integrations in~\eqn{master-eq} can be performed
analytically (cf.\ Eq.~(\ref{DvacDLA})).
In Ref.~\cite{Caucal:2018dla}, the remaining energy integrations were
performed numerically.
To gain more physical intuition, we now develop an analytic
approximation, which is valid when both the energy and angular
logarithms are larger than $1/\sqrt{\as}$.
We give here the main ingredients of the calculation and defer details
to Appendix~\ref{app:DLA}.

In the limit of interest, the $\delta$ contribution to
$T^{\textrm{vac}}$ (the second term in~(\ref{Tvac})) can be neglected
in both $T^{\textrm{vac}}$ factors in~\eqn{master-eq}, the Bessel
functions can be approximated by their (exponential) asymptotic
behaviour and the integrations can be evaluated in the saddle-point
approximation.

For definiteness, let us consider parameters such that
$\om_L(R) < \ktmin/R$, meaning that the hadronisation line
$\om\theta =\ktmin$ and the medium boundary
$\om_L(\theta)=2/(L\theta^2)$ intersect at
$\om_{\textrm{min}}=L\ktmin^2/2$.
In practice we are interested in the fragmentation function
at energies $\om$ within the range $\om_{\textrm{min}}\ll\om\ll
\om_c$.
The saddle points for $\omega_1$ and $\omega_2$ integrals
are respectively found to be (see Appendix~\ref{app:DLA})
\begin{align}\label{omSP}
 \omega_{1}^{\star}=\sqrt{\frac{p_{T0}(2\qhat)^{1/3}}{R^{4/3}}} = \sqrt{p_{T0}\omega_0(R)}
 \,,\qquad
 \omega_{2}^{\star}=\sqrt{\frac{2\omega}{L\theta^2}}=\sqrt{\om\om_L(\theta)}\,,
\end{align}
with $\omega_0(\theta)\equiv(2\qhat/\theta^4)^{1/3}$ such that
$\omega_0(R)$ is the lowest possible energy for a VLE inside
the medium.
 
Several conditions are needed for these saddle points to control the energy integrations.
First, the integration ranges must be wide enough, $p_{T0}\gg\omega_0(R)$ and
$\om_L(\theta)\gg\om$, to allow for large enough logarithmic
contributions. This translates into the following conditions:
\beq\label{lnPS}
\sqrt{\abar}\,\ln\frac{p_{T0}}{\omega_0(R)}\,\gtrsim\,1
\qquad\mbox{and}\qquad
\sqrt{\abar}\,\ln\frac{\om_L(\theta)}{\om}\,\gtrsim\,1\,.\eeq
Second, for $\omega_{1}^{\star}$ to be a genuine saddle point, it must remain smaller than $\omega_c$, 
meaning
\beq\label{eq:cdt-pt0-omc}
p_{T0} < \om_c \left(\frac{R}{\theta_c}\right)^{4/3}\,=\,\frac{\qhat^{5/3}L^4R^{4/3}}{2^{7/3}}\,.
\eeq
When this condition is satisfied\footnote{In the opposite situation,
  which would occur for sufficiently large $p_{T0}$, the dominating
  region in phase-space is the rectangular region at
  $\omega_c\le\omega_1\le p_{T0}$ and $\theta_c<\theta_1<R$; see
  Appendix~\ref{app:DLA} for details.}  (which is always the case for
us in practice), the integral over $\om_1$ is dominated by relatively
low-energy emissions with $\om_0(\theta) <\om_1<\om_c$, i.e. by the
triangular region of the ``inside medium'' phase-space with energies
below $\om_c$, see Fig.~\ref{Fig:LundPS}.

Third, energy conservation in Eq.~\eqref{master-eq} requires
$\omega_{2}^{\star}\le \omega_{1}^{\star}$ which implies a
$\theta$-dependent upper limit on $\om$.
When computing the fragmentation function using Eq.~(\ref{DfromT}), 
this condition must be satisfied for all the angles $\theta$ that are
integrated over, including lower bound
$\theta_{\textrm{min}}= \ktmin/\omega$.
This defines a critical energy $\omega_{cr}$, obtained for
$\theta=\theta_{\textrm{min}}$, below which the saddle point method works:
\begin{equation}\label{ocr}
  \omega < \omega_{cr}=\big(p_{T0}\om_0(R)\om_{\textrm{min}}\big)^{1/3}=
  \left(\frac{p_{T0}L\ktmin^2(2\hat{q})^{1/3}}{2R^{4/3}}\right)^{1/3}=
  \left(\frac{p_{T0}\ktmin^2}{R^2}\right)^{1/3}\left(\frac{R}{\theta_c}\right)^{2/9}.
\end{equation}

When the conditions in Eqs.~\eqref{lnPS}--\eqref{ocr} are satisfied, the 
saddle point method gives a meaningful approximation for the double differential gluon distribution
in  \eqn{master-eq}, which reads (see Appendix~\ref{app:DLA})
\begin{equation}
\label{Tsaddle}
 T_{i,\textrm{out}}(\om,\theta^2|p_{T0},R^2)\simeq
 \frac{\alpha_s C_i}{4 \pi}\exp\left\{\sqrt{\frac{3\abar}{2}}\,\ln
  \frac{p_{T0}}{\omega_0(R)}\right\}
  \exp\left\{\sqrt{\abar}\,\ln\frac{\omega_{L}(\theta)}{\omega}\right\}
\end{equation}
The first exponential comes from the integrations over $\theta^2_1$ and
$\omega_1$, i.e. over the ``inside'' region, and can be interpreted as the number of partonic
sources generated via VLEs.
The second exponential represents the number of gluons generated by
each of these sources via gluon cascades developing outside the
medium.  This simple factorisation between the ``inside'' and the
``outside'' jet dynamics holds strictly speaking only in the saddle
point approximation (and for energies $\om\le\omega_{cr}$) and is
ultimately a consequence of the colour decoherence which washes out
any correlation between the emission angles outside and inside the
medium.

Integrating~\eqn{fullT} over $\theta$ using Eq.~(\ref{DfromT}) we find
the fragmentation function for $\om\le\omega_{cr}$:\footnote{The
  respective contribution of the first term
  $\propto T_{i}^{\textrm{vac}}$ in \eqn{fullT}, that would be
  non-zero only for $\om>\om_0(R)$, is
  comparatively small, since it lacks the evolution outside the
  medium.}
\begin{align}
\label{frag-DLA}
\om D^{\textrm{med}}_i(\omega)\simeq\frac{\sqrt{\abar}C_i}{4C_A}\exp\left\{\sqrt{\abar}\left(\sqrt{\frac{3}{2}}\ln\,\frac{p_{T0}R^{4/3}}{(2\qhat)^{1/3}}+\ln\,\frac{2\omega}{\ktmin^2L}\right)\right\}.
\end{align}
The integration is dominated by the lower limit, $\theta=\ktmin/\omega$.
Since  ${2\omega}/{\ktmin^2L}=\om/\om_{\textrm{min}}\gg 1$, the second
logarithm in~\eqref{frag-DLA} is positive and $\om D^{\textrm{med}}_i(\omega)$  decreases
when decreasing $\om$.

\begin{figure}[t] 
  \centering
  \begin{subfigure}[t]{0.48\textwidth}
    \includegraphics[page=2,width=\textwidth]{ff.pdf}
    \caption{Fragmentation function at DLA}\label{fig:DLA-dec-D}
  \end{subfigure}
  \hfill
  \begin{subfigure}[t]{0.48\textwidth}
    \includegraphics[page=1,width=\textwidth]{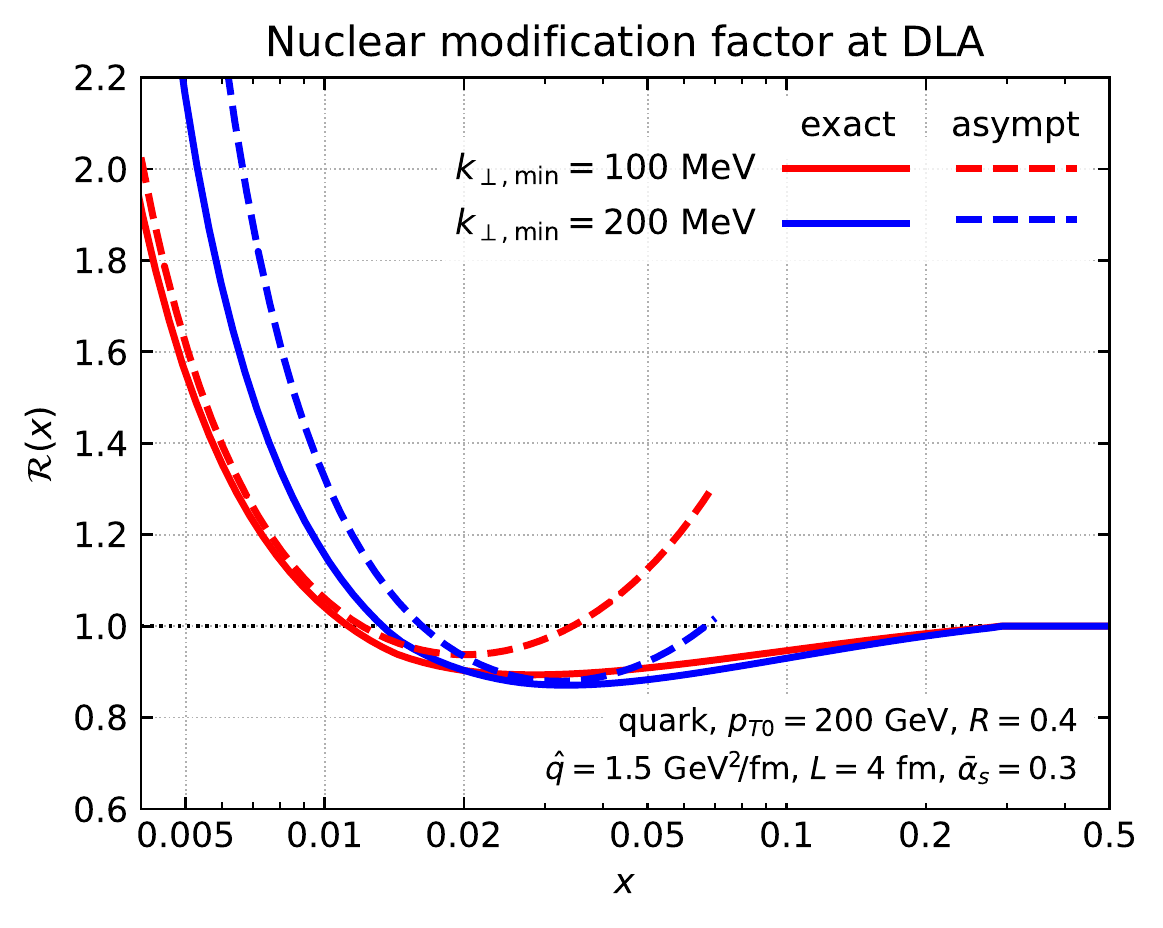}
    \caption{Nuclear modification $\mathcal{R}(x)$ at DLA}\label{fig:DLA-dec-R}
  \end{subfigure}
  \caption{\small Comparison of the exact calculation of fragmentation
    functions (solid lines) and the asymptotic approximations (dashed
    lines). }\label{Fig:DLA-dec}
\end{figure}

Our predictions are shown in Fig.~\ref{Fig:DLA-dec} for the
fragmentation function in Fig.~\ref{fig:DLA-dec-D} and the nuclear
modification factor $\mathcal{R}_i(x|p_{T0})$ in
Fig.~\ref{fig:DLA-dec-R}.
These plots compare the exact results at DLA based on Eq.~(\ref{Tvac}) and
(the numerical integration of) Eq.~(\ref{master-eq}) for the vacuum
and medium results respectively, to their asymptotic counterparts. The
latter are obtained by taking the asymptotic behaviour of~(\ref{Tvac})
in the vacuum case and by using the saddle-point approximation
Eq.~(\ref{frag-DLA}) for the medium results.
In Fig.~\ref{fig:DLA-dec-R} we consider two different values for the
IR cutoff $\ktmin$ (blue: $\ktmin=200$~MeV, red: $\ktmin=100$~MeV).
 Overall we see a good agreement, which is
moreover improving when $\ktmin$
decreases, i.e.\ when the phase-space increases and the saddle point
method becomes more reliable.

The fact that the ratio $\mathcal{R}_i(x|p_{T0})$ increases at small
$\om$ can be traced back to angular ordering and the associated
humpback plateau~\cite{Dokshitzer:1991wu}.
Unlike the double-differential gluon distribution \eqref{Tvac} which
keeps increasing when decreasing $\omega$ at fixed $\theta$, the
vacuum fragmentation function $\om D^{\textrm{vac}}_i(\omega)$ in
\eqn{DvacDLA} develops a maximum at
$\omega\simeq \om_{\rm hump}=(E\ktmin/R)^{1/2}$ and decreases very
fast for $\om$ below $\om_{\rm hump}$.
This is due to the fact that the angular phase-space at
$\ktmin/\om<\theta <R$ permitted by angular ordering shrinks to zero
when decreasing $\om$. For sufficiently small $\om$, namely such
that\footnote{The upper limit $p_{T0} \ktmin^2/R^2$ is smaller than
  $\omega_{cr}^3$ guaranteeing the validity of the saddle-point method.}
$\om^3\lesssim p_{T0} \ktmin^2/R^2$, the denominator
$\om D^{\textrm{vac}}_i(\omega)$ in the medium/vacuum ratio
$R_i(x|p_{T0})$ decreases faster with $1/\om$ than the respective
numerator $\om D^{\textrm{med}}_i(\omega)$ (see also
Fig.~\ref{fig:DLA-dec-D}), so the ratio is increasing.

\subsection{Beyond DLA: Monte-Carlo results}\label{sec:smallx-MC-results}

In this section we want to extend the DLA arguments from the previous
section to include all the ingredients in our physical picture of jet
quenching.
Our ultimate goal is to provide a deeper understand of the MC results
presented in Sect.~\ref{sec:MC}.

For this purpose, it is convenient to think in terms of the factorised
picture emerging from our DLA calculation which allows us to write (for $\om\le\om_{cr}$, cf. \eqn{ocr})
\begin{equation}
\label{factorization}
 \om D^{\textrm{med}}(\om)\simeq \mathcal{N}_{\textrm{in}}\times\Big(\om \frac{\dif N^{\textrm{out}}}{\dif \om}\Big)
\end{equation}
where $\mathcal{N}_{\textrm{in}}$ is the multiplicity of partonic sources produced by the
jet evolution inside the medium and $\om\dif N^{\textrm{out}}/\dif\om$ is the fragmentation function
generated outside the medium by any of these sources.
This picture is a consequence of colour decoherence which allows the
first out-of-medium emission to be emitted at any angle.
This factorisation is not expected to hold beyond DLA, but can still
be used for qualitative considerations.

Beyond DLA, several competing expects should be considered.
\texttt{(i)} 
VLEs are emitted with the full (DGLAP) splitting functions
(including energy conservation) and with a running coupling.
These effects are expected to reduce both factors in
\eqn{factorization}.
\texttt{(ii)} Adding the intra-jet MIEs enhances the
multiplicity $\mathcal{N}_{\textrm{in}}$ of the partonic sources.
\texttt{(iii)} Direct contributions of the MIEs to the fragmentation
function $D^{\textrm{med}}(\om)$ are also possible, but are expected
to be a small effect for the jet kinematics ($p_{T0}\sim 200$~GeV,
$x\le 0.02$) and medium parameters (see Table.~\ref{tab:parameters})
considered in this paper. Indeed, the relevant energies
$\omega\lesssim 2$~GeV are softer than the medium scale
$\ombr\sim 4$~GeV for multiple branching meaning that these MIEs would
be deviated outside the jet.
  
To test these expectations under realistic conditions, we
perform MC simulations for inclusive jets (using the full Born-level
hard spectrum) with $200\le p_T\le 251$ GeV and $|y|\le2.1$, and with
three different scenarios:
\texttt{(a)} the partons from the hard scattering are showered via VLEs only;
\texttt{(b)} the partons from the hard scattering are showered via both VLEs
and MIEs, but angular ordering is enforced all along the shower,
including for the first emission outside the medium (labelled ``no
decoherence'');
\texttt{(c)} the physical case where the partons from the hard scattering are showered via both VLEs and MIEs and the angle of the first emission outside the medium is unconstrained.

\begin{figure}[t] 
  \centering
  \begin{subfigure}[t]{0.48\textwidth}
    \includegraphics[page=1,width=\textwidth]{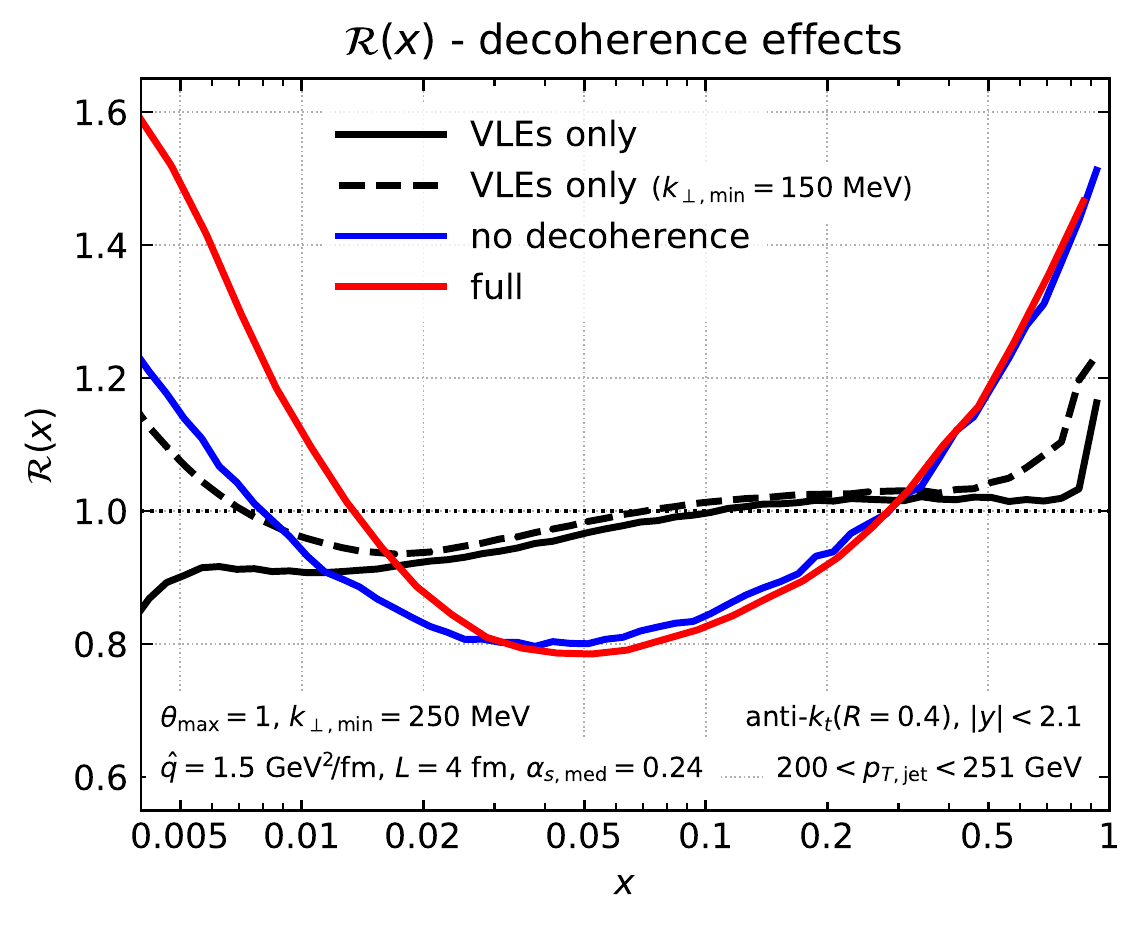}
    \caption{Different physical scenarios: with and without MIEs, and with and without
  violation of angular ordering.}\label{Fig:MC-dec-decoherence}
  \end{subfigure}
  \hfill
  \begin{subfigure}[t]{0.48\textwidth}
    \includegraphics[page=2,width=\textwidth]{plot-MC-decoherence.pdf}
    \caption{Comparison between quark- and gluon-initiated
      monochromatic jets and ``full'' jets including a  convolution with the initial, hard spectrum.}\label{Fig:MC-dec-spectrum}
  \end{subfigure}
  \caption{\small Nuclear effects on the fragmentation function at
    small $x$. Left figure: 3 different physical scenarios,
  }\label{Fig:MC-dec} 
\end{figure}

The MC results for $\mathcal{R}(x)$ are shown in
Fig.~\ref{Fig:MC-dec-decoherence} for each of these three setups.
The black curves correspond to setup \texttt{(a)} for two 2 different IR
cutoffs (solid: $\ktmin=200$~MeV, dashed: $\ktmin=150$~MeV).
compared to the DLA results in Fig.~\ref{fig:DLA-dec-R} the medium
enhancement is strongly reduced and can even be replaced by a
suppression for larger values of $\ktmin$.

Switching on MIEs leads to a robust nuclear
enhancement as visible from the blue curve which corresponds to setup
\texttt{(b)} with $\ktmin=200$~MeV.
This enhancement is even more pronounced for setup \texttt{(c)}
corresponding to the red curves in Fig.~\ref{Fig:MC-dec-decoherence}.
This new enhancement is easily associated with the fact that the first
``outside'' emission can be sourced by any ``inside'' emissions while
in setup \texttt{(b)} it can only be sourced by ``inside'' emissions
at larger angles.\footnote{For setup \texttt{(b)} the
  factorisation~(\ref{factorization}) is obviously violated as
  ``inside sources'' and ``outside emissions'' are correlated by
  angular ordering.}
Incidentally, the comparison between the blue and the red curves also
shows that the decoherence has no sizeable effects at $x\sim 1$.

For a more detailed understanding, we compare in
Fig.~\ref{Fig:MC-dec-spectrum} the results for $\mathcal{R}(x)$ with
the ratio $\mathcal{R}_i(x|p_{T0})$ corresponding to monochromatic
jets with $p_{T0}= 200$~GeV, for both quark-initiated ($i=q$, magenta,
dashed-dotted curve) and gluon-initiated ($i=g$, green, dashed, curve)
jets.
The small-$x$ enhancement appears to be stronger in the case where the
LP is a quark, rather than a gluon. Although this might look
surprising at first sight, one should recall that the
dominant $C_i$-dependence for monochromatic jets cancels out in the
medium/vacuum ratio $\mathcal{R}_i(x|p_{T0})$.
The differences between the quark and gluon curves visible in
Fig.~\ref{Fig:MC-dec-spectrum} is attributed to more subtle
sub-leading effects.
For example,  a gluon jet loses more energy than a quark jet via MIEs at large
angles and hence has a (slightly) smaller energy phase-space for
radiating outside the medium (and inside the jet).

\section{Jet fragmentation into subjets}\label{sec:frag-subjets}

The fragmentation function defined by Eq.~\eqref{frag-def} is not an 
infrared-and-collinear (IRC) safe observable.
It is sensitive to the details of hadronisation which is not included
in our present approach.
This translates in the strong dependence, observed in
Fig.~\ref{Fig:MCunphys}, on the cut-off scale $\ktmin$
which regulates the infrared behaviour of our partonic cascade.
This strong dependence on $\ktmin$ is also present in the analytic
calculations of sections~\ref{sec:x=1} and~\ref{sec:smallx}.

To circumvent this theoretical problem, we propose in this section a
different observable which uses subjets instead of individual hadrons
to characterise the jet fragmentation.
This observable is IRC-safe by construction and is therefore expected
to be less sensitive to non-perturbative effects in general and to
our $\ktmin$ cut-off in particular.
There are several ways to define a jet fragmentation function in terms
of subjets, e.g.\ using different jet algorithms or keeping different
branches of the clustering tree.
The definition we propose below relies on the Cambridge/Aachen
algorithm~\cite{Dokshitzer:1997in,Wobisch:1998wt}.
While other approaches, like those based on the $k_t$
algorithm~\cite{Catani:1993hr}, show a similar behaviour,
using the Cambridge/Aachen algorithm appears to be slightly more
sensitive to medium effects and easier to study analytically.

\subsection{Definition and leading-order estimate in the vacuum}

The fragmentation function $\mathcal{D}_{\textrm{sub}}(z)$ for jet
fragmentation into subjets is defined as follows. For a given jet with
transverse momentum $p_{T,\textrm{jet}}$, we iteratively decluster the
jet using the Cambridge/Aachen algorithm following the hardest branch
(in $p_T$).
At each step, this produces two subjets $p_1$ and $p_2$, with
$p_{T1}>p_{T2}$.
When the relative transverse momentum of the splitting, $k_\perp = p_{T2}\sqrt{\Delta
  y_{12}^2+\Delta\phi_{12}^2}$, is larger than a (semi-hard) cut-off
$k_{\perp,\textrm{cut}}$, we compute and record the splitting
fraction $z=\frac{p_{T2}}{p_{T1}+p_{T2}}$ of the splitting ($0 < z < 1/2$).
The procedure is iterated with the harder branch $p_1$ until it can no
longer be de-clustered.
The fragmentation function into subjets is then defined as the density
of subjets passing the $k_\perp>k_{\perp,\textrm{cut}}$ criterion
normalised by the total number of jets:\footnote{We use the notation
  $z$  for the splitting fraction to emphasise that it is defined w.r.t.
  the parent subjet, in contrast with the longitudinal momentum
  fraction $x$ used in the previous sections which is defined as a
  fraction of the total jet momentum $p_{T,\textrm{jet}}$.}
\begin{equation}\label{def-dsub}
 \mathcal{D}_{\textrm{sub}}(z) \equiv \frac{1}{N_\textrm{jets}}\frac{\dif N_{\textrm{sub}}}{\dif z}
\end{equation}

The cut-off scale $k_{\perp,\textrm{cut}}$ regulates the infrared
behaviour, guaranteeing that $\mathcal{D}_{\textrm{sub}}(z)$ be an
IRC-safe observable.
As long as $k_{\perp,\textrm{cut}}\gg \ktmin\sim \Lambda_{_{\rm QCD}}$
we therefore expect small non-perturbative effects and a small
dependence on the (non-physical) $\ktmin$ parameter.

Note that the definition is similar to measuring the Iterated Soft
Drop multiplicity~\cite{Frye:2017yrw} differentially in $z$.
It is also directly similar to the primary Lund-plane
density~\cite{Dreyer:2018nbf}, $\rho(\theta,k_\perp)$, integrated over all
angles $\theta$ satisfying the $k_{\perp,\textrm{cut}}$ condition at fixed
$x=k_\perp/(\theta p_{T,\text{jet}})$.

In the soft-and-collinear approximation, corresponding to the
double-logarithmic accuracy for $\mathcal{D}_{\textrm{sub}}(z)$, the
vacuum distribution is simply
\begin{align}\label{fragISDvac2}
  \mathcal{D}^{\textrm{vac}}_{\textrm{sub}}(z)
  & \simeq\left[\int_{0}^R\frac{\dif\theta}{\theta}\,
  \frac{2\alpha_s(z\theta p_{T,\textrm{jet}})}
  {\pi z}\Theta(z\theta p_{T,\textrm{jet}} -
  k_{\perp,\textrm{cut}})\right]
  \times\sum_{i=q,g} C_i\, f_i^{\textrm{vac}}(p_{T,{\textrm{jet}}}),\nn
  & \overset{\text{f.c.}}\simeq 
  \frac{2\alpha_s}{\pi z} \log\left(\frac{z Rp_{T,\textrm{jet}}}{k_{\perp,\textrm{cut}}} \right)
  \times\sum_{i=q,g} C_i\, f_i^{\textrm{vac}}(p_{T,{\textrm{jet}}}),
\end{align}
where
$f_{q(g)}^{\textrm{vac}}(p_{T,{\textrm{jet}}})$ is the Born-level
cross-section for quark (gluon) production with transverse momentum
$p_{T,{\textrm{jet}}}$ normalised to the total number of jets, as
defined in Eq.~\eqref{quenching-factor}.
The second line in the above equation gives the result for a
fixed-coupling approximation.

\subsection{Nuclear modification for $\mathcal{D}_{\textrm{sub}}(z)$: Monte-Carlo results}

In this section, we provide Monte Carlo results for the nuclear
modification factor for the fragmentation function into subjets,
defined as
$\mathcal{R}_{\textrm{sub}}(z)\equiv
\mathcal{D}^{\textrm{med}}_{\textrm{sub}}/\mathcal{D}^{\textrm{vac}}_{\textrm{sub}}$.

\begin{figure}[t] 
  \centering
  \includegraphics[page=1,width=0.48\textwidth]{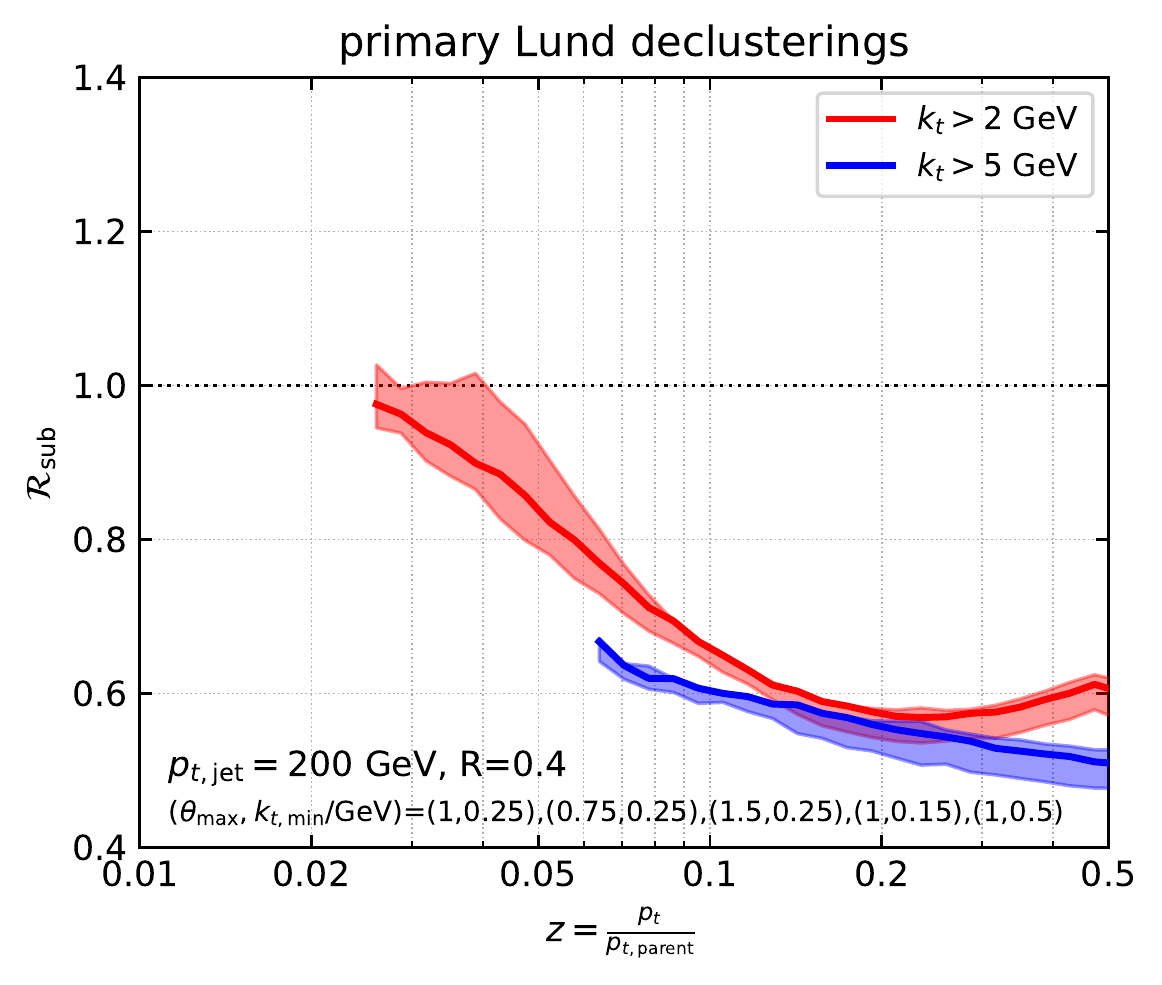}
  \hfill
   \includegraphics[page=2,width=0.48\textwidth]{plot-subjets.pdf}
  \caption{\small Monte Carlo
  results for the nuclear modification factor  $\mathcal{R}_{\textrm{sub}}(z)$
  for the fragmentation function into subjets, 
  for jets with $p_{T,{\textrm{jet}}}>200$~GeV (left)
  and $p_{T,{\textrm{jet}}}>500$~GeV (right) and for 2 values of the lower momentum
  cut-off $k_{\perp,\textrm{cut}}$ (2 and 5~GeV). The bands show the 
  variability of our results w.r.t. changes  in the ``unphysical'' parameters  around their
central values  $\theta_{\textrm{max}}=1$ and $\ktmin=250$~MeV.}
\label{Fig:fragsubjet} 
\end{figure}

\begin{figure}[t] 
  \centering
  \includegraphics[page=3,width=0.48\textwidth]{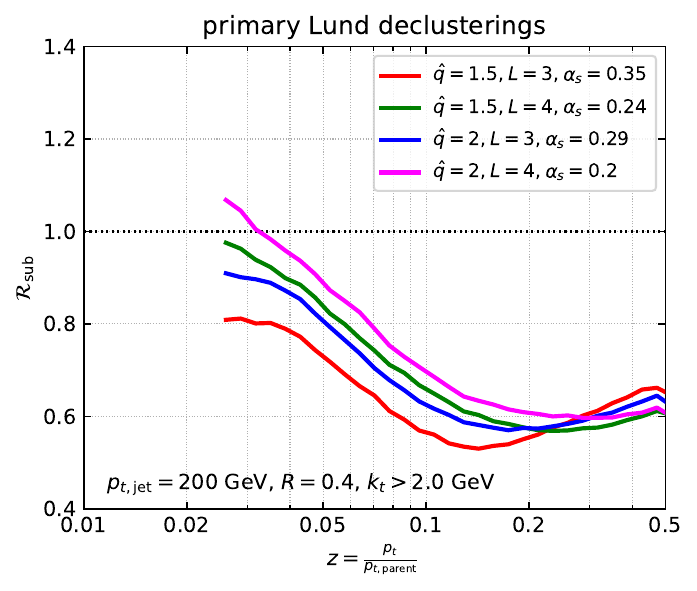}
  \hfill
   \includegraphics[page=5,width=0.48\textwidth]{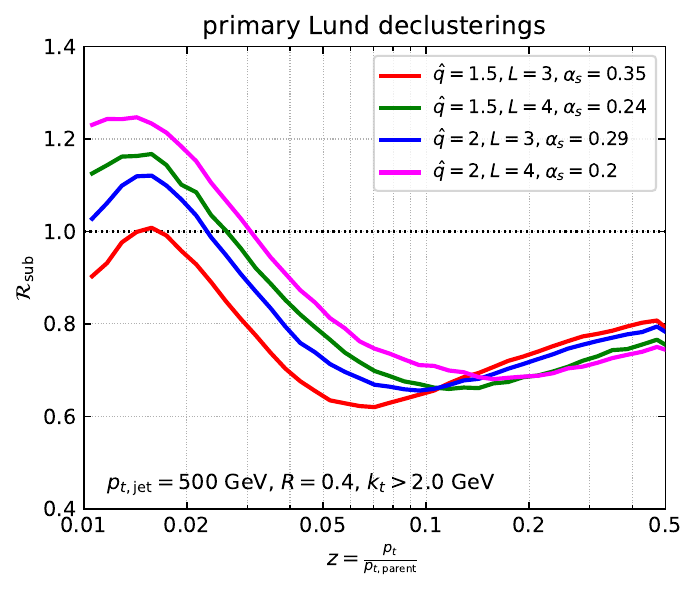}
  \caption{\small  Monte Carlo results for the nuclear modification factor 
  $\mathcal{R}_{\textrm{sub}}(z)$ for the values of the medium parameters that reproduce the
ATLAS $R_{AA}$ ratio (cf. Fig.~\ref{Fig:MC-pheno}), for the same two
ranges in $p_{T,\textrm{jet}}$ as in~Fig.~\ref{Fig:fragsubjet} and for
$k_{\perp,\textrm{cut}}=2$~GeV. The unphysical parameters are fixed to 
$\theta_{\textrm{max}}=1$ and $\ktmin=250$~MeV.
   }
\label{Fig:subjet-med} 
\end{figure}

As for the study of the jet fragmentation function $\mathcal{D}(x)$,
we first study the dependence of the the fragmentation function into
subjets, $\mathcal{D}_{\textrm{sub}}(z)$, on the non-physical
parameters $\theta_{\textrm{max}}$ and $\ktmin$ of our Monte Carlo.
This is shown in Fig.~\ref{Fig:fragsubjet} for two different jet $p_T$
cuts (200 and 500~GeV) and two different lower cut-offs
$k_{\perp,\textrm{cut}}$ (2 and 5~GeV).
The medium parameters are taken as their default values (cf.\
Table~\ref{tab:parameters}) and the non-physical parameters are varied
as for Fig.~\ref{Fig:MCunphys}.
As expected, the uncertainty bands in Fig.~\ref{Fig:fragsubjet} are
much smaller than what was observed in Fig.~\ref{Fig:MCunphys},
confirming that the (IRC-safe) fragmentation function into subjets
$\mathcal{D}_{\textrm{sub}}(z)$ is under much better perturbative
control than (the IRC-unsafe) $\mathcal{D}(x)$.

That said, we must keep in mind that taking $k_{\perp,\textrm{cut}}$
large-enough to guarantee $k_{\perp,\textrm{cut}}\gg \ktmin \sim
\Lambda_{\text{QCD}}$  also cuts some of the medium effects occurring
below this cut.
E.g., it removes the direct contributions to
$\mathcal{D}_{\textrm{sub}}(z)$ coming from MIEs with transverse
momenta $k_\perp\lesssim k_{\perp,\textrm{cut}}$.
One should therefore choose the free parameter
$k_{\perp,\textrm{cut}}$ such as to simultaneously minimise the effects
of hadronisation and highlight the interesting medium effects.

In Fig.~\ref{Fig:subjet-med}, we show the subjet fragmentation
function for the values of the medium parameters that reproduce the
ATLAS $R_{AA}$ ratio (cf. Fig.~\ref{Fig:MC-pheno}), for the same two
values of $p_{T,\textrm{jet}}$ as in~Fig.~\ref{Fig:fragsubjet} and for
$k_{\perp,\textrm{cut}}=2$~GeV.
Compared to Fig.~\ref{Fig:MC-pheno}, we notice that the curves are
less degenerate at small and intermediate values of $z$. Most
importantly, the dependence on the medium parameters is larger than
the uncertainty bands related to non-physical parameters shown
in~Fig.~\ref{Fig:fragsubjet}.

\subsection{Analytic studies of the nuclear effects}
\label{sec:DsubAnalytic}

In this section, we would like to disentangle, based on physics considerations
and simple analytic calculations, the various nuclear effects contributing to the behaviour
observed in the MC results in Fig.~\ref{Fig:subjet-med}.
To understand how  Eq.~\eqref{fragISDvac2} is affected by the medium, it is sufficient
  to consider jets made of a single splitting (i.e.\ two subjets) with
  $k_\perp\ge k_{\perp,\textrm{cut}}$.
For definiteness, all the numerical results shown in this subsection
correspond to $k_{\perp,\textrm{cut}}=2$~GeV.

\paragraph{Vetoed region.} When only VLEs are taken into account, the
leading medium effect is the vetoed region.
Its effect is straightforwardly
included in \eqn{fragISDvac2} by inserting the 
step-function $\Theta_{\notin\textrm{veto}}$ defined in
Eq.~\eqref{step-veto} within the integrand.
The largest $k_\perp$ in the vetoed region is $Q_s\equiv (2\hat{q} \om_c)^{1/4} =
(\hat q L)^{1/2}$ which is about 2.4~GeV for our default choice of
medium parameters.
The vetoed region has thus no effect for
$k_{\perp,\textrm{cut}}=5$~GeV and only a small effect for
$k_{\perp,\textrm{cut}}=2$~GeV (see
Fig.~\ref{Fig:LundPS} for an illustration).

\begin{figure}[t] 
  \centering
  \begin{subfigure}[t]{0.48\textwidth}
    \includegraphics[page=2,width=\textwidth]{frag-ISD-analytic.pdf}
    \caption{Analytic results}\label{Fig:fragISD-LL-analytic}
  \end{subfigure}
  \hfill
  \begin{subfigure}[t]{0.48\textwidth}
    \includegraphics[page=1,width=\textwidth]{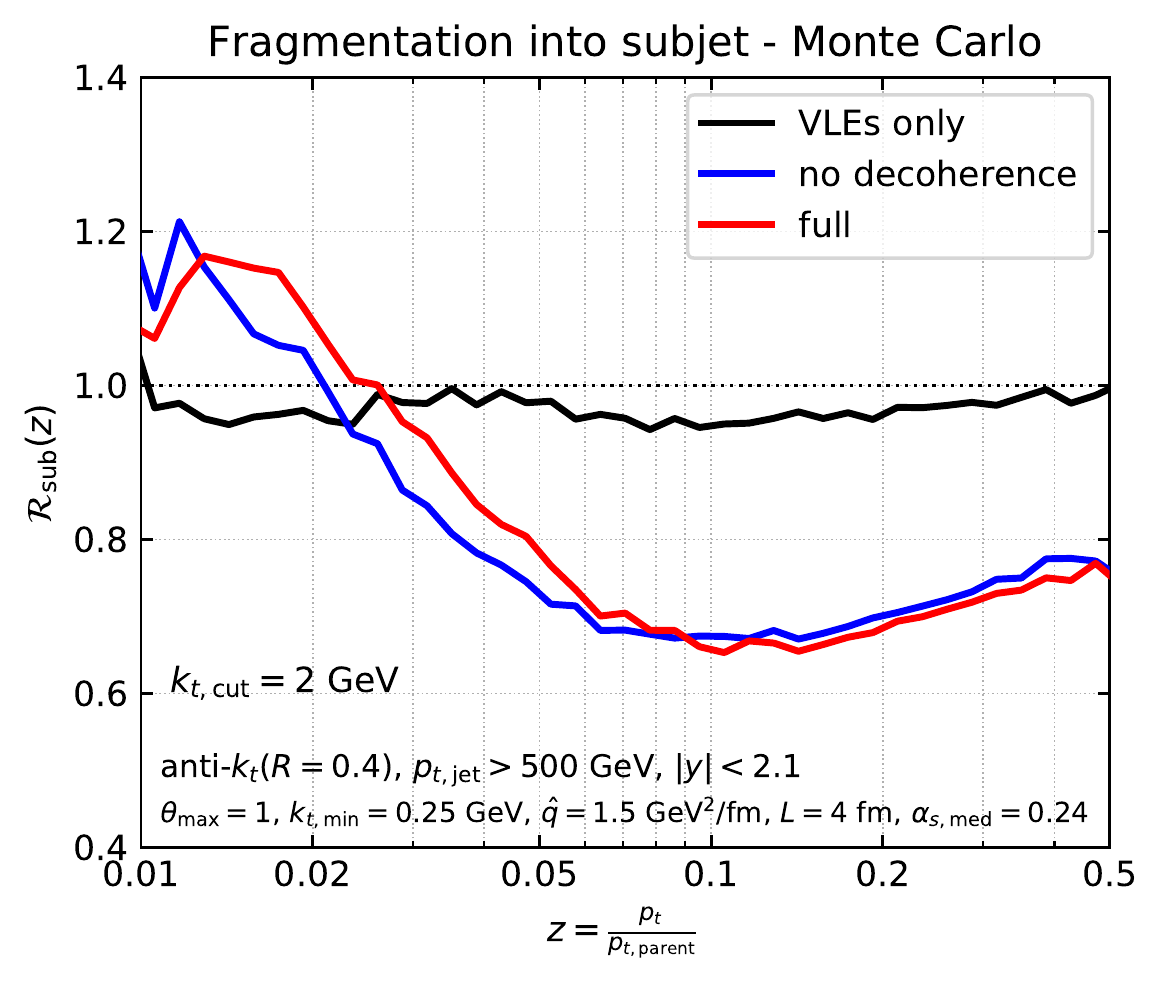}
    \caption{Results from Monte Carlo simulations}\label{Fig:fragISD-LL-mc}
  \end{subfigure}
  \caption{\small  Disentangling nuclear effects on the subjet fragmentation function.
  Left: analytic approximations illustrating the effects of the vetoed region, the energy
  loss at large angles, and the intra-jet MIEs. Right: MC calculations which illustrate
  the importance of  MIEs and the lack of sensitivity to violations of angular ordering.}
\label{Fig:fragISD-LL} 
\end{figure}

This is confirmed both by our analytic calculations, based on \eqn{fragISDvac2} 
with the additional constraint $\Theta_{\notin\textrm{veto}}$, 
and by MC simulations with only VLEs shown as the black curves in
Fig.~\ref{Fig:fragISD-LL}.
Of course,
one could enhance the effect of the vetoed region by decreasing the value of $k_{\perp,\textrm{cut}}$,
but this would also amplify the sensitivity of $\mathcal{D}_{\textrm{sub}}(z)$  
to the non-perturbative, soft, emissions.

Incidentally, the previous discussion also shows that, for the ranges
of $k_{\perp,\textrm{cut}}$ considered here, the VLEs which control
$\mathcal{D}_{\textrm{sub}}(z)$ do either occur in the ``inside''
region of the phase-space in Fig.~\ref{Fig:LundPS}, or at very small
angles $\theta\lesssim \theta_c$ in the ``outside'' region. They are
therefore not significantly affected by colour decoherence. To check
that, we have performed MC calculations with and without the effects
of decoherence (i.e.\ by enforcing or not angular ordering for the first
outside emission). The results, shown by the red and blue
curves in Fig.~\ref{Fig:fragISD-LL-mc}, respectively, are indeed
very close to each other.
 
\paragraph{Energy loss at large angles.}   
From the discussion in Sect.~\ref{sec:x=1}, we already know
that the energy loss by a (sub)jet via MIEs at large angles $\theta\gtrsim R$ may have two main effects
on a substructure observable such as  $\mathcal{D}_{\textrm{sub}}(z)$: \texttt{(i)} a shift between the measured
value $z$ of the splitting fraction and the respective value at the time of splitting, and  \texttt{(ii)} a bias introduced by the steeply falling initial spectrum which favours jets losing less energy 
than average jets, with the second effect being larger than the first one.
The same two effects are still at play for $\mathcal{D}_{\textrm{sub}}(z)$. As in the case of the
standard fragmentation function discussed  in Sect.~\ref{sec:x=1}, we expect the effects of the energy
loss to be more important for relatively large values $z\gtrsim 0.1$ of the splitting fraction. However,
their effects on $\mathcal{R}_{\textrm{sub}}(z)$ is opposite to those on $\mathcal{R}(x)$: unlike the hard-fragmenting jets, which lose {\it less} energy than
the average jets (leading to an enhancement in $\mathcal{R}(x)$ at $x\gtrsim 0.5$), the 
jets selected by $\mathcal{D}_{\textrm{sub}}(z)$ lose {\it more} energy than the average jets, so we expect
a {\it nuclear suppression}, $\mathcal{R}_{\textrm{sub}}(z)< 1$, at sufficiently large $z$.
The main reason for this larger energy loss is the following: the jets included in $\mathcal{D}_{\textrm{sub}}(z)$
involve at least two (relatively hard) subjets with  $z\gtrsim 0.1$ and $k_\perp > k_{\perp,\text{cut}}$.
For the typical values of $z$ and $k_\perp$, the angle $\theta\simeq k_\perp/p_{T2}$ between
these two subjets is larger than the critical angle $\theta_c$ characterising the angular resolution
of the plasma ($\theta_c\lesssim 0.06$, see Table~\ref{tab:parameters}).
Accordingly the two subjets lose energy independently from each other and the whole jet
loses more energy than a typical jet from the inclusive sample $N_\textrm{jets}$
\cite{Mehtar-Tani:2017ypq,Caucal:2019uvr} which also includes single-prong jets, as well as two-prong configurations with $\theta <
\theta_c$.

%


%
This discussion is in qualitative agreement with the MC results in
Fig.~\ref{Fig:fragISD-LL-mc}, except at very small $z$ where new
effects discussed below contribute. 
For a more quantitative argument, 
we notice that, if one neglects the
shift in the value of $z$, then the energy loss at large angles
affects only the quark- and gluon-jet ``fractions'' $f_i^{\textrm{med}}$
in~\eqn{fragISDvac2}. These should be computed following
Eq.~\eqref{quenching-factor}, with different energy losses in the
numerator and respectively the denominator.
In the numerator, $\mathcal{E}_i^{n=2}$ is the
energy loss of jets having two subjets with transverse
momentum balance $z$ and angle $\theta$ ($p_T\equiv p_{T,\textrm{jet}}$)
\begin{equation}
\label{incoh-subjet-eloss}
 \mathcal{E}_i^{n=2}(z,\theta)=\mathcal{E}_i((1-z)p_{T},R)+\mathcal{E}_g(z p_{T},R) \qquad
 \textrm{ if }(z,\theta)\in\textrm{ inside region,}
\end{equation}
whereas in the denominator, $ \mathcal{E}_i= \mathcal{E}_i(p_{T},R)$.
Using the energy loss as a function of $p_T$ and $R$ extracted from the MC
simulations in Ref.~\cite{Caucal:2019uvr} in
Eqs.~\eqref{quenching-factor} and~\eqn{fragISDvac2}, one obtains the
dashed, green, curve in Fig.~\ref{Fig:fragISD-LL-analytic}.
This indeed shows a nuclear suppression,
$\mathcal{R}_{\textrm{sub}}(z)<1$. The suppression is more pronounced
at large $z$, as anticipated,
 since the discrepancy (in terms of energy loss) between the special jets selected by 
$\mathcal{D}^{\textrm{med}}_{\textrm{sub}}(z)$ and the average jets increases with $z$.

\paragraph{Intra-jet MIEs.} A relatively hard subjet with $k_\perp > k_{\perp,\textrm{cut}}$ may also be 
created by a semi-hard MIE, with energy $\om\gtrsim\ombr$, which remains inside the jet.
To leading order, the respective contributions from VLEs and MIEs can be simply added together,
as in \eqn{Elossred}.
Compared to the latter, the calculation of
$\mathcal{D}^{\textrm{med}}_{\textrm{sub}}(z)$ must also keep the
information about the emission angle, in order to ensure the condition
$k_{\perp}>k_{\perp,\textrm{cut}}$.
We therefore write
\begin{equation}
\label{frag-isd-tot}
 \mathcal{D}^{\textrm{med}}_{\textrm{sub}}(z)=\left[
\int_{0}^R\dif\theta\left(\frac{2\alpha_s(k_\perp)}{\pi z \theta}\Theta_{\notin\textrm{veto}}+\sqrt{\frac{2\omega_c}{p_{T,\textrm{jet}}}}\frac{\alpha_{s,\textrm{med}}}{\pi z^{3/2}}\mathcal{P}_{\mathcal{B}}(z,\theta)\right)\Theta(k_\perp-k_{\perp,\textrm{cut}})\right]\sum_{i=q,g} C_i f_i^{\textrm{med}}
\end{equation}
where $k_\perp=z\theta p_{T,\textrm{jet}}$ and
$\mathcal{P}_{\mathcal{B}}(z,\theta)= 2\theta
\omega^2\Gamma(0,\omega^2\theta^2/Q_s^2)/Q_s^2$, with
$\omega\simeq zp_{T,\textrm{jet}}$ and $Q_s^2=\hat q L$, is the
angular distribution due to transverse momentum broadening after
emission, averaged over all the emission times between $0$ and
$L$~\cite{Mehtar-Tani:2016aco,Caucal:2019uvr}.
In writing \eqn{frag-isd-tot}, we have assumed for simplicity that the
energy loss at large angles is given by \eqn{incoh-subjet-eloss} for
both the vacuum-like and medium-induced emissions that generates the
subjets.
This rough approximation could be relaxed in practice,
but is sufficient for our illustrative purposes.
The distribution $\mathcal{P}_{\mathcal{B}}(z,\theta)$ for MIEs is
rather strongly peaked near $k_\perp\sim Q_s$ \cite{Caucal:2019uvr} so
its corresponding contribution to \eqn{frag-isd-tot} is expected to be
important only when $k_{\perp,\textrm{cut}}\lesssim Q_s$, in which
case it should be rapidly increasing at small $z$.
This is in agreement with the MC results in
Figs.~\ref{Fig:fragsubjet} and~\ref{Fig:subjet-med}, which show an
enhancement at small $z$ for $k_{\perp,\textrm{cut}}=2$~GeV and no
visible enhancement for $k_{\perp,\textrm{cut}}=5$~GeV. 
(Note that $Q_s^2$ vary between 4.5 and 8~GeV$^2$ for 
the different curves shown in these figures.)
  
\eqn{frag-isd-tot} includes all the medium effects discussed in this
section.
The red curve in Fig.~\ref{Fig:fragISD-LL-analytic} shows the result
of numerically evaluating the integral in~\eqn{frag-isd-tot}. The new
enhancement at small $z$ compared to the dashed, green, curve is due
to the intra-jet MIEs.
The overall behaviour agrees well with the full MC results shown in
Fig.~\ref{Fig:fragISD-LL-mc} as well as with
Figs.~\ref{Fig:fragsubjet} and~\ref{Fig:subjet-med}.

\section{Conclusions}
\label{sec:conclusions}

In this paper, we have studied the fragmentation of a jet propagating
through a dense quark-gluon plasma, using a recently-developed pQCD
framework in which the vacuum-like and the medium-induced branchings
 in the parton shower are factorised in time.
We have presented both numerical simulations, using a Monte
Carlo implementation of our framework, and semi-analytic calculations.

Our main conclusion is that this approach provides a good, qualitative and even
semi-quantitative, description for the main nuclear effects observed in the relevant 
data at the LHC: an enhancement in the jet fragmentation function at both small 
($x\ll 1$) and large ($x\gtrsim 0.5$) values for the parton longitudinal momentum
fraction $x= p_T/p_{T,\textrm{jet}}$.
This good agreement is obtained for values of the physical parameters that
characterise the medium ($\hat{q}$, $L$ and $\amed$) which were shown
in a previous study to agree with the jet measured nuclear modification factor $R_{AA}$.
Since the fragmentation function is not an infrared-and-collinear-safe
quantity in pQCD, our calculations show a strong dependence on the
kinematic cutoff $k_{\perp,\textrm{min}}$ which can be viewed as
playing the role of a confinement scale in our (parton-level)
framework.
Yet, insofar as $k_{\perp,\textrm{min}}$ is varied within reasonable
limits, our result remain in qualitative agreement with the LHC measurements.

The physical interpretation of our results is greatly facilitated by our analytic studies,
that we have separately developed using approximations valid either at
large $x$ or at  small $x$.
These studies have revealed that the nuclear effects visible in the medium/vacuum ratio for the
fragmentation function generally involve an interplay between several microscopic phenomena.
These phenomena can either change  the fragmentation pattern of a ``monochromatic'' jet
(i.e.\ a jet initiated by a leading parton of a given flavour and energy), 
or modify the proportion of ``monochromatic'' jets which contribute to the 
fragmentation function at a given value of $x$ (within the spectrum of
jets produced via hard scattering).

Specifically we have found that the partons contributing to the in-medium 
fragmentation function at small-$x$ are predominantly produced via VLEs and that their excess 
w.r.t.\ the vacuum is the combined result of two mechanisms amplifying each other: 
the enhanced angular phase-space available
to the first emission outside the medium (which, due to the colour decoherence of its
emitters, is not constrained by angular ordering) and the additional sources for soft VLEs
coming from relatively hard, intra-jet, MIEs. At small-$x$, the bias introduced by the
initial production spectrum, although numerically important, does not alter the overall
qualitative behaviour.

The situation at large $x$, $x\gtrsim 0.5$, is radically different. We have found that
the medium effects on the fragmentation function of monochromatic jets, although separately sizeable
and physically interesting, act in opposite directions leaving only a
small effect on the final result.
Their net effect is too small to be distinguished from the significantly 
larger nuclear enhancement generated by the 
bias introduced by the initial hard spectrum. This bias favours hard-fragmenting
jets initiated by a quark because they lose less energy towards the medium than the average jets.
One may be able to avoid, or at least reduce, this bias by looking at rare $\gamma$-jet,
or $Z$-jet events (where the energy of the vector boson offers an estimate
for the initial energy of the jet)  \cite{Chatrchyan:2012gt,Sirunyan:2017jic}, 
or by using the ``quantile'' strategy proposed in \cite{Brewer:2018dfs} in the analysis of
the nuclear effects on single jets. It would be interesting to check whether
such methods could give us a more direct, experimental,
access to the genuine modifications in the jet fragmentation function near $x=1$.

Given the difficulty to make accurate theoretical predictions for a quantity like the
jet fragmentation function, which is sensitive to the non-perturbative physics of the confinement,
we proposed alternative observables, infrared-and-collinear-safe by
construction, which can still be used for studies of
the in-medium jet fragmentation. Roughly speaking, these are quantities which characterise
the jet fragmentation into subjets where the ``subjets'' are sufficiently hard
to be well within the reach of perturbation theory.
We studied one specific example in which the subjets are generated via primary emissions by the leading parton,
with a relative transverse momentum larger than a (semi)hard cutoff
$k_{\perp,\textrm{cut}}$.
We have shown that by judiciously choosing the value of this cutoff, within the range
$k_{\perp,\textrm{min}}\ll k_{\perp,\textrm{cut}}<Q_s$, with $Q_s^2=\hat q L$,
one can minimise the sensitivity of the results to the infrared cutoff $k_{\perp,\textrm{min}}$, 
while still keeping some salient medium effects. It would be interesting to measure
this observable at the LHC and compare with our respective predictions in 
Figs.~\ref{Fig:fragsubjet} and~\ref{Fig:subjet-med}.

Whereas the use of infrared-and-collinear-safe observables should strongly reduce the sensitivity
of our calculations to the non-perturbative physics of hadronisation,
it would be interesting to supplement our framework with a model for
hadronisation (both in the vacuum and in the medium) and see how this
affects our description of the fragmentation function and its uncertainties.

Finally, the description of the medium in our framework needs to be
improved and this is our priority for the future.
Notably, we should allow for the longitudinal expansion of the quark-gluon
plasma and hence for time-dependent medium parameters.
We are currently working on that and our conclusions
  should hopefully be available in the near future.
We are also aiming at an improved theoretical description of the elastic collisions in the 
plasma and of their consequences in terms of momentum broadening, medium-induced
radiation, energy loss and colour decoherence. This should also allow us to include
the response of the medium to the jet propagation and thus have a better control on 
the small-$x$ region of the in-medium fragmentation function and as
well as on other
observables, like the jet shape and the jet radius ($R$) dependence of the nuclear
modification factor $R_{AA}$~\cite{CMS:2019btm,Haake:2019pqd,Pablos:2019ngg}.

\appendix

\section*{Acknowledgements}
The work of P.C.,  E.I. and G.S. is supported in part by the Agence Nationale de la Recherche project 
 ANR-16-CE31-0019-01.   The work of A.H.M.
is supported in part by the U.S. Department of Energy Grant \# DE-FG02-92ER40699.

\section{Expressions with running coupling}\label{sec:rc-effects}

Several results in this paper have been given in the fixed-coupling
approximation. For completeness, we give in this Appendix the
corresponding results including running coupling effects.
These are obtained by evaluating the strong coupling constant at the
scale of the transverse momentum $k_\perp$ of each emission with respect to its
emitter:
\begin{equation}\label{eq:alphas-kperp}
\alpha_s(k_\perp) = \frac{\alpha_s}{1+2\alpha_s\beta_0\ln\frac{k_\perp}{p_TR}},
\end{equation}
with $\alpha_s\equiv\alpha_s(p_TR)$ and
$\beta_0=\frac{11C_A-2n_f}{12\pi}$.

Defining $u\equiv \alpha_sL$ and $v\equiv \alpha_sL_0$ with
$L=\ln\frac{1}{1-x}$ and $L_0=\ln\frac{p_{T0}R}{k_{\perp,\text{min}}}$
($v> u$), the expressions for the NLL Sudakov exponents in the vacuum,
Eqs.~\ref{g1} and~\ref{g2}, become
\begin{align}
 g_{1,i}(u,v) & =\frac{C_i}{\pi
                \beta_0}\left[1-\ln\Big(\frac{1-2\beta_0u}{1-2\beta_0v}\Big)+\frac{\ln(1-2\beta_0u)}{2\beta_0u}\right],\\
 g_{2,i}(u,v) & =\gamma_E\frac{\partial u g_{1,i}}{\partial u}-\ln\bigg[\Gamma\Big(1-\frac{\partial u g_{1,i}}{\partial u}\Big)\bigg]+\frac{C_iB_i}{\pi \beta_0}\ln(1-2\beta_0v)\,,      
\end{align}
The details of the calculation of these functions in the vacuum are given in Appendix \ref{app:NLL}.

For the effects of the veto region, the expression corresponding to
Eq.~(\ref{eq:g1-veto}) and including running-coupling effects is
found to be
\begin{equation}\label{eq:g1-veto-rc}
  Lg_{1,i}^{\textrm{veto}}(u,v) = Lg_{1,i}(u,v)+\frac{2C_i}{\pi}\mathcal{A}_{\textrm{veto}}(L)
\end{equation}
where the logarithmic area of the veto region $\mathcal{A}_{\textrm{veto}}(L)$ is defined as:
\begin{equation}\label{eq:aveto-def}
\mathcal{A}_{\textrm{veto}}(L)=\int_{e^{-L}}^1 \frac{\dif z}{z}\int_{0}^{R}\frac{\dif\theta}{\theta}\alpha_s(z p_{T0}\theta)(1-\Theta_{\textrm{veto}})
\end{equation}
and $\Theta_{\textrm{veto}}$ is given by \eqref{step-veto}. Introducing the following function:
\begin{equation}
\mathcal{T}(x,y,z)\equiv \frac{y+zx}{z}\ln(1+\alpha_s\beta_0(y+zx))\,,
\end{equation}
the logarithmic area $\mathcal{A}_{\textrm{veto}}(L)$ reads:

\begin{align*}\label{eq:aveto}
  \mathcal{A}_{\textrm{veto}}(L) \overset{1-x<z_L}=
  &\frac{1}{2\beta_0}\Big[
    \mathcal{T}\big(\ln z_{0},0,2\big)
    -\mathcal{T}\big(\ln z_L,0,2\big)
   +\mathcal{T}\big(\ln z_c ,\tfrac{3}{2}\ln z_{0},\tfrac{1}{2}\big)
    -\mathcal{T}\big(\ln z_{0},\tfrac{3}{2}\ln z_{0},\tfrac{1}{2}\big)\\
  & -\mathcal{T}\big(\ln z_0,\ln z_L,1\big)
    +\mathcal{T}\big(\ln z_L,\ln z_L,1\big)
    -\mathcal{T}\big(\ln z_c,\ln z_L,1\big)
    +\mathcal{T}\big(\ln z_{0},\ln z_L,1\big)\Big]\\
  \overset{z_L<1-x<z_0}=
  &\frac{1}{2\beta_0}\big[
    \mathcal{T}\big(\ln z_{0},0,2\big)
    -\mathcal{T}\big({-L},0,2\big)
    +\mathcal{T}\big(\ln z_c,\tfrac{3}{2}\ln z_{0},\tfrac{1}{2}\big)
    -\mathcal{T}\big(\ln z_0,\tfrac{3}{2}\ln z_0,\tfrac{1}{2}\big)\\
  & -\mathcal{T}\big(\ln z_{0},\ln z_L,1\big)
    +\mathcal{T}\big({-L},\ln z_L,1\big)
    -\mathcal{T}\big(\ln z_c,\ln z_L,1\big)
    +\mathcal{T}\big(\ln z_0,\ln z_L,1\big)\Big]\\
  \overset{z_0<1-x<z_c}=
  &\frac{1}{2\beta_0}\big[
    \mathcal{T}\big(\ln z_c,\tfrac{3}{2}\ln z_0,\tfrac{1}{2}\big)
    -\mathcal{T}\big({-L},\tfrac{3}{2}\ln z_0,\tfrac{1}{2}\big)
    -\mathcal{T}\big(\ln z_c,\ln z_L,1\big)
    +\mathcal{T}\big({-L},\ln z_L,1\big)\Big)\Big]
\end{align*}
with $z_0\equiv\omega_0(R)/p_{T0}=(2\qhat/(p_{T0}^3R^4))^{1/3}$, $z_L\equiv\omega_L(R)/p_{T0}=2/(Lp_{T0}R^2)$ and $z_c\equiv\omega_c/p_{T0}$.

\section{Large $x$ jet fragmentation to NLL accuracy}
\label{app:NLL}

Eq.~\eqref{g2} can be deduced from the coherent branching algorithm
(also known as MLLA evolution equation \cite{Dokshitzer:1991wu}) which resums to all orders
leading and next-to-leading logarithms of the form
$-\alpha_s\ln(1-x)$.
Since the fragmentation function is not IRC safe, we introduce a lower transverse momentum cut-off $\ktmin$ for any resolvable splitting. The final result strongly depends on $\ktmin$ so we need to keep track of any $\ktmin$ dependence in the calculation.

To NLL accuracy, one can neglect the quark/gluon mixing terms. We
discuss this approximation at the end of this appendix.
We focus on quark-initiated jets and the generalisation to gluon-jets
is straightforward.
The MLLA equation for the quark cumulative fragmentation function
reduces to
\begin{equation}\label{MLLA}
  Q\frac{\partial \Sigma_q(x,Q)}{\partial Q}=\int_{0}^{1}\dif z\,K_{q}^{q}(z,k_\perp)\Big[\Sigma_q\Big(\frac{x}{z},zQ\Big)-\Sigma_q(x,Q)\Big]
\end{equation}
where the evolution variable is $Q=p_{T0}\theta$ to account for the ordering in the angle $\th$ of successive emissions and the kernel is
\begin{equation}
 K_q^{q}(z,k_\perp)=\frac{\alpha_s(k_\perp)}{\pi}P_{qq}(z)\Theta(k_\perp-\ktmin),\qquad P_{qq}=C_F\frac{1+z^2}{1-z}.
\end{equation}
The initial condition for \eqref{MLLA} is
$\Sigma_q(x,k_\perp=\ktmin)=\Theta(1-x)$. At NLL accuracy,
$k_\perp=z(1-z)Q\simeq (1-z)Q$ and $\Sigma_q(\frac{x}{z},zQ)\simeq
\Sigma_q(\frac{x}{z},Q)$ since the dominant contribution for $x\simeq 1$ comes from $z\simeq 1$. 

The standard way to solve Eq.~\eqref{MLLA} is to go to Mellin space
$\Sigma_q(x,Q)\rightarrow \tilde{\Sigma}_q(j,Q)$
where the integral in the r.h.s.\ becomes a product. 
In Mellin space, $x$ close to 1 corresponds to
$j\rightarrow\infty$, more precisely, $\ln(j)\sim-\ln(1-x)$, so we
keep all terms of the form $\alpha_s^n\ln(j)^n\sim 1$ in the exact
solution. Anticipating our resummed result, we note
$\lambda_j=\alpha_s\ln(j)$ and
$\lambda_0=\alpha_s\ln(p_{T0}R/\ktmin)=\alpha_sL_0$,
\begin{align}
 \ln(j\tilde{\Sigma}_q(j,p_{T0}R))&=\int_{Q_0}^{p_{T0}R}\frac{\dif Q'}{Q'}\,\int_{0}^{1}\dif z\,(z^{j}-1)K_q^{q}(z,Q')\\
 &=\frac{C_F}{\pi \beta_0}\Big[\ln(j)\Big(1-\ln\Big(\frac{1-2\beta_0\lambda_j}{1-2\beta_0\lambda_0}\Big)+\frac{\ln(1-2\beta_0\lambda_j)}{2\beta_0\lambda_j}\Big)\nonumber \\
 &\hspace{1cm}-\gamma_E\ln\Big(\frac{1-2\beta_0\lambda_j}{1-2\beta_0\lambda_0}\Big)+B_q\ln(1-2\beta_0\lambda_0)\Big]+O(\alpha_s\lambda_j^n,\alpha_s\lambda_0^n),\label{logDj}
\end{align}
where we used the standard trick
$z^{j}-1\simeq-\Theta(e^{-\gamma_E}/j-z)$ valid at NLL accuracy~\cite{Catani:1989ne} and we kept only the singular and finite part $B_q=-3/4$ of the quark splitting function when $z\simeq1$.
Eq.~\eqref{logDj} resums to all orders leading and next-to-leading logarithms of the form $\lambda_j$, $\lambda_0$. More explicitly,
\begin{align}
 \ln(j\tilde{\Sigma}^{\textrm{NLL}}_q(j,p_{T0}R))&=\ln(j)g_1(\lambda_j,\lambda_0)+f_2(\lambda_j,\lambda_0)\\
 g_1(u,v)&=\frac{C_F}{\pi \beta_0}\Big[1-\ln\Big(\frac{1-2\beta_0u}{1-2\beta_0v}\Big)+\frac{\ln(1-2\beta_0u)}{2\beta_0u}\Big]\\
 f_{2,q}(u,v)&=\frac{C_F}{\pi \beta_0}\Big[-\gamma_E\ln\Big(\frac{1-2\beta_0u}{1-2\beta_0v}\Big)+B_q\ln(1-2\beta_0v)\Big]
\end{align}
The final step is to calculate the inverse Mellin transform of \eqref{logDj}.
\begin{align}
 \Sigma_{q}(x)&=\frac{1}{2\pi i}\int_{\mathcal{C}}\frac{\dif j}{j}\,e^{-j\ln(x)}\Big(j\tilde{\Sigma}_{q}(j)\Big)=\frac{1}{2\pi i}\int_{\mathcal{C}}\dif u \,e^{u-\ln(u)+G_q[\ln(u)-\ln(-\ln(x))]}
\end{align}
where $\mathcal{C}$ is a contour parallel to the imaginary axis and $G_q[\ln(j)]\equiv\ln(j\tilde{\Sigma}_{q}(j))$.
For this, we Taylor-expand the function $G_q$ around $L=-\ln(-\ln(x))\simeq-\ln(1-x)$.
\begin{equation}
 G_q[L+\ln(u)]=G_q[L]+\ln(u)G_q'[L]+\sum_{k=2}^{\infty}\ln(u)^k\frac{G_q^{(k)}[L]}{k!}
\end{equation}
For $k\ge2$, $G_q^{(k)}[L]$ is certainly beyond NLL accuracy because the derivatives of $\alpha_s\beta_0 L$ with respect to $L$ bring always at least one extra $\alpha_s$ factor. Thus, we truncate the expansion up to the first derivative. Moreover, the derivative of $f_{2,q}(\alpha_sL,\alpha_sL_0)$ with respect to $L$ is also subleading.
Finally, using
\begin{equation}
 \frac{1}{2\pi i}\int_{\mathcal{C}}\dif u\, e^{u+x\ln(u)}=\frac{1}{\Gamma(-x)}
\end{equation}
one gets the following result for the cumulative distribution:
\begin{equation}
\label{cumulative}
 \Sigma^{\textrm{NLL}}_{q}(x,p_{T0}R)=\frac{e^{G_{q}[L]}}{\Gamma(1-G_q'[L])}=\frac{\exp\Big(Lg_1(\alpha_sL,\alpha_sL_0)+f_{2,q}(\alpha_sL,\alpha_sL_0)\Big)}{\Gamma\Big(1-\frac{\partial ug_1(u,\alpha_sL_0)}{\partial u}_{|u=\alpha_sL}\Big)}
\end{equation}
which is exactly \eqref{log-expansion}, \eqref{g1} and \eqref{g2}.

\paragraph{Sub-leading $j$ contributions and quark/gluon mixing
  terms.} Besides N$^2$LL contributions, we have neglected terms of
order $\mathcal{O}(\alpha_s^n\ln^n(j)/j)$ in formulas \eqref{MLLA} and
\eqref{logDj}. Among such terms, those associated with quark/gluon mixings
give sizeable numerical corrections to the NLL results, especially in
the gluon-jet case.
The main reason for this is that, even though the (power-suppressed)
probability for a gluon to split in a $q\bar q$ pair where the quark
carries most of the momentum ($x\sim 1$) is much
smaller than the probability to find a hard gluon, once such a
splitting occurs, the Sudakov appearing in~\eqref{cumulative} becomes
that of a quark, i.e.\ has a much smaller suppression because of the
colour factor $C_F< C_A$ appearing in the exponential.
In the inclusive fragmentation function, this becomes an increasingly
likely situation~\cite{DeGrand:1978te}.

Including all terms of order $O(\alpha_s^n\ln^n(j)/j)$ is beyond the scope of this simple analysis of the large $x$ behaviour of the fragmentation function. Instead, one can correct Eq.~\eqref{cumulative} for gluon jets with an additional piece $\Sigma_{g,\textrm{mix}}(x,p_{T0}R)$ describing the splitting of the gluon in a $q\bar{q}$ pair, with either the quark or the antiquark carrying a large fraction $x$ of the initial energy:
 \begin{align}
\label{sigmagqq}
 \Sigma_{g,\textrm{mix}}(x,p_{T0}R)&=\int_{0}^{1-x}\dif \xi\, P_g^q(\xi)\int_0^{R}\frac{\dif\theta}{\theta}\frac{\alpha_s(\xi p_{T0}\theta)}{\pi}\Theta(\xi p_{T0}\theta-\ktmin)\nonumber\\ 
 &\times \exp\Big(-\frac{2C_A}{\pi}\int_{\xi}^1\frac{\dif z}{z}\int_{\theta}^{R}\frac{\dif\theta'}{\theta'}\alpha_s(zp_{T0}\theta')\Theta(zp_{T0}\theta'-\ktmin)\Big)\nonumber\\ 
 &\times \exp\Big(-\frac{2C_F}{\pi}\int_{\xi}^1\frac{\dif z}{z}\int_{0}^{\theta}\frac{\dif\theta'}{\theta'}\alpha_s(zp_{T0}\theta')\Theta(zp_{T0}\theta'-\ktmin)\Big)
\end{align}
with $P_g^q(\xi)=2n_fT_R(\xi^2+(1-\xi)^2)\simeq 2n_fT_R$ since $\xi\le1-x\ll1$.\comment{In Mellin space, this is equivalent to solve the linear differential equation with an inhomogeneous term associated with the non-diagonal elements of the kernel matrix.}In Fig.~\ref{Fig:sigma-vac}, the analytical ``NLL'' curve for gluon jets is actually 
$\Sigma^{\textrm{NLL}}_{g}(x)+\Sigma_{g,\textrm{mix}}(x)$.

\section{Saddle-point method for in-medium intra-jet multiplicity at DLA}
\label{app:DLA}

Our starting point is Eq.~\eqref{master-eq},
assuming~$\omega_L(R)<\ktmin/R$.
For definiteness, we also assume $\th^2\ge\th_c^2$, although it turns out that our
conclusions remain valid for $\th^2\le \th_c^2$.
It is convenient to use logarithmic variables:
$x_1=\ln(p_{T0}/\om_1)$, $y_1=\ln(R^2/\th_1^2)$, $x_2=\ln(\om_2/\om)$,
$y_2=\ln(\th_2^2/\th^2)$ and $X\equiv\ln(p_{T0}/\om)$,
$Y\equiv\ln(R^2/\th^2)$. The energy scales $\omega_0(R)$ and
$\omega_L(R)$, related respectively to the inside and outside domains,
become $x_{0}\equiv\ln(p_{T0}/\om_0(R))$ and
$x_L\equiv\ln(p_{T0}/\om_L(R))$, and the logarithmic scale associated
with $\th_c^2$ is $y_c\equiv\ln(R^2/\th_c^2)=4(x_L-x_{0})/3$.
To get the leading asymptotic behaviour of $T_{i,\textrm{out}}(X,Y)$,
one can neglect the $\delta$ contribution to $T^{\textrm{vac}}$
in~(\ref{Tvac}) since it generates terms with at least one exponential
factor missing. We thus get
\begin{align}\label{master-3}
 T_{i,\textrm{out}}(X,Y)=\abar^3\int_{0}^{\min(X,x_{0})} dx_1\int_{0}^{\min(y_c,\frac{3}{2}(x_{0}-x_1))}dy_1&\int_{0}^{\min(X-x_1,X+Y-x_L)}dx_2\int_{0}^{X+Y-x_L-x_2}dy_2 \nonumber\\
 &\hspace{1.3cm}\textrm{I}_0(2\sqrt{\abar x_1 y_1})\,\textrm{I}_0(2\sqrt{\abar x_2 y_2})
 \end{align}
The integral over $y_1$ and $y_2$ can be performed exactly using the the following relation:
\begin{equation}\label{angular-integral}
  \int_0^s dy\, \textrm{I}_0(2\sqrt{\abar x y})
  =\sqrt{\frac{s}{\abar x}}\textrm{I}_1(2\sqrt{\abar x s})
  \overset{\abar xs\gg 1}{\simeq}\sqrt{\frac{s}{\abar x}}\frac{\exp(2\sqrt{\abar x s})}{\sqrt{4\pi\sqrt{\abar x s}}}.
\end{equation}
Using \eqref{angular-integral}, one gets
\begin{align} \label{master-4}
 T_{i,\textrm{out}}(X,Y)=\abar^3\int_{0}^{\min(X,x_{0})} dx_1&\int_{0}^{\min(X-x_1,X+Y-x_L)}dx_2\,R_1(x_1)R_2(x_2) \nonumber\\
 &\hspace{2.cm} e^{2\sqrt{\abar x_1 \min(y_c,\frac{3}{2}(x_{0}-x_1))}} e^{2\sqrt{\abar x_2 (X+Y-x_L-x_2)}}
\end{align}
with the two non-exponential functions
\begin{equation}
 R_1(x_1)=\frac{1}{\sqrt{4\pi}}\frac{(\min(y_c,\frac{3}{2}(x_{0}-x_1)))^{1/4}}{(\abar x_1)^{3/4}},\qquad
 R_2(x_2)=\frac{1}{\sqrt{4\pi}}\frac{(X+Y-x_L-x_2)^{1/4}}{(\abar x_2)^{3/4}}
\end{equation}

The $x_2$ integrations cannot be performed exactly so we use the
saddle-point approximation:
\begin{equation}
  \int_{x_1}^{x_2} dx\, f(x)e^{Mg(x)}
  \overset{M\to\infty}\simeq\sqrt{ \frac{2\pi}{-Mg''(x^\star)}}f(x^\star)e^{Mg(x^\star)},
\end{equation}
where the saddle point $x^\star$ is the \textit{maximum} of $g(x)$
between $x_1$ and $x_2$. This formula is valid as long as
$x_1<x^\star<x_2$.

Setting $M_2\equiv(X+Y-x_L)=\ln(\omega_L(\theta)/\omega)$ and
integrating over $x_2/M_2$, one get
\begin{equation}
 \mathcal{N}_{\textrm{out}} \equiv \abar\int_{0}^{\min(X-x_1,X+Y-x_L)}dx_2\, R_2(x_2)e^{2\sqrt{\abar x_2 (X+Y-x_L-x_2)}}
 \overset{\sqrt{\abar}M_2\rightarrow\infty}{\simeq} \frac{1}{2}e^{\sqrt{\abar}M_2}.
\end{equation}
The corresponding saddle point is
$x_2^\star=M_2/2=\ln(\sqrt{\omega_L(\theta)/\omega})$ so that the saddle-point approximation is valid if $x_2^\star<X-x_1$. This gives the condition $x_1\le X-x_2^\star$ in the first integral, in order to ensure energy conservation along the cascade.

Calling $\mathcal{N}_{\textrm{med}}$ the remaining integral over
$x_1$, which is truly a gluon multiplicity \textit{inside} the medium,
we are left with:
\begin{equation}
 \mathcal{N}_{\textrm{med}}\equiv\abar\int_{0}^{\min(x_{0},X-x_2^\star)} dx_1\,R_1(x_1)e^{2\sqrt{\abar x_1 \min(y_c,\frac{3}{2}(x_{0}-x_1))}}.
\end{equation}
Since $\min(X-x_2^\star,x_{0})>x_c\equiv\ln(p_{T0}/\om_c)$, the
integral can be split into two pieces: $x_1<x_c$ where
$\min(y_c,3(x_{0}-x_1)/2)=y_c$ and $x_1>x_c$ where
$\min(y_c,3(x_{0}-x_1)/2)=3(x_{0}-x_1)/2$. The first piece is
calculated exactly, and we use again the saddle point method to
evaluate the second piece, assuming
$x_0=\ln(p_{T0}/\omega_0(R))\to \infty$. We get (using $x_1'=x_1/x_0$)
\begin{align} 
 \mathcal{N}_{med}&=\int_{0}^{x_c}dx_1\abar \sqrt{\frac{y_c}{\abar x_1}}\textrm{I}_1(2\sqrt{\abar x_1 y_c})+\abar\int_{x_c}^{\min(X-x_2^\star,x_{0})}dx_1\,R(x_1)e^{2\sqrt{\abar x_1\frac{3}{2}(x_{0}-x_1)}}
 \nonumber \\*[0.2cm]
 &=-1+\textrm{I}_0(2\sqrt{\abar x_c y_c})\,+ \abar^{1/4}\sqrt{\frac{x_0}{4\pi}}\int_{x_c/x_0}^{\textrm{min}(1,(X-x_2^\star)/x_0)}\frac{dx_1'}{x_1'^{1/2}}\Big(\frac{3(1-x_1')}{2x_1}\Big)^{1/4} \,
 e^{2x_0\sqrt{\frac{3}{2}\abar x'_1(1-x'_1)}}\nonumber \\*[0.2cm]
 &\overset{\sqrt{\abar}x_0\rightarrow\infty}{\sim} \frac{e^{2\sqrt{\abar x_c y_c}}}{\sqrt{4\pi\sqrt{\abar x_c y_c}}}+\frac{1}{2}e^{\sqrt{\frac{3\abar}{2}}x_{0}}\, \label{Nmed-calc}
\end{align}
The first term in equation \eqref{Nmed-calc} is subleading due to the
square root in the argument and in the denominator. Thus, the leading
term for $\mathcal{N}_{med}$ comes from the ``inside-medium'' region
with $\om_1\le\om_c$.\footnote{That is why we can trust our final
  result for $T(\om,\th^2)$ even for $\th^2\le\th_c^2$.}

The saddle point of the integral over $x_1$ is $x_1^\star=x_{0}/2=\ln(\sqrt{p_{T0}/\omega_0(R)})$ so our estimation for $\mathcal{N}_{med}$ is valid only if $x_c<x_1^\star<X-x_2^\star$. The condition $x_c<x_1^\star$ leads to the condition~\eqref{eq:cdt-pt0-omc}. The condition $x_1^\star<X-x_2^\star$ leads to the condition~(\ref{ocr}), when $x_2^\star=\ln(\sqrt{\omega_L(\theta)/\omega})$ is evaluated at its largest value, that is when $\th=\th_{\textrm{min}}\equiv\ktmin/\omega$.

We have thus demonstrated that when both $\sqrt{\abar}x_0\equiv
\sqrt{\abar}\ln(p_{T0}/\omega_0(R))$ and
$\sqrt{\abar}(X+Y-x_L)\equiv\sqrt{\abar}\ln(\omega_L(\theta)/\omega)$
are large and $X>x_1^\star+x_2^\star$, i.e. $\omega<\omega_{cr}$, we have
\begin{equation}\label{Tout-asymptot} 
 T_{i,\textrm{out}}(X,Y)\sim\frac{\abar}{4}\exp\left[\sqrt{\abar}\left(X+Y-x_L+\sqrt{\frac{3}{2}}x_{0}\right)\right],
\end{equation}
which is precisely formula \eqref{Tsaddle}.

From~\eqref{Tout-asymptot} and~\eqref{fullT}, one deduces the
asymptotic DLA behaviour of the small-$x$ fragmentation function by
integrating $T_i(\omega,\th^2|p_{T0},R^2)$ over $\th^2$ between
$\ktmin^2/\omega^2$ and $R^2$. The leading contribution comes from the lower limit of this integral or, in logarithmic units, from the upper bound $2(x_{\textrm{max}}-X)$ on the integral on $Y$, 
with $x_{\textrm{max}}=\ln(p_{T0}R/\ktmin)$. This reproduces~\eqref{frag-DLA} in logarithmic units:
\begin{equation}
 D^{\textrm{med}}_i(X)
 =\int_0^{2(x_{\textrm{max}}-X)}\dif Y\,T_{i}(X,Y)
\simeq \frac{\sqrt{\abar}C_i}{4C_A}\exp\left[\sqrt{\abar}\Big(-X + 2 x_{\textrm{max}} - x_L+\sqrt{\frac{3}{2}}x_{0}\Big)\right] \label{Dmed-asymptot}
\end{equation}

Finally, the asymptotic form of the ratio $\mathcal{R}_i(X)\equiv\,D^{\textrm{med}}_i(X)/D_i^{\textrm{vac}}(X)$ is obtained from \eqref{Dmed-asymptot} and \eqref{DvacDLA}, using again the asymptotic form of $\textrm{I}_1(x)$ at
 large $x$:
 \begin{align}
 D_i^{\textrm{vac}}(X)
 & \simeq\,\frac{C_i}{\sqrt{4\pi}C_A}\left[\frac{2\abar(x_{\textrm{max}}-X)}{X^3}\right]^{1/4}\exp\Big(2\sqrt{2\abar X(x_{\textrm{max}}-X)}\Big)\label{Dvac-approx}\\
  \mathcal{R}_i(X)& \sim\frac{\sqrt{\abar\pi}}{2}e^{\sqrt{\frac{3}{2}}x_{0}-x_L}\left[\frac{X^3}{2\abar(x_{\textrm{max}}-X)}\right]^{1/4}\exp\left[\sqrt{\abar}\big(\sqrt{X}-\sqrt{2(x_{\textrm{max}}-X)}\big)^2\right].\label{Rfinal}
 \end{align}
 From~\eqref{Dvac-approx}, one can estimate the position of the maximum $x_{\textrm{hump}}$ of $D_i^{\textrm{vac}}(X)$. Neglecting the non-exponential prefactor,  one finds ${\dif D_i^{\textrm{vac}}}/{\dif X}
\propto x_{\textrm{max}}-2X$,
so that the $x_{\textrm{hump}}\simeq x_{\textrm{max}}/2$ and $\om_{\textrm{hump}}\simeq \sqrt{p_{T0}\ktmin/R}$. For $X\ge x_{\textrm{hump}}$ i.e. $\om\le\om_{\textrm{hump}}$, the derivative is negative, hence $D_i^{\textrm{vac}}(\omega)$ decreases when $\omega$ decreases.
Similarly, one can study the variation of $\mathcal{R}_i(X)$ from the exponential factor alone:
\begin{equation}
\frac{\dif \mathcal{R}_i}{\dif X}\simeq\frac{\abar\sqrt{\pi}}{2}e^{\sqrt{\frac{3}{2}}x_{0}-x_L}\frac{\big(\sqrt{2X}+\sqrt{x_{\textrm{max}}-X}\big)\big(\sqrt{X}-\sqrt{2(x_{\textrm{max}}-X)}\big)}{\sqrt{X(x_{\textrm{max}}-X)}}e^{\sqrt{\abar}\big(\sqrt{X}-\sqrt{2(x_{\textrm{max}}-X)}\big)^2}
\end{equation}
The derivative is positive when $\sqrt{X}-\sqrt{2(x_{\textrm{max}}-X)}\ge0$ i.e. when $X\ge 2x_{\textrm{max}}/3$. Hence, for $\om\lesssim(p_{T0}\ktmin^2/R^2)^{1/3}$, the ratio $R_i(\omega)$ increases when $\om$ decreases.


\providecommand{\href}[2]{#2}\begingroup\raggedright\begin{thebibliography}{10}

\bibitem{Aaboud:2018hpb}
{\bfseries ATLAS} Collaboration, M.~Aaboud {\em et al.}, ``{Measurement of jet
  fragmentation in Pb+Pb and $pp$ collisions at $\sqrt{s_{NN}} = 5.02$ TeV with
  the ATLAS detector},''
  \href{http://dx.doi.org/10.1103/PhysRevC.98.024908}{{\em Phys. Rev.}
  {\bfseries C98} no.~2, (2018) 024908},
\href{http://arxiv.org/abs/1805.05424}{{\ttfamily arXiv:1805.05424 [nucl-ex]}}.

\bibitem{Spousta:2015fca}
M.~Spousta and B.~Cole, ``{Interpreting single jet measurements in Pb $+$ Pb
  collisions at the LHC},''
  \href{http://dx.doi.org/10.1140/epjc/s10052-016-3896-0}{{\em Eur. Phys. J. C}
  {\bfseries 76} no.~2, (2016) 50},
  \href{http://arxiv.org/abs/1504.05169}{{\ttfamily arXiv:1504.05169
  [hep-ph]}}.

\bibitem{Casalderrey-Solana:2016jvj}
J.~Casalderrey-Solana, D.~Gulhan, G.~Milhano, D.~Pablos, and K.~Rajagopal,
  ``{Angular Structure of Jet Quenching Within a Hybrid Strong/Weak Coupling
  Model},'' \href{http://dx.doi.org/10.1007/JHEP03(2017)135}{{\em JHEP}
  {\bfseries 03} (2017) 135}, \href{http://arxiv.org/abs/1609.05842}{{\ttfamily
  arXiv:1609.05842 [hep-ph]}}.

\bibitem{Tachibana:2017syd}
Y.~Tachibana, N.-B. Chang, and G.-Y. Qin, ``{Full jet in quark-gluon plasma
  with hydrodynamic medium response},''
  \href{http://dx.doi.org/10.1103/PhysRevC.95.044909}{{\em Phys. Rev. C}
  {\bfseries 95} no.~4, (2017) 044909},
  \href{http://arxiv.org/abs/1701.07951}{{\ttfamily arXiv:1701.07951
  [nucl-th]}}.

\bibitem{KunnawalkamElayavalli:2017hxo}
R.~Kunnawalkam~Elayavalli and K.~C. Zapp, ``{Medium response in JEWEL and its
  impact on jet shape observables in heavy ion collisions},''
  \href{http://dx.doi.org/10.1007/JHEP07(2017)141}{{\em JHEP} {\bfseries 07}
  (2017) 141}, \href{http://arxiv.org/abs/1707.01539}{{\ttfamily
  arXiv:1707.01539 [hep-ph]}}.

\bibitem{Chen:2017zte}
W.~Chen, S.~Cao, T.~Luo, L.-G. Pang, and X.-N. Wang, ``{Effects of jet-induced
  medium excitation in $\gamma$-hadron correlation in A+A collisions},''
  \href{http://dx.doi.org/10.1016/j.physletb.2017.12.015}{{\em Phys. Lett. B}
  {\bfseries 777} (2018) 86--90},
  \href{http://arxiv.org/abs/1704.03648}{{\ttfamily arXiv:1704.03648
  [nucl-th]}}.

\bibitem{Casalderrey-Solana:2018wrw}
J.~Casalderrey-Solana, Z.~Hulcher, G.~Milhano, D.~Pablos, and K.~Rajagopal,
  ``{Simultaneous description of hadron and jet suppression in heavy-ion
  collisions},'' \href{http://dx.doi.org/10.1103/PhysRevC.99.051901}{{\em Phys.
  Rev.} {\bfseries C99} no.~5, (2019) 051901},
\href{http://arxiv.org/abs/1808.07386}{{\ttfamily arXiv:1808.07386 [hep-ph]}}.

\bibitem{Casalderrey-Solana:2019ubu}
J.~Casalderrey-Solana, G.~Milhano, D.~Pablos, and K.~Rajagopal, ``{Modification
  of Jet Substructure in Heavy Ion Collisions as a Probe of the Resolution
  Length of Quark-Gluon Plasma},''
  \href{http://dx.doi.org/10.1007/JHEP01(2020)044}{{\em JHEP} {\bfseries 01}
  (2020) 044}, \href{http://arxiv.org/abs/1907.11248}{{\ttfamily
  arXiv:1907.11248 [hep-ph]}}.

\bibitem{Caucal:2018dla}
P.~Caucal, E.~Iancu, A.~H. Mueller, and G.~Soyez, ``{Vacuum-like jet
  fragmentation in a dense QCD medium},''
  \href{http://dx.doi.org/10.1103/PhysRevLett.120.232001}{{\em Phys. Rev.
  Lett.} {\bfseries 120} (2018) 232001},
\href{http://arxiv.org/abs/1801.09703}{{\ttfamily arXiv:1801.09703 [hep-ph]}}.

\bibitem{Caucal:2019uvr}
P.~Caucal, E.~Iancu, and G.~Soyez, ``{Deciphering the $z_g$ distribution in
  ultrarelativistic heavy ion collisions},''
  \href{http://dx.doi.org/10.1007/JHEP10(2019)273}{{\em JHEP} {\bfseries 10}
  (2019) 273},
\href{http://arxiv.org/abs/1907.04866}{{\ttfamily arXiv:1907.04866 [hep-ph]}}.

\bibitem{MehtarTani:2010ma}
Y.~Mehtar-Tani, C.~A. Salgado, and K.~Tywoniuk, ``{Antiangular Ordering of
  Gluon Radiation in QCD Media},''
  \href{http://dx.doi.org/10.1103/PhysRevLett.106.122002}{{\em Phys. Rev.
  Lett.} {\bfseries 106} (2011) 122002},
\href{http://arxiv.org/abs/1009.2965}{{\ttfamily arXiv:1009.2965 [hep-ph]}}.

\bibitem{MehtarTani:2011tz}
Y.~Mehtar-Tani, C.~A. Salgado, and K.~Tywoniuk, ``{Jets in QCD Media: from
  Color Coherence to Decoherence},''
  \href{http://dx.doi.org/10.1016/j.physletb.2011.12.042}{{\em Phys. Lett.}
  {\bfseries B707} (2012) 156--159},
\href{http://arxiv.org/abs/1102.4317}{{\ttfamily arXiv:1102.4317 [hep-ph]}}.

\bibitem{CasalderreySolana:2011rz}
J.~Casalderrey-Solana and E.~Iancu, ``{Interference Effects in Medium-Induced
  Gluon Radiation},'' \href{http://dx.doi.org/10.1007/JHEP08(2011)015}{{\em
  JHEP} {\bfseries 08} (2011) 015},
\href{http://arxiv.org/abs/1105.1760}{{\ttfamily arXiv:1105.1760 [hep-ph]}}.

\bibitem{Mehtar-Tani:2014yea}
Y.~Mehtar-Tani and K.~Tywoniuk, ``{Jet (de)coherence in PbPb collisions at the
  LHC},'' \href{http://dx.doi.org/10.1016/j.physletb.2015.03.041}{{\em Phys.
  Lett.} {\bfseries B744} (2015) 284--287},
\href{http://arxiv.org/abs/1401.8293}{{\ttfamily arXiv:1401.8293 [hep-ph]}}.

\bibitem{Baier:2000sb}
R.~Baier, A.~H. Mueller, D.~Schiff, and D.~Son, ``{'Bottom up' thermalization
  in heavy ion collisions},''
  \href{http://dx.doi.org/10.1016/S0370-2693(01)00191-5}{{\em Phys.Lett.}
  {\bfseries B502} (2001) 51--58},
\href{http://arxiv.org/abs/hep-ph/0009237}{{\ttfamily arXiv:hep-ph/0009237
  [hep-ph]}}.

\bibitem{Jeon:2003gi}
S.~Jeon and G.~D. Moore, ``{Energy loss of leading partons in a thermal QCD
  medium},'' \href{http://dx.doi.org/10.1103/PhysRevC.71.034901}{{\em
  Phys.Rev.} {\bfseries C71} (2005) 034901},
\href{http://arxiv.org/abs/hep-ph/0309332}{{\ttfamily arXiv:hep-ph/0309332
  [hep-ph]}}.

\bibitem{Blaizot:2013hx}
J.-P. Blaizot, E.~Iancu, and Y.~Mehtar-Tani, ``{Medium-induced QCD cascade:
  democratic branching and wave turbulence},''
  \href{http://dx.doi.org/10.1103/PhysRevLett.111.052001}{{\em Phys.Rev.Lett.}
  {\bfseries 111} (2013) 052001},
\href{http://arxiv.org/abs/1301.6102}{{\ttfamily arXiv:1301.6102 [hep-ph]}}.

\bibitem{Blaizot:2013vha}
J.-P. Blaizot, F.~Dominguez, E.~Iancu, and Y.~Mehtar-Tani, ``{Probabilistic
  picture for medium-induced jet evolution},''
  \href{http://dx.doi.org/10.1007/JHEP06(2014)075}{{\em JHEP} {\bfseries 1406}
  (2014) 075},
\href{http://arxiv.org/abs/1311.5823}{{\ttfamily arXiv:1311.5823 [hep-ph]}}.

\bibitem{Baier:1996kr}
R.~Baier, Y.~L. Dokshitzer, A.~H. Mueller, S.~Peigne, and D.~Schiff,
  ``{Radiative Energy Loss of High Energy Quarks and Gluons in a Finite-Volume
  Quark-Gluon Plasma},''
  \href{http://dx.doi.org/10.1016/S0550-3213(96)00553-6}{{\em Nucl. Phys.}
  {\bfseries B483} (1997) 291--320},
\href{http://arxiv.org/abs/hep-ph/9607355}{{\ttfamily arXiv:hep-ph/9607355}}.

\bibitem{Baier:1996sk}
R.~Baier, Y.~L. Dokshitzer, A.~H. Mueller, S.~Peigne, and D.~Schiff,
  ``{Radiative Energy Loss and P(T)-Broadening of High Energy Partons in
  Nuclei},'' \href{http://dx.doi.org/10.1016/S0550-3213(96)00581-0}{{\em Nucl.
  Phys.} {\bfseries B484} (1997) 265--282},
\href{http://arxiv.org/abs/hep-ph/9608322}{{\ttfamily arXiv:hep-ph/9608322}}.

\bibitem{Zakharov:1996fv}
B.~G. Zakharov, ``{Fully Quantum Treatment of the Landau-Pomeranchuk-Migdal
  Effect in QED and QCD},'' \href{http://dx.doi.org/10.1134/1.567126}{{\em JETP
  Lett.} {\bfseries 63} (1996) 952--957},
\href{http://arxiv.org/abs/hep-ph/9607440}{{\ttfamily arXiv:hep-ph/9607440}}.

\bibitem{Zakharov:1997uu}
B.~G. Zakharov, ``{Radiative Energy Loss of High Energy Quarks in Finite-Size
  Nuclear Matter and Quark-Gluon Plasma},''
  \href{http://dx.doi.org/10.1134/1.567389}{{\em JETP Lett.} {\bfseries 65}
  (1997) 615--620},
\href{http://arxiv.org/abs/hep-ph/9704255}{{\ttfamily arXiv:hep-ph/9704255}}.

\bibitem{Baier:1998kq}
R.~Baier, Y.~L. Dokshitzer, A.~H. Mueller, and D.~Schiff, ``{Medium-Induced
  Radiative Energy Loss: Equivalence Between the Bdmps and Zakharov
  Formalisms},'' \href{http://dx.doi.org/10.1016/S0550-3213(98)00546-X}{{\em
  Nucl. Phys.} {\bfseries B531} (1998) 403--425},
\href{http://arxiv.org/abs/hep-ph/9804212}{{\ttfamily arXiv:hep-ph/9804212}}.

\bibitem{Fister:2014zxa}
L.~Fister and E.~Iancu, ``{Medium-induced jet evolution: wave turbulence and
  energy loss},'' \href{http://dx.doi.org/10.1007/JHEP03(2015)082}{{\em JHEP}
  {\bfseries 03} (2015) 082},
\href{http://arxiv.org/abs/1409.2010}{{\ttfamily arXiv:1409.2010 [hep-ph]}}.

\bibitem{Aaboud:2018twu}
{\bfseries ATLAS} Collaboration, M.~Aaboud {\em et al.}, ``{Measurement of the
  nuclear modification factor for inclusive jets in Pb+Pb collisions at
  $\sqrt{s_\mathrm{NN}}=5.02$ TeV with the ATLAS detector},''
  \href{http://dx.doi.org/10.1016/j.physletb.2018.10.076}{{\em Phys. Lett.}
  {\bfseries B790} (2019) 108--128},
\href{http://arxiv.org/abs/1805.05635}{{\ttfamily arXiv:1805.05635 [nucl-ex]}}.

\bibitem{Cacciari:2008gp}
M.~Cacciari, G.~P. Salam, and G.~Soyez, ``{The anti-$k_t$ jet clustering
  algorithm},'' \href{http://dx.doi.org/10.1088/1126-6708/2008/04/063}{{\em
  JHEP} {\bfseries 04} (2008) 063},
\href{http://arxiv.org/abs/0802.1189}{{\ttfamily arXiv:0802.1189 [hep-ph]}}.

\bibitem{Cacciari:2011ma}
M.~Cacciari, G.~P. Salam, and G.~Soyez, ``{FastJet User Manual},''
  \href{http://dx.doi.org/10.1140/epjc/s10052-012-1896-2}{{\em Eur. Phys. J.}
  {\bfseries C72} (2012) 1896},
\href{http://arxiv.org/abs/1111.6097}{{\ttfamily arXiv:1111.6097 [hep-ph]}}.

\bibitem{Milhano:2015mng}
J.~G. Milhano and K.~C. Zapp, ``{Origins of the di-jet asymmetry in heavy ion
  collisions},'' \href{http://dx.doi.org/10.1140/epjc/s10052-016-4130-9}{{\em
  Eur. Phys. J. C} {\bfseries 76} no.~5, (2016) 288},
  \href{http://arxiv.org/abs/1512.08107}{{\ttfamily arXiv:1512.08107
  [hep-ph]}}.

\bibitem{Chesler:2015nqz}
P.~M. Chesler and K.~Rajagopal, ``{On the Evolution of Jet Energy and Opening
  Angle in Strongly Coupled Plasma},''
  \href{http://dx.doi.org/10.1007/JHEP05(2016)098}{{\em JHEP} {\bfseries 05}
  (2016) 098}, \href{http://arxiv.org/abs/1511.07567}{{\ttfamily
  arXiv:1511.07567 [hep-th]}}.

\bibitem{Rajagopal:2016uip}
K.~Rajagopal, A.~V. Sadofyev, and W.~van~der Schee, ``{Evolution of the jet
  opening angle distribution in holographic plasma},''
  \href{http://dx.doi.org/10.1103/PhysRevLett.116.211603}{{\em Phys. Rev.
  Lett.} {\bfseries 116} no.~21, (2016) 211603},
  \href{http://arxiv.org/abs/1602.04187}{{\ttfamily arXiv:1602.04187
  [nucl-th]}}.

\bibitem{Chien:2015hda}
Y.-T. Chien and I.~Vitev, ``{Towards the understanding of jet shapes and cross
  sections in heavy ion collisions using soft-collinear effective theory},''
  \href{http://dx.doi.org/10.1007/JHEP05(2016)023}{{\em JHEP} {\bfseries 05}
  (2016) 023}, \href{http://arxiv.org/abs/1509.07257}{{\ttfamily
  arXiv:1509.07257 [hep-ph]}}.

\bibitem{Catani:1992ua}
S.~Catani, L.~Trentadue, G.~Turnock, and B.~R. Webber, ``{Resummation of large
  logarithms in e+ e- event shape distributions},''
\href{http://dx.doi.org/10.1016/0550-3213(93)90271-P}{{\em Nucl. Phys.}
  {\bfseries B407} (1993) 3--42}.

\bibitem{Mehtar-Tani:2017ypq}
Y.~Mehtar-Tani and K.~Tywoniuk, ``{Radiative energy loss of neighboring
  subjets},'' \href{http://dx.doi.org/10.1016/j.nuclphysa.2018.09.041}{{\em
  Nucl. Phys.} {\bfseries A979} (2018) 165--203},
\href{http://arxiv.org/abs/1706.06047}{{\ttfamily arXiv:1706.06047 [hep-ph]}}.

\bibitem{Dokshitzer:1991wu}
Y.~L. Dokshitzer, V.~A. Khoze, A.~H. Mueller, and S.~I. Troian, ``{Basics of
  perturbative QCD},''. Gif-sur-Yvette, France. Ed. Frontieres (1991) 274 p.

\bibitem{Dokshitzer:1997in}
Y.~L. Dokshitzer, G.~D. Leder, S.~Moretti, and B.~R. Webber, ``{Better jet
  clustering algorithms},''
  \href{http://dx.doi.org/10.1088/1126-6708/1997/08/001}{{\em JHEP} {\bfseries
  08} (1997) 001},
\href{http://arxiv.org/abs/hep-ph/9707323}{{\ttfamily arXiv:hep-ph/9707323
  [hep-ph]}}.

\bibitem{Wobisch:1998wt}
M.~Wobisch and T.~Wengler, ``{Hadronization corrections to jet cross-sections
  in deep inelastic scattering},'' in {\em {Monte Carlo generators for HERA
  physics. Proceedings, Workshop, Hamburg, Germany, 1998-1999}}, pp.~270--279.
\newblock 1998.
\newblock
\href{http://arxiv.org/abs/hep-ph/9907280}{{\ttfamily arXiv:hep-ph/9907280
  [hep-ph]}}.
\newblock

\bibitem{Catani:1993hr}
S.~Catani, Y.~L. Dokshitzer, M.~Seymour, and B.~Webber, ``{Longitudinally
  invariant $K_t$ clustering algorithms for hadron hadron collisions},''
  \href{http://dx.doi.org/10.1016/0550-3213(93)90166-M}{{\em Nucl. Phys. B}
  {\bfseries 406} (1993) 187--224}.

\bibitem{Frye:2017yrw}
C.~Frye, A.~J. Larkoski, J.~Thaler, and K.~Zhou, ``{Casimir Meets Poisson:
  Improved Quark/Gluon Discrimination with Counting Observables},''
  \href{http://dx.doi.org/10.1007/JHEP09(2017)083}{{\em JHEP} {\bfseries 09}
  (2017) 083},
\href{http://arxiv.org/abs/1704.06266}{{\ttfamily arXiv:1704.06266 [hep-ph]}}.

\bibitem{Dreyer:2018nbf}
F.~A. Dreyer, G.~P. Salam, and G.~Soyez, ``{The Lund Jet Plane},''
  \href{http://dx.doi.org/10.1007/JHEP12(2018)064}{{\em JHEP} {\bfseries 12}
  (2018) 064},
\href{http://arxiv.org/abs/1807.04758}{{\ttfamily arXiv:1807.04758 [hep-ph]}}.

\bibitem{Mehtar-Tani:2016aco}
Y.~Mehtar-Tani and K.~Tywoniuk, ``{Groomed jets in heavy-ion collisions:
  sensitivity to medium-induced bremsstrahlung},''
  \href{http://dx.doi.org/10.1007/JHEP04(2017)125}{{\em JHEP} {\bfseries 04}
  (2017) 125},
\href{http://arxiv.org/abs/1610.08930}{{\ttfamily arXiv:1610.08930 [hep-ph]}}.

\bibitem{Chatrchyan:2012gt}
{\bfseries CMS} Collaboration, S.~Chatrchyan {\em et al.}, ``{Studies of jet
  quenching using isolated-photon+jet correlations in PbPb and $pp$ collisions
  at $\sqrt{s_{NN}}=2.76$ TeV},''
  \href{http://dx.doi.org/10.1016/j.physletb.2012.11.003}{{\em Phys. Lett. B}
  {\bfseries 718} (2013) 773--794},
  \href{http://arxiv.org/abs/1205.0206}{{\ttfamily arXiv:1205.0206 [nucl-ex]}}.

\bibitem{Sirunyan:2017jic}
{\bfseries CMS} Collaboration, A.~M. Sirunyan {\em et al.}, ``{Study of Jet
  Quenching with $Z+\text{jet}$ Correlations in Pb-Pb and $pp$ Collisions at
  ${\sqrt{s}}_{NN}=5.02\text{ }\text{ }\mathrm{TeV}$},''
  \href{http://dx.doi.org/10.1103/PhysRevLett.119.082301}{{\em Phys. Rev.
  Lett.} {\bfseries 119} no.~8, (2017) 082301},
  \href{http://arxiv.org/abs/1702.01060}{{\ttfamily arXiv:1702.01060
  [nucl-ex]}}.

\bibitem{Brewer:2018dfs}
J.~Brewer, J.~G. Milhano, and J.~Thaler, ``{Sorting out quenched jets},''
  \href{http://dx.doi.org/10.1103/PhysRevLett.122.222301}{{\em Phys. Rev.
  Lett.} {\bfseries 122} no.~22, (2019) 222301},
  \href{http://arxiv.org/abs/1812.05111}{{\ttfamily arXiv:1812.05111
  [hep-ph]}}.

\bibitem{CMS:2019btm}
{\bfseries CMS} Collaboration, ``{Measurement of Jet Nuclear Modification
  Factor in PbPb Collisions at $\sqrt{s_{NN}}$ = 5.02 TeV with CMS},''
 \href{http://cds.cern.ch/record/2698506}{{\ttfamily Report number:CMS-PAS-HIN-18-014}}.

\bibitem{Haake:2019pqd}
{\bfseries ALICE} Collaboration, R.~Haake, ``{Machine Learning based jet
  momentum reconstruction in Pb-Pb collisions measured with the ALICE
  detector},'' in {\em {2019 European Physical Society Conference on High
  Energy Physics}}.
\newblock 9, 2019.
\newblock \href{http://arxiv.org/abs/1909.01639}{{\ttfamily arXiv:1909.01639
  [nucl-ex]}}.

\bibitem{Pablos:2019ngg}
D.~Pablos, ``{Jet Suppression From a Small to Intermediate to Large Radius},''
  \href{http://dx.doi.org/10.1103/PhysRevLett.124.052301}{{\em Phys. Rev.
  Lett.} {\bfseries 124} no.~5, (2020) 052301},
  \href{http://arxiv.org/abs/1907.12301}{{\ttfamily arXiv:1907.12301
  [hep-ph]}}.

\bibitem{Catani:1989ne}
S.~Catani and L.~Trentadue, ``{Resummation of the QCD Perturbative Series for
  Hard Processes},'' \href{http://dx.doi.org/10.1016/0550-3213(89)90273-3}{{\em
  Nucl.\ Phys.\ B} {\bfseries 327} (1989) 323--352}.

\bibitem{DeGrand:1978te}
T.~A. DeGrand, ``{Structure Functions of Quarks, Gluons, and Hadrons in Quantum
  Chromodynamics},'' \href{http://dx.doi.org/10.1016/0550-3213(79)90452-8}{{\em
  Nucl.\ Phys.\ B} {\bfseries 151} (1979) 485--517}.

\end{thebibliography}\endgroup

\providecommand{\href}[2]{#2}\begingroup\raggedright\endgroup

\end{document}